\documentclass[aps,pra,reprint]{revtex4-1}

\usepackage{amsfonts}
\usepackage{amsmath}
\usepackage{amsthm}
\usepackage{amscd}
\usepackage{amssymb}
\usepackage{subfigure}
\usepackage{amsxtra}
\usepackage{gensymb}
\usepackage{bm}
\usepackage{lipsum}        % bold math
\usepackage{bbm}
\usepackage{graphicx}
\usepackage{epstopdf}
\usepackage{color}
\usepackage{mathtools}
\usepackage{times}
\usepackage{mdframed}
\usepackage{array}
\usepackage{mathrsfs}
\usepackage{enumerate}
\usepackage[colorlinks=true,citecolor=blue,urlcolor=black]{hyperref}

\def\bi#1\ei {\begin{itemize}#1\end{itemize}}
\def\bn#1\en {\begin{enumerate}#1\end{enumerate}}
\def\bea#1\eea {\begin{align}#1\end{align}}
\def\bean#1\eean {\begin{align*}#1\end{align*}}
\def\ben#1\een {\begin{equation*}#1\end{equation*}}
\def\be#1\ee {\begin{equation}#1\end{equation}}
\def\bes#1\ees {\begin{equation}\begin{split}#1\end{split}\end{equation}}
\def\bear#1\eear {\begin{eqnarray}#1\end{eqnarray}}
\def\bear#1\eear {\begin{eqnarray*}#1\end{eqnarray*}}
\emergencystretch=\maxdimen
\newcommand{\beq}{\begin{equation}}
\newcommand{\eeq}{\end{equation}}

\begin{document}

\title{Finite-key security analysis for quantum key distribution with leaky sources}

\author{Weilong Wang$^1$}
\email{wwang@com.uvigo.es}
\author{Kiyoshi Tamaki$^2$}
\author{Marcos Curty$^1$}
\affiliation{$^1$EI~Telecomunicaci\'on,~Department~of~Signal~Theory~and~Communications, University~of~Vigo,~Vigo~E-36310,~Spain \\
$^2$Graduate School of Science and Engineering for Education, University of Toyama, Gofuku 3190, Toyama 930-8555, Japan}

\begin{abstract}
Security proofs of quantum key distribution (QKD) typically assume that the devices of the legitimate users are perfectly shielded from the eavesdropper. This assumption is, however, very hard to meet in practice, and thus the security of current QKD implementations is not guaranteed. Here, we fill this gap by providing a finite-key security analysis for QKD which is valid against arbitrary information leakage from the state preparation process of the legitimate users. For this, we extend the techniques introduced in~\cite{tamaki2016decoy} to the finite-key regime, and we evaluate the security of a leaky decoy-state BB84 protocol with biased basis choice, which is one of the most implemented QKD schemes today. Our simulation results demonstrate the practicability of QKD over long distances and within a reasonable time frame given that the legitimate users' devices are sufficiently isolated.
\end{abstract}
\maketitle

\section{Introduction}
\begin{figure*}[!t]
  \includegraphics*[scale=0.54]{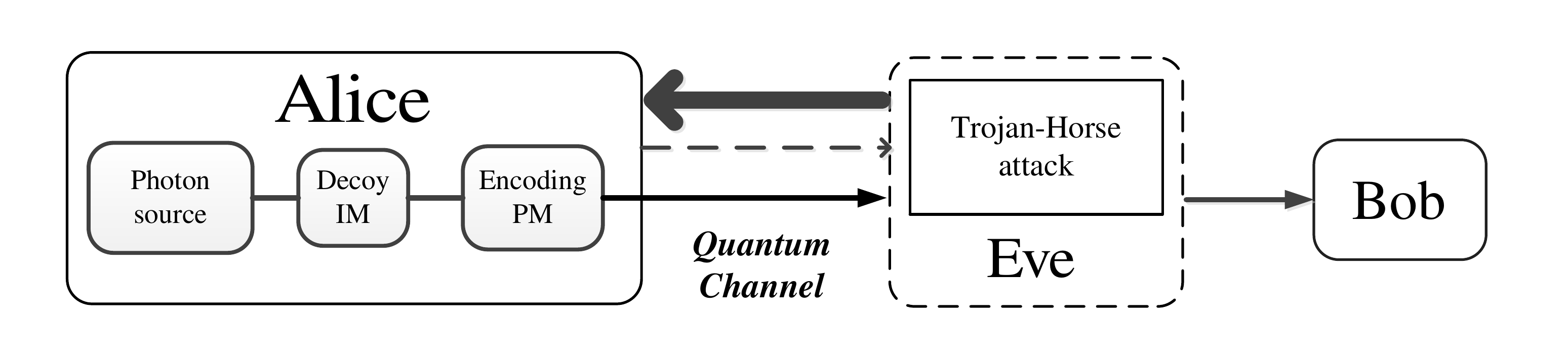}
 \caption{ \footnotesize { The sender Alice has a photon source to generate light pulses. She
uses an IM to generate decoy states and a PM for state encoding. Eve has full control of the quantum channel (thin solid arrow). In an active THA, Eve injects bright light pulses (thick solid arrow) into Alice's transmitter and then she measures the back-reflected light (thin dashed arrow}) to extract information about Alice's state preparation process.} \label{Tha}
\end{figure*}

Quantum key distribution (QKD)~\cite{bennett1984quantum} is undoubtedly the principal application of quantum cryptography nowadays, as in theory it can offer information-theoretic secure communications~\cite{scarani2009security, lo2014secure}. Unfortunately, however, real-life implementations of QKD do not typically fulfill the assumptions which are made in the security proofs and thus their security cannot be guaranteed. Indeed, one key assumption in most security proofs of QKD is that the devices of the legitimate users (commonly known as Alice and Bob) are perfectly isolated from the eavesdropper (Eve), which means that they do not leak any unwanted information to a channel. This strong assumption is, however, very difficult (if not impossible) to assure in practice. For instance, Eve could perform a so-called Trojan-horse attack (THA)~\cite{gisin2006trojan, jain2015risk} against Alice and/or Bob's devices to obtain side-channel information about their internal settings for each transmitted signal. For example, Eve could try to learn, say, the basis information and the intensity setting used by Alice to generate each of her signals in a decoy-state based QKD protocol. For this, Eve could inject bright light pulses into Alice's transmitter and then measure the back-reflected light to extract information about Alice's state preparation process. This situation is illustrated in Fig.~\ref{Tha}. Also, Eve might obtain information about the internal functioning of Alice and Bob's devices by passively monitoring, for instance, their power consumption or electromagnetic radiation.

This problem has been analysed recently in~\cite{lucamarini2015practical,tamaki2016decoy}. More precisely, the authors of~\cite{lucamarini2015practical} evaluated the security of a QKD system in the presence of information leakage from Alice's phase modulator (PM), which is used to encode the bit and basis information of the generated signals. A key observation here is that, in this scenario, the joint state of Alice's transmitted signals and Eve's back-reflected light from her THA is not basis-independent but it depends on Alice's basis choice. This means that the security of the system can be analysed using the techniques introduced in~\cite{lo2007security}. More recently, these seminal results have been generalised, and a formalism to prove the security of decoy-state QKD~\cite{hwang2003quantum, lo2005decoy, wang2005beating} in the presence of arbitrary information leakage from both the PM and the intensity modulator (IM), which is used to select the intensity setting for each emitted signal, has been introduced~\cite{tamaki2016decoy}. In so doing, it is now possible to quantify the amount of device isolation that is needed to achieve a certain performance ({\it i.e.}, a certain secret key rate at a given distance) with a realistic leaky QKD system.

While the results in~\cite{lucamarini2015practical,tamaki2016decoy} constitute an important step toward guaranteeing the security of quantum communication systems in the presence of information leakage, both analyses consider the asymptotic scenario where Alice sends Bob an infinite number of light pulses. This means that these results cannot be directly applied to real-life QKD implementations, where Alice sends Bob only a finite number of signals and they distill finite-length keys~\cite{tomamichel2012tight,hayashi2012concise,curty2014finite,lim2014concise,mizutani2015finite}. In this work, we fill this gap and we extend the general framework introduced in~\cite{tamaki2016decoy} to the finite-key scenario. For this, we present a finite-key parameter estimation method which can be applied in the presence of information leakage. In particular, and for concreteness, we consider a biased basis choice decoy-state QKD protocol~\cite{lo2005efficient,wei2013decoy} with three-intensity settings. This is one of the most implemented QKD schemes today~\cite{zhao2006experimental,peng2007experimental,schmitt2007experimental,yuan2007unconditionally,
rosenberg2007long,liu2010decoy,frohlich2017long}. Note, however, that our results could be straightforwardly adapted as well to analyze the security of other decoy-state based QKD systems.

In addition, we shall consider information leakage from both the IM and the PM of Alice's transmitter. The former implies that a key assumption of the decoy-state method is violated, as now the yield of an $n$-photon signal could depend on the intensity setting used by Alice to generate it. As a result, we have that the security analysis cannot be based on the typical counterfactual scenario where the intensity setting for each transmitted signal is selected by Alice $a~posteriori$, that is, after Bob has already detected all the incoming signals. To solve this problem, we use the trace distance argument introduced in~\cite{tamaki2016decoy}, which relates the $n$-photon yields (as well as the error rates) associated to pulses generated with different intensity settings, in combination with Azuma's inequality~\cite{azuma1967weighted}. This inequality allows us to tackle statistical fluctuations in a finite-key regime while guaranteeing security against general attacks. To include the effect of information leakage from the PM in the security analysis, we apply the quantum coin idea introduced in~\cite{gottesman2004security,koashi2005simple} to the finite key regime. For this, we modify slightly the classical post-processing steps of the QKD protocol such that the security proof can go through. More precisely, we now include a random data post-selection step before the sifting step (see Sec.~\ref{secII}). This way, we can quantify the amount of device isolation which is required to provide a certain performance, as a function of the total number of signals transmitted. As expected, the amount of isolation is inversely proportional to the number of signals transmitted. That is, for a certain intensity of Eve's back-reflected light, the resulting performance of the QKD system improves when Alice transmits more signals ({\it i.e.}, when the post-processing data block sizes increase).

The paper is organised as follows. In Sec.~\ref{secII}, we provide a brief summary of the assumptions made in the security analysis and describe the decoy-state BB84 protocol that we consider in our security analysis. Then, in Sec.~\ref{secIII} we present a finite-key parameter estimation method to determine, in the presence of information leakage from both the IM and the PM, the parameters which are needed to evaluate the secret key rate formula. Next, for illustration purpose, in Sec.~\ref{secIV} we consider three particular examples of THAs, and we evaluate the resulting secure secret key rate as a function of the isolation of the legitimate users' devices and the total number of signals transmitted. Finally, in Sec.~\ref{secV} we summarise what has been achieved. Also, the paper contains a few appendices which contain a summary of the notation used in this paper, as well as the calculations that we use to derive the results presented in the main text, together with some additional simulation results which provide further insights into the effect of information leakage in a finite-key scenario.

\section{Assumptions and decoy-state BB84 Protocol}\label{secII}

In the standard BB84 protocol with practical light sources emitting phase-randomized weak coherent pulses (WCPs), the so-called  photon-number-splitting attack~\cite{huttner1995quantum, brassard2000limitations} allows Eve to obtain full information about the part of the key generated from the multi-photon pulses without inducing any disturbance in the signal transmission. Decoy-state QKD~\cite{hwang2003quantum, lo2005decoy, wang2005beating} has been proposed to protect QKD schemes against this attack. In a typical implementation of the decoy-state BB84 QKD scheme with phase-randomized WCPs, Alice sends Bob mixtures of Fock states with different mean photon numbers, which can be described as
\begin{equation}
{\rho ^{{\gamma ^j}}} = \sum\limits_{n = 0}^\infty  {p_n^j\left| n \right\rangle \left\langle n \right|,}
\end{equation}
where $p_{n}^{j} = {\left( {{\gamma ^j}} \right)^n}{e^{ -  {{\gamma ^j} } }}/ {n!} $ is the probability that the optical pulse sent by Alice contains $n$ photons given that she selects the intensity $\gamma^j$, and $|n\rangle$ denotes a Fock state with $n$ photons. A key assumption of the decoy-state method is that both the yield and the error rate of an $n$-photon state, $|n\rangle$, are independent of the intensity setting selected by Alice. Here, the yield of $|n\rangle$ refers to the conditional probability of observing a detection event at Bob's measurement device given that Alice sends him such a state. The independence of the yields on the intensity setting selected by Alice arises from the fact that given an $n$-photon state, Eve cannot determine with certainty the intensity setting with which the state was generated, as $|n\rangle$ does not contain any information about the intensity setting. However, as pointed out in~\cite{tamaki2016decoy}, this assumption is no longer true in the presence of information leakage from the source. Since in such a scenario the state Alice sends is not in a single mode anymore, Eve might obtain partial information about Alice's intensity setting choice. This latter situation is discussed in more detail in Sec. III.

Although, in general, Alice can use an arbitrary number of intensity settings to prepare her decoy states in the protocol, for simplicity and without loss of generality, in this work we shall consider, just as an example, a three-intensity decoy-state BB84 protocol with a biased basis choice~\cite{lo2005efficient,wei2013decoy}, which is the most implemented solution for long-distance QKD experiments ~\cite{zhao2006experimental,peng2007experimental,schmitt2007experimental,yuan2007unconditionally,rosenberg2007long,liu2010decoy,frohlich2017long}. In addition, we shall assume that Alice and Bob distill a secret key only from those events where both of them select the Z basis and Alice selects the signal intensity, which typically corresponds to the largest intensity. Also, we shall consider a non-iterative sifting strategy which protects the protocol against the sifting attack~\cite{pfister2016sifting}. We remark, however, that there exist iterative sifting strategies that can also guarantee the security of QKD against the sifting attack, like the one introduced in~\cite{tamaki2018security} and they could be used here as well.

Before describing the protocol, we first present briefly the assumptions on the user's devices.
\begin{enumerate}[(1)]
  \item The global phase of each coherent pulse generated by Alice is randomized.
  \item The joint state of all the pulses generated by Alice is in a tensor product. That is, there is no correlation among pulses in the absence of information leakage.
  \item The coherent light generated by Alice does not suffer from intensity fluctuations.
  \item Perfect encoding of the bit and basis information. That is, the phase modulation values applied to the coherent pulses generated by Alice's source are exactly $\{0, \pi/2, \pi, 3\pi/2\}$.
  \item Alice's phase modulator modulates only the phase of the pulses, and her intensity modulator modulates only the intensity of the pulses. The state of a pulse emitted by Alice's source is in a single mode. That is, in the absence of information leakage, Alice's modulators do not introduce side-channels. (Note that in the presence of information leakage, the state of Alice's pulses is not in a single mode anymore.)
  \item The efficiency of Bob's detectors is basis independent for any received pulses.
\end{enumerate}

Bellow, we describe the protocol that we consider in more detail. Note, however, that the analysis presented in Sec.~III can be adapted as well to other prescriptions of the decoy-state BB84 protocol.

The protocol we consider includes an unusual step where Alice and Bob probabilistically post-select part of their data (see step 3 in the description of the protocol below), which is necessary for the finite-key security analysis to go through. Indeed, this step guarantees that the actual protocol is equivalent to the fictitious protocol that we consider in order to estimate the phase error rate in Sec III.~B~2.
\\
\\*
\textbf{\emph{Protocol}}
\begin{enumerate}
  \item \emph{State preparation.} The first two steps of the protocol are repeated $N$ times. In each round, Alice randomly selects a bit value 0 or 1. Then she probabilistically selects a basis $\chi_{\rm A} \in\{\rm Z,~\rm X\}$ with probabilities $p_{\rm Z}$ and $p_{\rm X}=1-p_{\rm Z}$, respectively, and an intensity $\gamma^j$ where $j \in \{\rm s,v,w\}$ and $\gamma^{\rm s}>\gamma^{\rm v}>\gamma^{\rm w}\geq 0$, with probability $p_{j}$. Then she prepares a phase-randomised WCP of intensity $\gamma^j$ in the chosen basis state of $\chi_{\rm A}$ and sends it to Bob via a quantum channel.
  \item \emph{Measurement.} Bob selects a basis $\chi_{\rm B} \in\{\rm Z,~\rm X\}$ with probabilities $p_{\rm Z}$ and $p_{\rm X}$, respectively, to measure the state received from Alice and records the outcome.
  \item \emph{Random data post-selection:} Once the $N$ rounds of quantum transmission and measurement have finished, Bob announces in which rounds he got a click event. For each click event, Alice selects a fictitious basis ${\rm Z}_{\rm c}$ or ${\rm X}_{\rm c}$ with probability $p_{\rm Z_{A_c}}$ and $p_{\rm X_{A_c}}=1-p_{\rm Z_{A_c}}$, respectively, and she announces her selection.
\item \emph{Sifting:} If Alice's choice was the ${\rm X}_{\rm c}$ basis, Bob declares his measurement basis choice but Alice does not declare her state preparation basis choice and then they discard the corresponding data. If Alice's choice was the ${\rm Z}_{\rm c}$ basis, both Alice and Bob declare their basis choices, that is, Alice declares the basis that she chose to prepare the state and Bob declares his measurement basis choice, and Alice also announces her intensity setting via an authenticated classical channel. We denote by $Z^j$ ($X^j$) the set of indexes that identifies the click events where Alice chose the ${\rm Z}_{\rm c}$ basis and the intensity $\gamma^j$ and both Alice and Bob chose the basis Z (X). If the sifting conditions $| Z^j | \ge N^j_{\rm Z}$ and $| X^j |\ge N^j_{\rm X}$ are satisfied for all $j$, where $N^j_{\rm Z}$ and $N^j_{\rm X}$ are predetermined threshold values, Alice and Bob proceed to execute the following steps of the protocol~\footnote{Note that Alice and Bob could also probabilistically post-select $N^j_{\rm Z}$ ($N^j_{\rm X}$) events from each set indexed by $Z^j$ ($X^j$) to fix the sizes of the sifted data sets. This might be convenient for some post-processing steps of the protocol, like information reconciliation and privacy amplification. Here, for simplicity, we do not consider such a post-selection. However, note that our analysis could be adapted as well to the situation where the the sizes of the sifted data sets are fixed $\it a~priori$.} If the sifting conditions are not satisfied, the protocol aborts.
  \item \emph{Parameter estimation.} Alice and Bob estimate a lower bound, which we denote by $N^{L}_{{\rm click,0,\gamma^s}|\rm Z}~( N^{L}_{{\rm click,1,\gamma^s}|\rm Z})$, on the number of vacuum (single-photon) click events in the sifted key data identified by the set $Z^{\rm s}$. Also they use all the data indexed by the sets $Z^j$ and $X^j$ to estimate an upper bound on the single-photon phase error rate, which we denote by $e^{\rm U}_{\rm ph}$, of the sifted key data indexed by $Z^{\rm s}$.
  \item {\it Information reconciliation and privacy amplification.} Alice and Bob perform an error correction step for a predetermined quantum bit error rate, QBER, which we denote by $E_{\rm Z}^{\rm s}$. Then, they perform an error verification step, in which Alice computes a hash of length ${\log _2}\frac{1}{{{\varepsilon _{{\rm{cor}}}}}}$ (where $\varepsilon _{{\rm{cor}}}$ is the correctness parameter of the protocol~\cite{ben2005universal,renner2005universally}) of the corrected sifted key data indexed by $Z^{\rm s}$. For this, Alice uses a random universal$_{\rm 2}$ hash function and she sends Bob both the hash value and the hash function to verify that their corrected keys are indeed identical. If this error verification step is successful, they perform a privacy amplification step by applying a random universal$_{\rm 2}$ hash function to the corrected sifted key data indexed by $Z^{\rm s}$ to distill a shorter bit string of length $\ell$ that constitutes the final secret key.
\end{enumerate}

By using the finite-key security analysis introduced in~\cite{mizutani2015finite}, we have that the length $\ell$ of the secret key is lower bounded by
\begin{equation}
\begin{array}{*{20}{l}}
\ell \geq & N^{L}_{{\rm click,0,\gamma^s}|\rm Z} + N^{L}_{{\rm click,1,\gamma^s}|\rm Z} \left[1 - H\left({e^{\rm U}_{\rm ph}}\right)\right] \\
&- leak{_{\rm EC}} - {\log _2}\frac{2}{{{\varepsilon^2 _{{\rm{sec}}}} - \varepsilon }} - {\log _2}\frac{2}{{{\varepsilon _{{\rm{cor}}}}}} ,\label{key}
\end{array}
\end{equation}
where $H( x ) =  - x{\log _2}( x ) - ( {1 - x} ){\log _2}( {1 - x} )$ is the binary Shannon entropy function, $leak_{\rm EC}$ is the amount of syndrome information revealed by Alice in the error correction step of the protocol, $\varepsilon _{{\rm{sec}}}$ is the secrecy parameter of the protocol~\cite{ben2005universal,renner2005universally}, and $\varepsilon  \leq 1 - {\varepsilon _{\rm Z,0}}{\varepsilon _{\rm Z,1}}{\varepsilon _{\rm ph,1}}$, where $\varepsilon _{\rm Z,0}$, $\varepsilon _{\rm Z,1}$ and $\varepsilon _{\rm ph,1}$ are defined as the success probabilities associated to the estimation of the parameters $N^{L}_{{\rm click,0,\gamma^s}|\rm Z}$, $N^{L}_{{\rm click,1,\gamma^s}|\rm Z}$ and $e^{\rm U}_{\rm ph}$, respectively. That is, $\varepsilon$ is the probability that at least one of the estimations $N^{L}_{{\rm click,0,\gamma^s}|\rm Z}$, $N^{L}_{{\rm click,1,\gamma^s}|\rm Z}$ and $e^{\rm U}_{\rm ph}$ is incorrect. The main goal of the next section is to estimate the parameters $N^{L}_{{\rm click,0,\gamma^s}|\rm Z}$, $N^{L}_{{\rm click,1,\gamma^s}|\rm Z}$ and $e^{\rm U}_{\rm ph}$ in the presence of information leakage.

\section{Parameter estimation in the presence of information leakage}\label{secIII}
For concreteness we shall assume that Eve launches an active THA against the main optical components of the decoy-state transmitter, which are the IM that generates pulses with different intensities, and the PM that encodes the bit and basis information into the pulses. We remark, however, that our analysis could also be applied to any passive information leakage scenario. More precisely, we shall consider that Eve sends bright light pulses into Alice's transmitter and then she measures the back-reflected light to obtain partial information about the quantum state emitted by Alice each given time. For simplicity of analysis, we assume that Eve launches THAs against the IM and PM separately. Importantly, as we will see in detail below, a THA against the IM affects the estimation of the parameters $N^{L}_{{\rm click,0,\gamma^s}|\rm Z}$, $N^{L}_{{\rm click,1,\gamma^s}|\rm Z}$ and $e^{\rm U}_{\rm ph}$, while a THA against the PM only has effect on the estimation of the parameter $e^{\rm U}_{\rm ph}$.

\subsection{Estimation of the parameters $N^{L}_{{\rm click,0,\gamma^s}|\rm Z}$ and $N^{L}_{{\rm click,1,\gamma^s}|\rm Z}$}
With her THA against the IM, Eve could obtain partial information about Alice's intensity setting choice in each round of the protocol. This violates a key assumption of the decoy-state method and, as a consequence, the typical procedure~\cite{ma2005practical} to estimate the parameters $N^{L}_{{\rm click,0,\gamma^s}|\rm Z}$ and $N^{L}_{{\rm click,1,\gamma^s}|\rm Z}$ needs to be modified in order to take this effect into account. For this we first review the results introduced in~\cite{tamaki2016decoy} which analyze this scenario in the asymptotic case where the number of signals transmitted is infinite. These seminal results provide a relationship between the expected number of click events for different intensity settings. Then we extend this analysis to the finite-key regime by mainly using Azuma's inequality~\cite{azuma1967weighted}.

Moreover, for simplicity and for the moment, we shall assume that $p_{\rm Z_{A_c}}=1$, $\it i.e.$, we disregard the `random data post-selection' step of the protocol. This is possible because this step is not needed to analyze the information leakage from the IM but it is only needed for the analysis of the information leakage from the PM.

\subsubsection{Asymptotic Case}
The key idea is rather simple. Suppose, for instance, that Alice has $m$ different intensity setting choices and let us denote the intensity set by $S=\{\gamma^1,\gamma^2,...,\gamma^m\}$. Also, let $S_{\rm 1}$ and $S_{\rm 2}$ be any two non-empty disjoint subsets of $S$ ($\it i.e.$, $S_{\rm 1}\neq\emptyset$, $S_{\rm 2}\neq\emptyset$ where $\emptyset$ denotes the empty set, and $S_{\rm 1}\cap S_{\rm 2}=\emptyset$). In a THA against the IM, Eve first prepares a probe system $E_{\rm p}$ (which might be entangled with an ancillary system, $E_{\rm a}$, which could be, for instance, a quantum memory in her laboratory) and then she sends this system to Alice. Afterwards, she performs a joint measurement on the back-reflected light from $E_{\rm p}$, which we denote by $E'_{\rm p}$, together with all the optical pulses emitted by Alice and the system $E_{\rm a}$ to try to obtain information about Alice's intensity setting choices.

By using the trace distance argument~\cite{nielsen2000quantum}, it is easy to show that when Alice sends Bob an $n$-photon state in the basis ${\chi}$ in the $i$th trial and Bob measures the incoming signal also in the ${\chi}$ basis with $\chi\in\{\rm Z,~\rm X\}$, we have that~\cite{tamaki2016decoy}
\begin{equation}
\left| {{\rm{P}}{{\rm{r}}^i}\left( {{\rm{click}}\left|n,{S_{\rm{1}}}, \chi  \right.} \right) - {\rm{P}}{{\rm{r}}^i}\left( {{\rm{click}}\left|n,{S_2}, \chi  \right.} \right)} \right| \le D_{n,{S_{\rm{1}}},{S_{\rm{2}}},\chi }^i,\label{DS1S2}
\end{equation}
where ${{\rm{P}}{{\rm{r}}^i}\left( {{\rm{click}}\left|n,{S_{\alpha}}, \chi  \right.} \right)}$ is the conditional probability that Bob's detectors click in the $i$th trial given that Alice selects an intensity in $S_{\alpha}$ (with $\alpha \in \{1,~2\}$) and sends Bob an $n$-photon pulse and Bob selects the $\chi$ basis. Denote by $\rho _{n,\chi}^{{\gamma ^l},i}$ the normalized state which is the joint state of Alice's $n$-photon pulse given that she selects the intensity $\gamma^l$ and the basis $\chi$ in the $i$th trial, and Eve's systems $E_{\rm a}$ and $E'_{\rm p}$. Then we have
\begin{equation}
\rho _{n,\chi }^{{S_\alpha },i} =\frac{1}{\cal N} \sum\limits_{l|{\gamma ^l} \in {S_\alpha }} {\left( {{p_l}p_n^l\rho _{n,\chi }^{{\gamma ^l},i}} \right)},
\end{equation}
where $p_n^l$ is the conditional probability that Alice emits a pulse with $n$ photons given that she chooses the intensity setting $l$ and $\mathcal{N}=\sum_{l}{{p_l}p_n^l}$ is the normalization factor. That is, the normalized state $\rho _{n,\chi}^{{S_{\alpha},i} }$ denotes the joint state of Alice's $n$-photon pulse in the $i$th trial when she selects an intensity in the subset $S_{\alpha}$ and the basis $\chi$, and Eve's systems $E_{\rm a}$ and $E'_{\rm p}$. The parameter $D_{n,{S_{\rm{1}}},{S_{\rm{2}},\chi}}^i$, on the other hand, denotes the trace distance between the states $\rho _{n,\chi}^{{S_{\rm{1}}},i}$ and $\rho _{n,\chi}^{{S_{\rm{2}}},i}$ and it is given by
\begin{equation}
D_{n,{S_{\rm{1}}},{S_{\rm{2}}},\chi }^i: = \frac{1}{2}{\rm{Tr}}\left[ {\sqrt {{{\left(\rho _{n,\chi }^{{S_{\rm{1}}},i} - \rho _{n,\chi }^{{S_{\rm{2}}},i}\right)}^2}} } \right].
\end{equation}

For simplicity and without loss of generality, from now on we focus on the three-intensity case which we have described in the previous section. This means, in particular, that we can assume that $S_{\rm 1}=\{\gamma^j\}$ and $S_{\rm 2}=\{\gamma^k,\gamma^l\}$ with $j,k,l \in \{\rm s,v,w\}$ and where $k$ might be equal to $l$. Then Eq.~(\ref{DS1S2}) can be rewritten as
\begin{eqnarray} \label{DJS2}
&&\left| {{\rm{P}}{{\rm{r}}^i}\left( {{\rm{click}}\left| {n,{S_{\rm{1}}},\chi } \right.} \right) - {\rm{P}}{{\rm{r}}^i}\left( {{\rm{click}}\left| {n,{S_2},\chi } \right.} \right)} \right| \nonumber \\
&&= |{\rm{P}}{{\rm{r}}^i}\left( {{\rm{click}}\left| {n,{\gamma ^j},\chi } \right.} \right) - [{q_{nkl}}{\rm{P}}{{\rm{r}}^i}\left( {{\rm{click}}\left| {n,{\gamma ^k},\chi } \right.} \right)\nonumber \\
&&+(1 - {q_{nkl}}){\rm{P}}{{\rm{r}}^i}\left( {{\rm{click}}\left| {n,{\gamma ^l},\chi } \right.} \right)]| \nonumber \\
&&\le D_{n,\{ {\gamma ^j}\} ,\{ {\gamma ^k},{\gamma ^l}\} ,\chi }^i,
\end{eqnarray}
for $k,l\neq j$ and where we have used the fact that ${\rho _{n,\chi }^{{S_{\rm{2}}},i} }=q_{nkl}{\rho _{n,\chi }^{{\gamma ^k},i} }+(1-q_{nkl}){\rho _{n,\chi }^{{\gamma ^l},i}}$ with $q_{nkl}: = {p_k}p_n^{{k}}/( {{p_k}p_n^{{k}} + {p_l}p_n^{{l}}}) $. By multiplying both sides of Eq.~(\ref{DJS2}) by ${{p_j}p_n^j}$ and taking the sum over $i = \{ 1,2,...,{N_\chi }\}$ with $N_{\chi}$ being the number of events where Alice sends a pulse in the $\chi$ basis and Bob measures it also in the $\chi$ basis, we obtain
 \begin{eqnarray}\label{THA}
&&{\bigg|\sum\limits_{i = 1}^{{N_\chi }} {{\rm{P}}{{\rm{r}}^i}\left( {{\rm{click}},n,{\gamma ^j}\left| \chi  \right.} \right)}  - {p_j}p_n^j\sum\limits_{i = 1}^{{N_\chi }} \bigg[ {q_{nkl}}}\nonumber \\
&&{ \times \frac{{{\rm{P}}{{\rm{r}}^i}\left( {{\rm{click}},n,{\gamma ^k}\left| \chi  \right.} \right)}}{{{p_k}p_n^k}} + \left( {1 - {q_{nkl}}} \right)\frac{{{\rm{P}}{{\rm{r}}^i}\left( {{\rm{click}},n,{\gamma ^l}\left| \chi  \right.} \right)}}{{{p_l}p_n^l}}\bigg]\bigg|}\nonumber\\
 &&\le {{p_j}p_n^j{N_\chi }{D_{n,\{ {\gamma ^j}\} ,\{ {\gamma ^k},{\gamma ^l}\} ,\chi }},}
\end{eqnarray}
where ${\rm Pr}^i\left( {{\rm click},n,\gamma ^{j}} \left| \chi  \right.\right)$ denotes the conditional probability that in the $i$th trial Alice selects the intensity $\gamma^j$ and sends Bob an $n$-photon pulse, and Bob's detectors click given that both Alice and Bob select the $\chi$ basis, and \begin{equation}
{D_{n,\{{\gamma ^j}\},\{ {\gamma ^k},{\gamma ^l}\} ,\chi }} = \frac{1}{{{N_\chi }}}\sum\limits_{i = 1}^{{N_\chi }} {D_{n,\{{\gamma ^j}\},\{ {\gamma ^k},{\gamma ^l}\} ,\chi }^i}.
\end{equation}

The quantity ${\sum_{i = 1}^{N_{\chi}} {{\rm Pr}^i\left( {{\rm click},n,\gamma ^{j}} \left| \chi  \right.\right)}}$ represents the conditional expected number of events where Alice selects the intensity $\gamma^j$ and sends Bob an $n$-photon pulse, and Bob's detectors click given that both Alice and Bob select the $\chi$ basis. We will denote it by $\mathcal{E}_{{\rm{click}},n,{\gamma ^j}|\chi}$ and, with this notation, Eq.~(\ref{THA}) has the following form:
\begin{eqnarray}\label{ND}
&&\Big| \mathcal{E}_{{\rm{click}},n,{\gamma ^j}|\chi} - \Big[ q_{nkl}\frac{p_jp_n^j}{p_kp_n^k}\mathcal{E}_{{\rm{click}},n,{\gamma ^k}|\chi}\nonumber \\
&&+ \left( {1 - {q_{nkl}}} \right)\frac{p_jp_n^j}{p_lp_n^l}\mathcal{E}_{{\rm{click}},n,{\gamma ^l}|\chi} \Big] \Big|\nonumber \\
&&\le p_jp_n^jN_{\chi}{D_{n,\{ {\gamma ^j}\} ,\{ {\gamma ^k},{\gamma ^l}\} ,\chi }}.
\end{eqnarray}

From Eq. (\ref{ND}), it is now straightforward to obtain the expressions that relate the expected number of click events associated to different intensity settings. For this, let us first consider the case where $l=k$. That is, here Eve wants to discriminate between any pair of possible intensity settings. In this case Eq. (\ref{ND}) can be rewritten as follows:
\begin{equation}\label{YD}
\left| \mathcal{E}_{{\rm{click}},n,{\gamma ^j}|\chi} - \frac{p_jp_n^j}{p_kp_n^k}\mathcal{E}_{{\rm{click}},n,{\gamma ^k}|\chi} \right| \le p_jp_n^jN_{\chi}{D_{n,\{ {\gamma ^j}\} ,\{ {\gamma ^k}\} ,\chi }},
\end{equation}
where
\begin{equation}
\begin{array}{*{20}{l}}
{D_{n,\{ {\gamma ^j}\} ,\{ {\gamma ^k}\} ,\chi }} &= \frac{1}{{{N_\chi }}}\sum\limits_{i = 1}^{{N_\chi }} {D_{n,\{ {\gamma ^j}\} ,\{ {\gamma ^k}\} ,\chi }^i} \\
&: = \frac{1}{{2{N_\chi }}}\sum\limits_{i = 1}^{{N_\chi }} {{\rm{Tr}}} \left[ \sqrt{\left({\rho _{n,\chi }^{{\gamma ^j},i}-\rho _{n,\chi }^{{\gamma ^k},i}}\right)^2} \right].
\end{array}
\end{equation}

Eq~(\ref{YD}) can be equivalently written as:
\begin{equation}\label{YDel}
\mathcal{E}_{{\rm{click}},n,{\gamma ^j}|\chi} = \frac{p_jp_n^j}{p_kp_n^k}\mathcal{E}_{{\rm{click}},n,{\gamma ^k}|\chi} + {\Delta_{\chi,n} ^{jk}},
\end{equation}
where ${\Delta_{\chi,n} ^{jk}}$ lies in an interval $\left[ { - p_jp_n^jN_{\chi}{D_{n,\{ {\gamma ^j}\} ,\{ {\gamma ^k}\} ,\chi }},~p_jp_n^jN_{\chi}{D_{n,\{ {\gamma ^j}\} ,\{ {\gamma ^k}\} ,\chi }}} \right]$. Note that the situation where Alice's transmitters are perfectly shielded from Eve corresponds to the case where  ${\Delta_{\chi,n} ^{jk}}=0$ and then, as expected, $\mathcal{E}_{{\rm{click}},n,{\gamma ^j}|\chi} = \frac{p_jp_n^j}{p_kp_n^k}\mathcal{E}_{{\rm{click}},n,{\gamma ^k}|\chi}$ for all $j,k \in\{\rm{s,v,w}\}$.

Similarly, by taking, for instance, $\left\{j={\rm v},~k=\rm s\right\}$ and $\left\{j={\rm w},~k=\rm s\right\}$, we obtain the following two equations:
\begin{equation}
\begin{array}{l} \label{Ys}
\mathcal{E}_{{\rm{click}},n,{\gamma ^{\rm{v}}}|\chi} =\frac{p_{\rm v}p_n^{\rm v}}{p_{\rm s}p_n^{\rm s}} \mathcal{E}_{{\rm{click}},n,{\gamma ^{\rm{s}}}|\chi} + {\Delta_{\chi,n} ^{\rm vs}},\\
\mathcal{E}_{{\rm{click}},n,{\gamma ^{\rm{w}}}|\chi} = \frac{p_{\rm w}p_n^{\rm w}}{p_{\rm s}p_n^{\rm s}}\mathcal{E}_{{\rm{click}},n,{\gamma ^{\rm{s}}}|\chi} + {\Delta_{\chi,n} ^{\rm ws}}.
\end{array}
\end{equation}
That is, we can relate $\mathcal{E}_{{\rm{click}},n,{\gamma ^{\rm{v}}}|\chi}$ and $\mathcal{E}_{{\rm{click}},n,{\gamma ^{\rm{w}}}|\chi}$ to $\mathcal{E}_{{\rm{click}},n,{\gamma ^{\rm{s}}}|\chi}$ by means of the deviation terms ${\Delta_{\chi,n} ^{\rm vs}}$ and ${\Delta_{\chi,n} ^{\rm ws}}$ which arise from information leaked by the IM. We remark that by using different combinations of $j$ and $k$, one can obtain more constraints which are similar to those given by Eq. (\ref{Ys}). In any case, our simulation results suggest that the contribution of these additional constraints is negligible once one has already taken into account those imposed by Eq. (\ref{Ys}).

To conclude the analysis in the asymptotic scenario, let us now consider the case where $k\neq l$. In this situation, it is easy to show that Eq. (\ref{ND}) implies the following:
\begin{equation}
\begin{array}{*{20}{l}}\label{Ysvw}
\mathcal{E}_{{\rm{click}},n,{\gamma ^{\rm{s}}}|\chi}= &{q_{n\rm vw}}\frac{p_{\rm s}p_n^{\rm s}}{p_{\rm v}p_n^{\rm v}}\mathcal{E}_{{\rm{click}},n,{\gamma ^{\rm{v}}}|\chi}\\
 &+ \left( {1 - {q_{n\rm vw}}} \right)\frac{p_{\rm s}p_n^{\rm s}}{p_{\rm w}p_n^{\rm w}}\mathcal{E}_{{\rm{click}},n,{\gamma ^{\rm{w}}}|\chi} + {\Delta_{\chi,n} ^{\rm svw}},\\
\mathcal{E}_{{\rm{click}},n,{\gamma ^{\rm{v}}}|\chi}= &{q_{n\rm sw}}\frac{p_{\rm v}p_n^{\rm v}}{p_{\rm s}p_n^{\rm s}}\mathcal{E}_{{\rm{click}},n,{\gamma ^{\rm{s}}}|\chi}\\
 &+ \left( {1 - {q_{n\rm sw}}} \right)\frac{p_{\rm v}p_n^{\rm v}}{p_{\rm w}p_n^{\rm w}}\mathcal{E}_{{\rm{click}},n,{\gamma ^{\rm{w}}}|\chi} + {\Delta_{\chi,n} ^{\rm vsw}},\\
\mathcal{E}_{{\rm{click}},n,{\gamma ^{\rm{w}}}|\chi}= &{q_{n\rm sv}}\frac{p_{\rm w}p_n^{\rm w}}{p_{\rm s}p_n^{\rm s}}\mathcal{E}_{{\rm{click}},n,{\gamma ^{\rm{s}}}|\chi}\\
& + \left( {1 - {q_{n\rm sv}}} \right)\frac{p_{\rm w}p_n^{\rm w}}{p_{\rm v}p_n^{\rm v}}\mathcal{E}_{{\rm{click}},n,{\gamma ^{\rm{v}}}|\chi} + {\Delta_{\chi,n} ^{\rm wsv}},
\end{array}
\end{equation}
where $\Delta_{\chi,n} ^{jkl}$ lies in an interval $\left[-p_jp_n^jN_{\chi}D_{n,\{ {\gamma ^j}\} ,\{ {\gamma ^k},{\gamma ^l}\} ,\chi },~p_jp_n^jN_{\chi}D_{n,\{ {\gamma ^j}\} ,\{ {\gamma ^k},{\gamma ^l}\} ,\chi }\right]$.

By combining equations (\ref{Ys}) and (\ref{Ysvw}), and by taking also into account the bounds on all the parameters $\Delta_{\chi,n} ^{jkl}$, we find that these deviation parameters have to fulfill the following conditions:
\begin{eqnarray}\label{Dsvw}
&&\left| {{q_{n{\rm{vw}}}}\frac{{{p_{\rm{s}}}p_n^{\rm{s}}}}{{{p_{\rm{v}}}p_n^{\rm{v}}}}\Delta _{\chi ,n}^{{\rm{vs}}} + \left( {1 - {q_{n{\rm{vw}}}}} \right)\frac{{{p_{\rm{s}}}p_n^{\rm{s}}}}{{{p_{\rm{w}}}p_n^{\rm{w}}}}\Delta _{\chi ,n}^{{\rm{ws}}}} \right|\nonumber\\
 &&\le {p_{\rm{s}}}p_n^{\rm{s}}{N_\chi }{D_{n,\{ {\gamma ^{\rm{s}}}\} ,\{ {\gamma ^{\rm{v}}},{\gamma ^{\rm{w}}}\} ,\chi }},\nonumber\\
&&\left| {\Delta _{\chi ,n}^{{\rm{vs}}} - \left( {1 - {q_{n{\rm{sw}}}}} \right)\frac{{{p_{\rm{v}}}p_n^{\rm{v}}}}{{{p_{\rm{w}}}p_n^{\rm{w}}}}\Delta _{\chi ,n}^{{\rm{ws}}}} \right| \nonumber\\ &&\le {p_{\rm{v}}}p_n^{\rm{v}}{N_\chi }{D_{n,\{ {\gamma ^{\rm{v}}}\} ,\{ {\gamma ^{\rm{s}}},{\gamma ^{\rm{w}}}\} ,\chi }},\nonumber\\
&&\left| {\Delta _{\chi ,n}^{{\rm{ws}}} - \left( {1 - {q_{n{\rm{sv}}}}} \right)\frac{{{p_{\rm{w}}}p_n^{\rm{w}}}}{{{p_{\rm{v}}}p_n^{\rm{v}}}}\Delta _{\chi ,n}^{{\rm{vs}}}} \right| \nonumber\\
&&\le {p_{\rm{w}}}p_n^{\rm{w}}{N_\chi }{D_{n,\{ {\gamma ^{\rm{w}}}\} ,\{ {\gamma ^{\rm{s}}},{\gamma ^{\rm{v}}}\} ,\chi }}.
\end{eqnarray}

In the asymptotic limit where Alice sends an infinite number of pulses to Bob, thanks to Azuma's inequality~\cite{azuma1967weighted}, the actual numbers converge to the expected numbers of events. However, in the finite key regime, there is some deviation between these two quantities, and this is analyzed in the next section.

\subsubsection{Finite-key Regime}

The analysis from the previous section can be easily extended to the realistic finite-key scenario where Alice sends Bob a finite number $N$ of pulses. For this, we use Azuma's inequality~\cite{azuma1967weighted}. This inequality allows us to relate the probability of certain events to the actual number of such events in a finite scenario even when there are arbitrary correlations between different trials due to Eve's actions. It states that if a sequence of random variables satisfies the martingale and the bounded difference conditions, then
\begin{equation}\label{Azuma}
{\mathcal{E}_\lambda } \equiv \sum^N_{i=1}{\rm Pr}^{i}(\lambda|\lambda_{\overrightarrow {i - 1}  })= {{N_\lambda }} + {\delta _\lambda },
\end{equation}
where ${\mathcal{E}_\lambda }$ denotes the expected number of times that the event `$\lambda$' occurs in $N$ trials, ${\rm Pr}^{i}(\lambda|\lambda_{\overrightarrow {i - 1}})$ is the conditional probability to observe the event `$\lambda$' in the $i$th trial given the results of the first $i-1$ trials, ${{N_\lambda }} $ represents the actual number of times that the event `$\lambda$' occurs in $N$ trials in an actual experiment, and the parameter $\delta_\lambda$ denotes the deviation term between the expected number and the actual number of times that the event `$\lambda$' occurs due to statistical fluctuations. Importantly, according to Azuma's inequality we have that the quantity $\delta_\lambda$ lies in an interval $[ { - \Delta_\lambda,\widehat{\Delta} _\lambda} ]$ except for a small error probability $\varepsilon_\lambda+\widehat{\varepsilon}_\lambda$, where the bounds $\Delta _\lambda$ and $\widehat{\Delta} _\lambda$ are given by $\Delta _\lambda = {f}( {{N },\varepsilon _\lambda} )$ and $\widehat{\Delta} _\lambda = {f}( {{N},\widehat{\varepsilon} _\lambda})$, respectively, with the function ${f}( {x,y} ) = \sqrt {2x\ln {1/y}}$. We refer the reader to Appendix A for more details about Azuma's inequality.

To use Azuma's inequality in our analysis, let `$\lambda$' be the event where Alice selects the intensity $\gamma^j$ and sends Bob a state, and Bob's detectors click given that both Alice and Bob select the $\chi$ basis. Then, from Eq.~(\ref{Azuma}) we have that
\begin{equation}\label{D}
\begin{array}{*{20}{l}}
{\mathcal{E}_{{\rm{click}},{\gamma ^j}|\chi}} &\equiv \sum\limits_{i = 1}^{{N_\chi }} {{\rm{P}}{{\rm{r}}^i}( {{\rm{click}},{\gamma ^j}\left|\chi,\lambda_{\overrightarrow {i - 1}} \right.} )} \\
& =  N_{{\rm{click}},{\gamma ^j}|\chi}  + \delta _\chi ^j.
\end{array}
\end{equation}
Here we restrict ourselves to the $actual$ number of trials, $N_{\chi}$, where both Alice and Bob select the $\chi$ basis. The quantity ${\rm Pr}^i({{\rm click},\gamma ^{j}}|\chi,\lambda_{\overrightarrow {i - 1}})$ denotes the conditional probability that in the $i$th trial Alice selects the intensity $\gamma^j$ and sends Bob a state, and Bob's detectors click given that both Alice and Bob select the $\chi$ basis and conditional on the outcomes obtained in the first $i-1$ trials, and $ N_{{\rm{click}},{\gamma ^j}|\chi} $ is the actual number of such events in $N_{\chi}$ trials. The parameter $\delta_{\chi}^j$ lies in an interval $[ { - \Delta _{\chi} ^j,\widehat{\Delta} _{\chi} ^j} ]$ except for a small error probability $\varepsilon_{\chi}^j+\widehat{\varepsilon}_{\chi}^j$ where the bounds $ \Delta _{\chi} ^j$ and $\widehat{\Delta} _{\chi} ^j$ can be directly calculated from Azuma's inequality.

On the other hand, we also have that
\begin{equation}\label{Nn}
\begin{array}{*{20}{l}}
 \mathcal{E}_{{\rm{click}},n,{\gamma ^j}|\chi}&\equiv \sum\limits_{i = 1}^{{N_\chi }} {{\rm{P}}{{\rm{r}}^i}( {{\rm{click}},n,{\gamma ^j}\left| \chi,\lambda_{\overrightarrow {i - 1}} \right.} )}\\
 &={ N_{{\rm{click}},n,{\gamma ^j}|\chi}  + \delta _{\chi ,n}^j},
\end{array}
\end{equation}
where $ N_{{\rm{click}},n,{\gamma ^j}|\chi} $ represents the actual number of events where Alice selects the intensity $\gamma ^{j}$ and sends Bob an $n$-photon pulse, and Bob's detectors click within $N_{\chi}$ trials given that both Alice and Bob select the $\chi$ basis, and $\delta _{\chi,n} ^j$ denotes a deviation term which lies in an interval $[ { - \Delta _{\chi,n} ^j,\widehat{\Delta} _{\chi,n} ^j} ]$ except for a small error probability $\varepsilon_{\chi,n}^j+\widehat{\varepsilon}_{\chi,n}^j$.

From Eqs (\ref{D}) and (\ref{Nn}), we have therefore that
\begin{equation}\label{Dn}
\begin{array}{*{20}{l}}
\mathcal{E}_{{\rm{click}},{\gamma ^j}|\chi}  &= \sum\limits_{n = 0}^\infty  {\mathcal{E}_{{\rm{click}},n,{\gamma ^j}|\chi} } \\&= \sum\limits_{n = 0}^\infty  { N_{{\rm{click}},n,{\gamma ^j}|\chi} }  + \sum\limits_{n = 0}^\infty  {\delta _{\chi,n} ^j}.
\end{array}
\end{equation}
That is, $ N_{{\rm{click}},{\gamma ^j}|\chi} =\sum\limits_{n = 0}^\infty  { N_{{\rm{click}},n,{\gamma ^j}|\chi} }$ and $\delta_{\chi}^j=\sum\limits_{n = 0}^\infty  {\delta _{\chi,n} ^j}$.

Given the above linear equations that relate the expected number of events to the actual number of them, we can estimate the parameters $N^{L}_{{\rm click,0,\gamma^s}|\rm Z}$ and $N^{L}_{{\rm click,1,\gamma^s}|\rm Z}$ by solving a linear optimization problem. More precisely, if we combine Eqs~(\ref{Ys}) and (\ref{Dn}), we obtain
\begin{equation}\label{Nsvw}
\begin{array}{*{20}{l}}
{\mathcal{E}_{{\rm{click}},{\gamma ^{\rm{s}}}|\chi}}& = &\sum\limits_{n = 0}^\infty  {{\mathcal{E}_{{\rm{click}},n,{\gamma ^{\rm{s}}}|\chi}}}\\&=&\sum\limits_{n = 0}^\infty  \left( N_{{\rm{click}},n,{\gamma ^{\rm s}}|\chi} +\delta _{\chi,n} ^{\rm{s}}\right)  ,\\
{\mathcal{E}_{{\rm{click}},{\gamma ^{\rm{v}}}|\chi}}& = &\sum\limits_{n = 0}^\infty  {{\mathcal{E}_{{\rm{click}},n,{\gamma ^{\rm{v}}}|\chi}}}  \\&=& \sum\limits_{n = 0}^\infty  {\left( {\frac{{{p_{\rm{v}}}p_n^{\rm{v}}}}{{{p_{\rm{s}}}p_n^{\rm{s}}}}{\mathcal{E}_{{\rm{click}},n,{\gamma ^{\rm{s}}}|\chi}} + \Delta _{\chi ,n}^{{\rm{vs}}}} \right)}\\
&=&\sum\limits_{n = 0}^\infty  \left( \frac{{{p_{\rm{v}}}p_n^{\rm{v}}}}{{{p_{\rm{s}}}p_n^{\rm{s}}}}{\mathcal{E}_{{\rm{click}},n,{\gamma ^{\rm{s}}}|\chi}}\right) + \Delta _{\chi}^{{\rm{vs}}}\\&=&\sum\limits_{n = 0}^\infty   \frac{{{p_{\rm{v}}}p_n^{\rm{v}}}}{{{p_{\rm{s}}}p_n^{\rm{s}}}}\left( N_{{\rm{click}},n,{\gamma ^{\rm s}}|\chi} +\delta _{\chi,n} ^{\rm{s}}\right) + \Delta _{\chi}^{{\rm{vs}}},\\
{\mathcal{E}_{{\rm{click}},{\gamma ^{\rm{w}}}|\chi}}& =& \sum\limits_{n = 0}^\infty  {\mathcal{E}_{{\rm{click}},n,{\gamma ^{\rm{w}}}|\chi}} \\&=& \sum\limits_{n = 0}^\infty  {\left( {\frac{{{p_{\rm{w}}}p_n^{\rm{w}}}}{{{p_{\rm{s}}}p_n^{\rm{s}}}}{\mathcal{E}_{{\rm{click}},n,{\gamma ^{\rm{s}}}|\chi}} + \Delta _{\chi ,n}^{{\rm{ws}}}} \right)}  \\
&=&\sum\limits_{n = 0}^\infty  \left( \frac{{{p_{\rm{w}}}p_n^{\rm{w}}}}{{{p_{\rm{s}}}p_n^{\rm{s}}}}{\mathcal{E}_{{\rm{click}},n,{\gamma ^{\rm{s}}}|\chi}}\right) + \Delta _{\chi}^{{\rm{ws}}}\\&=&\sum\limits_{n = 0}^\infty   \frac{{{p_{\rm{w}}}p_n^{\rm{w}}}}{{{p_{\rm{s}}}p_n^{\rm{s}}}}\left( N_{{\rm{click}},n,{\gamma ^{\rm s}}|\chi} +\delta _{\chi,n} ^{\rm{s}}\right) + \Delta _{\chi}^{{\rm{ws}}},
\end{array}
\end{equation}
where $\Delta _{\chi}^{{k{\rm s}}} = \sum\limits_{n = 0}^\infty  {\Delta _{{\chi},n}^{{k{\rm s}}}}$ with $k\in\{\rm v,w\}$. Finally, by combining these results with Eq~(\ref{D}), we obtain the following linear constraints:
\begin{equation}\label{LProg}
\begin{array}{*{20}{l}}
&{N_{{\rm{click}},{\gamma ^{\rm s}}|\chi} } =  \sum\limits_{n = 0}^\infty  {\left( { N_{{\rm{click}},n,{\gamma ^{\rm s}}|\chi}  + \delta _{\chi ,n}^{\rm{s}}} \right) - \delta _{\chi}^{\rm{s}}} ,\\
&{ N_{{\rm{click}},{\gamma ^{\rm v}}|\chi} }\\
&={\sum\limits_{n = 0}^\infty  {\frac{{{p_{\rm{v}}}p_n^{\rm{v}}}}{{{p_{\rm{s}}}p_n^{\rm{s}}}}} \left( { N_{{\rm{click}},n,{\gamma ^{\rm s}}|\chi}  + \delta _{\chi ,n}^{\rm{s}}} \right) + \Delta _\chi ^{{\rm{vs}}} - \delta _{\chi}^{\rm{v}},}\\
&{N_{{\rm{click}},{\gamma ^{\rm w}}|\chi}} \\
&={\sum\limits_{n = 0}^\infty  {\frac{{{p_{\rm{w}}}p_n^{\rm{w}}}}{{{p_{\rm{s}}}p_n^{\rm{s}}}}} \left( {N_{{\rm{click}},n,{\gamma ^{\rm s}}|\chi} + \delta _{\chi ,n}^{\rm{s}}} \right) + \Delta _\chi ^{{\rm{ws}}} - \delta _{\chi}^{\rm{w}}},\\
\end{array}
\end{equation}
where the bounds on the parameters $\delta _{\chi ,n}^{\rm{s}}$ and $\delta _{\chi}^{l}$ with $l\in \{\rm s,v,w\}$ can be calculated by using Azuma's inequality and the bounds on the parameters $\Delta _\chi ^{k{\rm s}}$ are given by $-\sum_{n = 0}^{\infty}{p_{k}p_n^{k}N_{\chi}D_{n,k,\rm {s}}}\leq \Delta_{{\chi}}^{k{\rm s}}\leq \sum_{n = 0}^{\infty}{p_{k}p_n^{k}N_{\chi}D_{n,k,\rm {s}}}$. Importantly, Eq.~(\ref{LProg}) relates the actual observed quantities $N_{{\rm{click}},{\gamma ^{j}}|\chi}$ to the quantities that we want to estimate, $N_{{\rm{click}},n,{\gamma ^{\rm s}}|\chi}$. For instance, to obtain $N^{L}_{{\rm click,0,\gamma^s}|\rm Z}$ ($N^{L}_{{\rm click,1,\gamma^s}|\rm Z}$), we set $\rm \chi=Z$ and solve a linear program that minimizes $ N_{{\rm{click}},0,{\gamma ^{\rm s}}|\chi} $ ($ N_{{\rm{click}},1,{\gamma ^{\rm s}}|\chi} $) given basically the constraints imposed by Eq.~(\ref{LProg}). Such a linear program can be solved either analytically or numerically. In this work, we employ the numerical approach. For this, since there are an infinite number of unknown variables in Eq. (\ref{LProg}), it is necessary to reduce the variables to a finite set. More technical details about how this is done can be found in Appendix B.

\subsection{Estimation of the parameter $e^{\rm U}_{\rm ph}$}
The definition of the phase error rate, $e_{\rm ph}$, is given by
\begin{equation}\label{eph}
{e_{{\rm{ph}}}} = \frac{{{{N_{{\rm{phase~error}}}}}}}{{ N_{{\rm{click}},1,{\gamma ^{\rm s}}|\rm Z} }} \le \frac{N_{\rm{phase~error}}^{\rm{U}}}{{{{N^{L}_{{\rm click,1,\gamma^s}|\rm Z}}}}} \equiv e_{{\rm{ph}}}^{\rm{U}},
\end{equation}
where $N_{{\rm{phase~error}}}$ denotes the number of phase errors in the set indexed by $Z{\rm_ 1^s}$, $\it i.e.$, within the single-photon click events where Alice uses the signal intensity setting and both Alice and Bob select the Z basis. The parameter $N_{\rm{phase~error}}^{\rm{U}}$ denotes an upper bound on $N_{\rm{phase~error}}$. A phase error is defined as the fictitious error that Alice and Bob would observe in a fictitious scenario (where Alice prepares the single-photon states in the Z basis by first generating a bipartite entangled state and then measuring the ancillary system of such a state in the Z basis) if Alice measured her ancillary system in the X basis instead of using the Z basis, and Bob also measured the incoming signal in the X basis, rather than in the Z basis. This will be further explained below.

To estimate $N_{\rm{phase~error}}^{\rm{U}}$ we will evaluate two possible scenarios. In the first one we consider information leakage only from the IM, and in the second scenario we consider information leakage from both the IM and the PM.

\subsubsection{Information leakage only from the IM}
Here we follow the same procedure used in~\cite{tomamichel2012tight,curty2014finite}. That is, we first estimate a lower bound on the number of single-photon click events in the X basis with intensity $\gamma^{\rm s}$, which we denote by $N^{L}_{{\rm click,1,\gamma^s}|\rm X}$, and an upper bound on the corresponding number of errors in the single-photon states in the X basis with intensity $\gamma^{\rm s}$, which we denote by $N^{U}_{{\rm error,1,\gamma^s}|\rm X}$. These estimations can be done by using a similar method like that used to calculate $N^{L}_{{\rm click,0,\gamma^s}|\rm Z}$ and $N^{L}_{{\rm click,1,\gamma^s}|\rm Z}$. Next we apply a random sampling argument. Specifically, from $N^{L}_{{\rm click,1,\gamma^s}|\rm X}$, $N^{U}_{{\rm error,1,\gamma^s}|\rm X}$ and the previously estimated quantity $N^{L}_{{\rm click,1,\gamma^s}|\rm Z}$ we can estimate an upper bound on the number of phase errors in the data indexed by the set $ Z^{\rm s}_{\rm 1} $ to obtain $N_{\rm{phase~error}}^{\rm U}$. This is possible because here we assume that the single-photon emissions are basis independent.

More precisely, $N^{L}_{{\rm click,1,\gamma^s}|\rm X}$ can be derived by using the same linear program which is used to calculate $N^{L}_{{\rm click,1,\gamma^s}|\rm Z}$ after replacing all the parameters and variables in the Z basis with those in the X basis. To derive $N^{U}_{{\rm error,1,\gamma^s}|\rm X}$, we now focus on the error events instead of the click events. In so doing, it is straightforward to show that one can obtain the following equations which are similar to those given by Eq.~(\ref{LProg}):
\begin{equation}
\begin{array}{*{20}{l}}  \label{LproE}
{N_{{\rm error,\gamma^s}|\rm X}} = &{\sum\limits_{n = 0}^\infty  {\left( {N_{{{\rm error},n,\gamma^s}|\rm X} + {\delta _{{\rm{E}}_{{\rm X} ,n}^{\rm{s}}}}} \right) - {\delta _{{\rm{E}}_{\rm{X}}^{\rm{s}}}}} ,}\\
{N_{{{\rm error},\gamma^v}|\rm X}} = &\sum\limits_{n = 0}^\infty  {\frac{{{p_{\rm{v}}}p_n^{\rm{v}}}}{{{p_{\rm{s}}}p_n^{\rm{s}}}}} \left( {N_{{{\rm error},n,\gamma^s}|\rm X} + {\delta _{{\rm{E}}_{{\rm X} ,n}^{\rm{s}}}}} \right)\\& + \Delta _{\rm X} ^{{\rm{vs}}} - {\delta _{{\rm{E}}_{\rm{X}}^{\rm{v}}}},\\
{N_{{{\rm error},\gamma^w}|\rm X}}= &\sum\limits_{n = 0}^\infty  {\frac{{{p_{\rm{w}}}p_n^{\rm{w}}}}{{{p_{\rm{s}}}p_n^{\rm{s}}}}} \left( {N_{{{\rm error},n,\gamma^s}|\rm X}+ {\delta _{{\rm{E}}_{{\rm X} ,n}^{\rm{s}}}}} \right) \\&+ \Delta _{\rm X} ^{{\rm{ws}}} - {\delta _{{\rm{E}}_{\rm{X}}^{\rm{w}}}}.
\end{array}
\end{equation}
where $N_{{{\rm error},\gamma^j}|\rm X}$ denotes the actual number of events where Alice sends Bob a pulse with intensity $\gamma^j$ and they obtain an error given that both Alice and Bob select the X basis. $N_{{{\rm error},n,\gamma^j}|\rm X}$ denotes the actual number of events where Alice sends Bob an $n$-photon pulse with intensity $\gamma^j$ and they obtain an error given that both Alice and Bob select the X basis, and $\delta _{{\rm E}_{\rm{X} }^j}$, $\delta _{{\rm E}_{\rm{X},n }^j}$ are the corresponding deviation terms, respectively, which can be bounded by using Azuma's inequality. Then, $N^{U}_{{\rm error,1,\gamma^s}|\rm X}$ can be obtained by solving a linear program that maximizes the value of $N_{{\rm error,1,\gamma^s}|\rm X}$ given the set of linear constraints imposed by Eq.~(\ref{LproE}).

As already mentioned above, from the parameters $N^{L}_{{\rm click,1,\gamma^s}|\rm Z}$, $N^{L}_{{\rm click,1,\gamma^s}|\rm X}$ and $N^{U}_{{\rm error,1,\gamma^s}|\rm X}$, one can use a random sampling argument to derive $e^{\rm U}_{\rm ph}$. See Appendix B for further details.

\subsubsection{Information leakage from the IM and the PM}

In principle, a THA against the PM allows Eve to learn partial information about Alice's basis choice each given time. That is, the outgoing states from by Alice's transmitter (which include the states emitted by Alice together with the back-reflected light from Eve's attack) are now basis dependent and thus they carry information about the particular basis used by Alice to prepare them. In this situation, we cannot use the techniques used in the previous section for estimating the phase error rate. The security of QKD with basis dependent states has been analyzed in a previous work~\cite{lo2007security}, where the authors used the idea of a quantum coin~\cite{gottesman2004security,koashi2005simple} to relate the balance of such a coin to the basis dependence of the signals prepared by Alice. In this section, we apply the same idea to estimate $N_{\rm{phase~error}}^{\rm{U}}$. Like previously, we first review the results introduced in~\cite{tamaki2016decoy,lucamarini2015practical} which are valid in the asymptotic limit of large $N$. Next, we adapt them to the finite-key regime by mainly using Azuma's inequality~\cite{azuma1967weighted}.
\\
\\*
\noindent{\it Asymptotic Case:}
For ease of illustration, let us first consider a scenario where Alice's source is a single-photon source. As already mentioned above, in the presence of a THA against the PM, the joint state of the single-photon states emitted by Alice together with Eve's back-reflected light could be basis dependent. We shall denote the joint state of Alice's single-photon signals and Eve's systems given that Alice selects the Z (X) basis in the $i$th trial by ${{{\left|{\Psi_{\rm{Z}}^i}\right\rangle}_{{{\rm{A}}_{\rm{q}}},{{\rm{A}}_{\rm{p}}},{{\rm{A}}_{\rm{a}}},{{\rm{E}}_{\rm{a}}},
{\rm{E}}'_{\rm{p}}}}}$ (${{{\left|{\Psi_{\rm{X}}^i}\right\rangle}_{{{\rm{A}}_{\rm{q}}},{{\rm{A}}_{\rm{p}}},{{\rm{A}}_{\rm{a}}},{{\rm{E}}_{\rm{a}}},
{\rm{E}}'_{\rm{p}}}}}$)~\cite{tamaki2016decoy}. Here, ${\rm A}_{\rm q}$ denotes a virtual qubit, which contains Alice's bit value choice, ${\rm A}_{\rm p}$ represents Alice's photonic system that she sends to Bob via a quantum channel and ${\rm A}_{\rm a}$ is an additional ancillary system in Alice's hands to account for the loss in her transmitter. For example, these states could have the following form:
\begin{equation}
\begin{array}{*{20}{l}}
{{{\left| {\Psi _{\rm{Z}}^i} \right\rangle }_{{{\rm{A}}_{\rm{q}}},{{\rm{A}}_{\rm{p}}},{{\rm{A}}_{\rm{a}}},{{\rm{E}}_{\rm{a}}},{{\rm{E}}^\prime }_{\rm{p}}}} }=&\frac{1}{{\sqrt 2 }}\bigg( {{\left| 0 \right\rangle }_{{{\rm{A}}_{\rm{q}}}}}{{\left| {\varphi _{{\rm{0Z}}}^i} \right\rangle }_{{{\rm{A}}_{\rm{p}}},{{\rm{A}}_{\rm{a}}},{{\rm{E}}_{\rm{a}}},{{\rm{E}}^\prime }_{\rm{p}}}}\\
 &+ {{\left| 1 \right\rangle }_{{{\rm{A}}_{\rm{q}}}}}{{\left| {\varphi _{{\rm{1Z}}}^i} \right\rangle }_{{{\rm{A}}_{\rm{p}}},{{\rm{A}}_{\rm{a}}},{{\rm{E}}_{\rm{a}}},{{\rm{E}}^\prime }_{\rm{p}}}} \bigg),\\
{{{\left| {\Psi _{\rm{X}}^i} \right\rangle }_{{{\rm{A}}_{\rm{q}}},{{\rm{A}}_{\rm{p}}},{{\rm{A}}_{\rm{a}}},{{\rm{E}}_{\rm{a}}},{{\rm{E}}^\prime }_{\rm{p}}}} }=&\frac{1}{{\sqrt 2 }}\bigg( {{\left|  +  \right\rangle }_{{{\rm{A}}_{\rm{q}}}}}{{\left| {\varphi _{{\rm{0X}}}^i} \right\rangle }_{{{\rm{A}}_{\rm{p}}},{{\rm{A}}_{\rm{a}}},{{\rm{E}}_{\rm{a}}},{{\rm{E}}^\prime }_{\rm{p}}}} \\
&+ {{\left|  -  \right\rangle }_{{{\rm{A}}_{\rm{q}}}}}{{\left| {\varphi _{{\rm{1X}}}^i} \right\rangle }_{{{\rm{A}}_{\rm{p}}},{{\rm{A}}_{\rm{a}}},{{\rm{E}}_{\rm{a}}},{{\rm{E}}^\prime }_{\rm{p}}}} \bigg),
\end{array}
\end{equation}
where ${{{|{\varphi _{{b\rm{\chi_A}}}^i}\rangle }_{{{\rm{A}}_{\rm{p}}},{{\rm{A}}_{\rm{a}}},{{\rm{E}}_{\rm{a}}},{{\rm{E}}' }_{\rm{p}}}}}$  with $b \in \{0,1\}$ denotes the joint the state of the systems ${{\rm{A}}_{\rm{p}}}$, ${{\rm{A}}_{\rm{a}}}$, ${{\rm{E}}_{\rm{a}}}$, and ${{\rm{E}}' }_{\rm{p}}$ for Alice's choice of the bit value `$b$' and basis `$\chi_{\rm A}$' in the $i$th trial.

Then, as already explained above, a phase error is the fictitious error that Alice and Bob would observe if Alice measured the system ${\rm A}_{\rm q}$ in the X basis and Bob measured the incoming signal also in the X basis given that Alice prepared the state ${{{\left|{\Psi_{\rm{Z}}^i}\right\rangle}_{{{\rm{A}}_{\rm{q}}},{{\rm{A}}_{\rm{p}}},{{\rm{A}}_{\rm{a}}},{{\rm{E}}_{\rm{a}}},
{\rm{E}}'_{\rm{p}}}}}$. To estimate this quantity we use a fictitious protocol. A sketch of this fictitious protocol is as follows (see Appendix. D for a more detailed description of each step of the fictitious protocol and its equivalence to the actual protocol). In particular, in the $i$th trial we assume that Alice prepares the state ${\left| {{\Psi ^i}} \right\rangle _{{{\rm{A}}_c},{{\rm{A}}_{\rm{q}}},{{\rm{A}}_{\rm{p}}},{{\rm{A}}_{\rm{a}}},{{\rm{E}}^\prime }_{\rm{p}},{{\rm{E}}_{\rm{a}}}}} = \sqrt {{p_{{{\rm{Z}}_{\rm{A}}}}}} {\left| 0 \right\rangle _{{{\rm{A}}_{\rm{c}}}}}{\left| {\Psi _{\rm{Z}}^i} \right\rangle _{{{\rm{A}}_{\rm{q}}},{{\rm{A}}_{\rm{p}}},{{\rm{A}}_{\rm{a}}},{{\rm{E}}^\prime }_{\rm{p}},{{\rm{E}}_{\rm{a}}}}} + \sqrt {{p_{{{\rm{X}}_{\rm{A}}}}}} {\left| 1 \right\rangle _{{{\rm{A}}_{\rm{c}}}}}{\left| {\Psi _{\rm{X}}^i} \right\rangle _{{{\rm{A}}_{\rm{q}}},{{\rm{A}}_{\rm{p}}},{{\rm{A}}_{\rm{a}}},{{\rm{E}}^\prime }_{\rm{p}},{\rm{E}}_{\rm{a}}}},$ where ${\rm A}_{\rm c}$ is a so-called quantum coin~\cite{gottesman2004security,koashi2005simple}. She keeps the systems ${\rm A}_{\rm c}$, $ {\rm A}_{\rm q}$ and $ {\rm A}_{\rm a}$ in her hands, and she sends system $ {\rm A}_{\rm p}$ to Bob. Bob performs a quantum nondemolition (QND) measurement on the received signal and determines if this signal will produce a click in his measurement device. For each click event determined by the QND measurement, Bob performs an X basis measurement on the received state and Alice also measures her system $\rm {A_q}$ in the X basis. Besides, Alice selects the ${\rm Z}_{\rm A_c}$ or ${\rm X}_{\rm A_c}$ basis with probabilities $p_{\rm Z_{A_c}}$ and $p_{\rm X_{A_c}}$, respectively, to measure the quantum coin in the selected basis. Importantly, these steps can be executed such that the actual and fictitious protocols are equivalent from Eve's point of view.

By applying the Bloch sphere bound in terms of probabilities~\cite{tamaki2003unconditionally} to this virtual scenario, one obtains
\begin{equation}\label{X error}
\begin{array}{*{20}{l}}
&&{1 - 2{{\Pr }^i}\left( {{{\rm{X}}_{{{\rm{A}}_{\rm{c}}}}} =  - \left| {{\rm{click}}{\rm{,sb}}{\rm{,X - error}},{{\rm{X}}_{{{\rm{A}}_{\rm{c}}}}}} \right.} \right)}\\
&& \le 2\sqrt {{{\Pr }^i}\left( {{{\rm{Z}}_{{{\rm{A}}_{\rm{c}}}}} = 1\left| {{\rm{click}}{\rm{,sb}}{\rm{,X - error}},{{\rm{Z}}_{{{\rm{A}}_{\rm{c}}}}}} \right.} \right)}\\
 &&\times\sqrt{ {1 - {{\Pr }^i}\left( {{{\rm{Z}}_{{{\rm{A}}_{\rm{c}}}}} = 1\left| {{\rm{click}}{\rm{,sb}}{\rm{,X - error}},{{\rm{Z}}_{{{\rm{A}}_{\rm{c}}}}}} \right.} \right)} } ,
\end{array}
\end{equation}
\begin{equation}\label{No Xerror}
\begin{array}{*{20}{l}}
&&{1 - 2{{\Pr }^i}\left( {{{\rm{X}}_{{{\rm{A}}_{\rm{c}}}}} =  - \left| {{\rm{click}}{\rm{,sb}}{\rm{,No}}\;{\rm{X - error}},{{\rm{X}}_{{{\rm{A}}_{\rm{c}}}}}} \right.} \right)}\\
&&\le 2\sqrt {{{\Pr }^i}\left( {{{\rm{Z}}_{{{\rm{A}}_{\rm{c}}}}} = 1\left| {{\rm{click}}{\rm{,sb}}{\rm{,No}}\;{\rm{X - error}},{{\rm{Z}}_{{{\rm{A}}_{\rm{c}}}}}} \right.} \right)}\\
 &&\times\sqrt{ {1 - {{\Pr }^i}\left( {{{\rm{Z}}_{{{\rm{A}}_{\rm{c}}}}} = 1\left| {{\rm{click}}{\rm{,sb}}{\rm{,No}}\;{\rm{X - error}},{{\rm{Z}}_{{{\rm{A}}_{\rm{c}}}}}} \right.} \right)} },
\end{array}
\end{equation}
where ${{\Pr }^i}\left( {{{\rm{X}}_{{{\rm{A}}_{\rm{c}}}}} =  - \left| {{\rm{click}}{\rm{,sb}},\;{\rm{X - error}},{{\rm{X}}_{{{\rm{A}}_{\rm{c}}}}}} \right.} \right)$ is the conditional probability that the X basis measurement result on the quantum coin is `$-$' in the $i$th trial given that Bob obtains a click in his measurement device, Alice and Bob select the same basis (sb) for the state preparation and measurement, respectively, Bob's measurement outcome differs from that obtained by Alice when she measures her system ${\rm A}_{\rm q}$, which we shall call an X basis error, and Alice performs the $\rm X_{A_c}$ basis measurement on the quantum coin. The other conditional probabilities that appear in Eqs (\ref{X error}) and (\ref{No Xerror}) are defined similarly.

Next, we multiply Eqs.~(\ref{X error}) and~(\ref{No Xerror}) by ${{{\Pr }^i}\left( {{{\rm{Z}}_{{{\rm{A}}_{\rm{c}}}}}\left| {{\rm{click}}} \right.} \right)}{{\Pr }^i}\left( { {\rm{sb}},{\rm{X - error}}\left| {{\rm{click}},{{\rm{X}}_{{{\rm{A}}_{\rm{c}}}}}} \right.} \right)$ and ${{{\Pr }^i}\left( {{{\rm{Z}}_{{{\rm{A}}_{\rm{c}}}}}\left| {{\rm{click}}} \right.} \right)}{{\Pr }^i}\left( { {\rm{sb}}{\rm{,No}}\;{\rm{X - error}}\left| {{\rm{click}},{{\rm{X}}_{{{\rm{A}}_{\rm{c}}}}}} \right.} \right)$, respectively, and add both terms together. The result is given by

\begin{widetext}
\begin{equation}
\begin{array}{*{20}{l}} \label{PZac}
&{{{\Pr }^i}\left( {{{\rm{Z}}_{{{\rm{A}}_{\rm{c}}}}}\left| {{\rm{click}}} \right.} \right) - 2{{\Pr }^i}\left( {{{\rm{Z}}_{{{\rm{A}}_{\rm{c}}}}}\left| {{\rm{click}}} \right.} \right){{\Pr }^i}\left( {{{\rm{X}}_{{{\rm{A}}_{\rm{c}}}}} =  - \left| {{\rm{click}},{{\rm{X}}_{{{\rm{A}}_{\rm{c}}}}}} \right.} \right)}\\
 &\le2\sqrt {{{\Pr }^i}\left( {{{\rm{Z}}_{{{\rm{A}}_{\rm{c}}}}} = 1,{\rm{X - error}},{{\rm{Z}}_{{{\rm{A}}_{\rm{c}}}}}\left| {{\rm{click}}} \right.} \right)
{{{\Pr }^i}\left( {{{\rm{Z}}_{{{\rm{A}}_{\rm{c}}}}} = 0,{\rm{X - error}},{{\rm{Z}}_{{{\rm{A}}_{\rm{c}}}}}\left| {{\rm{click}}} \right.} \right)} } \\
 &+2\sqrt {{{\Pr }^i}\left( {{{\rm{Z}}_{{{\rm{A}}_{\rm{c}}}}} = 1,{\rm{No}}\;{\rm{X - error}},{{\rm{Z}}_{{{\rm{A}}_{\rm{c}}}}}\left| {{\rm{click}}} \right.} \right)
 {{{\Pr }^i}\left( {{{\rm{Z}}_{{{\rm{A}}_{\rm{c}}}}} = 0,{\rm{No}}\;{\rm{X - error}},{{\rm{Z}}_{{{\rm{A}}_{\rm{c}}}}}\left| {{\rm{click}}} \right.} \right)}}.
\end{array}
\end{equation}
\end{widetext}
To obtain the expression above, we have taken into account that the events `${{{\rm{X}}_{{{\rm{A}}_{\rm{c}}}}}}$' and `${{{\rm{Z}}_{{{\rm{A}}_{\rm{c}}}}}}$' are independent of the events `click' and `sb', and the event `click' is independent of the event `sb' as well, and we also have removed the common factor `${\Pr ^i}\left( {{\rm{sb}}} \right)$' that appears on both sides. More details about this calculation can be found in Appendix D.

To relate the joint probabilities in Eq.~(\ref{PZac}) to the expected numbers of events, we take the sum over $i \in \{1,2,...,N_{\rm click}\}$, where $N_{\rm click}$ is the number of clicks. Due to the concavity of the square root function, we have that
\begin{widetext}
\begin{equation}
\begin{array}{*{20}{l}}
&\sum\limits_{i = 1}^{{N_{{\rm{click}}}}} {{{\Pr }^i}\left( {{{\rm{Z}}_{{{\rm{A}}_{\rm{c}}}}}\left| {{\rm{click}}} \right.} \right)} -2 \sum\limits_{i = 1}^{{N_{{\rm{click}}}}} {{{\Pr }^i}\left( {{{\rm{Z}}_{{{\rm{A}}_{\rm{c}}}}}\left| {{\rm{click}}} \right.} \right){{\Pr }^i}\left( {{{\rm{X}}_{{{\rm{A}}_{\rm{c}}}}} =  - \left| {{\rm{click}},{{\rm{X}}_{{{\rm{A}}_{\rm{c}}}}}} \right.} \right)} \\
 &\le2\sqrt {\sum\limits_{i = 1}^{{N_{{\rm{click}}}}} {{{\Pr }^i}\left( {{{\rm{Z}}_{{{\rm{A}}_{\rm{c}}}}} = 1,{\rm{X - error}},{{\rm{Z}}_{{{\rm{A}}_{\rm{c}}}}}\left| {{\rm{click}}} \right.} \right)}
\sum\limits_{i = 1}^{{N_{{\rm{click}}}}} {{{\Pr }^i}\left( {{{\rm{Z}}_{{{\rm{A}}_{\rm{c}}}}} = 0,{\rm{X - error}},{{\rm{Z}}_{{{\rm{A}}_{\rm{c}}}}}\left| {{\rm{click}}} \right.} \right)} } \\
 &+2\sqrt {\sum\limits_{i = 1}^{{N_{{\rm{click}}}}} {{{\Pr }^i}\left( {{{\rm{Z}}_{{{\rm{A}}_{\rm{c}}}}} = 1,{\rm{No}}\;{\rm{X - error}},{{\rm{Z}}_{{{\rm{A}}_{\rm{c}}}}}\left| {{\rm{click}}} \right.} \right)} \sum\limits_{i = 1}^{{N_{{\rm{click}}}}} {{{\Pr }^i}\left( {{{\rm{Z}}_{{{\rm{A}}_{\rm{c}}}}} = 0,{\rm{No}}\;{\rm{X - error}},{{\rm{Z}}_{{{\rm{A}}_{\rm{c}}}}}\left| {{\rm{click}}} \right.} \right)} } .
\end{array}\label{N ph}
\end{equation}
\end{widetext}
If we denote the expected number of times that event `$\lambda$' occurs after $N_{\rm click}$ trials by $\mathcal{E}_{\lambda}$, Eq.~(\ref{N ph}) can be rewritten as
\begin{equation}
\begin{array}{*{20}{l}}
&{p_{{{\rm{Z}}_{{{\rm{A}}_{\rm{c}}}}}}}{N_{{\rm{click}}}} - 2\frac{{{p_{{{\rm{Z}}_{{{\rm{A}}_{\rm{c}}}}}}}}}{{{p_{{{\rm{X}}_{{{\rm{A}}_{\rm{c}}}}}}}}}{\mathcal{E}_{{{\rm{X}}_{{{\rm{A}}_{\rm{c}}}}} =  - }}\\
 &\le 2\sqrt {{\mathcal{E}_{{\rm{X}},{\rm{X - error}}}}{\mathcal{E}_{{\rm{Z}},{\rm{X - error}}}}} \\
  &+ 2\sqrt {{\mathcal{E}_{{\rm{X}},{\rm{No}}\;{\rm{X - error}}}}{\mathcal{E}_{{\rm{Z}},{\rm{No}}\;{\rm{X - error}}}}}\label{n phase},
\end{array}
\end{equation}
where we have also used the fact that
\begin{equation}
\begin{array}{*{20}{l}}
&\sum\limits_{i = 1}^{{N_{{\rm{click}}}}} {{{\Pr }^i}\left( {{{\rm{Z}}_{{{\rm{A}}_{\rm{c}}}}}\left| {{\rm{click}}} \right.} \right)=p_{\rm Z_{A_c}}\sum\limits_{i = 1}^{{N_{{\rm{click}}}}}1=p_{{{\rm{Z}}_{{{\rm{A}}_{\rm{c}}}}}}}{N_{{\rm{click}}}},\\
&\sum\limits_{i = 1}^{{N_{{\rm{click}}}}} {{{\Pr }^i}\left( {{{\rm{Z}}_{{{\rm{A}}_{\rm{c}}}}}\left| {{\rm{click}}} \right.} \right){{\Pr }^i}\left( {{{\rm{X}}_{{{\rm{A}}_{\rm{c}}}}} =  - \left| {{\rm{click}},{{\rm{X}}_{{{\rm{A}}_{\rm{c}}}}}} \right.} \right)}\\
&={\rm{ }}\sum\limits_{i = 1}^{{N_{{\rm{click}}}}}{{p_{{{\rm{Z}}_{{{\rm{A}}_{\rm{c}}}}}}}}\frac{{{{\Pr }^i}\left( {{{\rm{X}}_{{{\rm{A}}_{\rm{c}}}}} =  - ,{{\rm{X}}_{{{\rm{A}}_{\rm{c}}}}}\left| {{\rm{click}}} \right.} \right)}}{{{{\Pr }^i}\left( {{{\rm{X}}_{{{\rm{A}}_{\rm{c}}}}}\left| {{\rm{click}}} \right.} \right)}}\\
&=\frac{{{p_{{{\rm{Z}}_{{{\rm{A}}_{\rm{c}}}}}}}}}{{{p_{{{\rm{X}}_{{{\rm{A}}_{\rm{c}}}}}}}}}\sum\limits_{i = 1}^{{N_{{\rm{click}}}}} {{{\Pr }^i}\left( {{{\rm{X}}_{{{\rm{A}}_{\rm{c}}}}} =  - ,{{\rm{X}}_{{{\rm{A}}_{\rm{c}}}}}\left| {{\rm{click}}} \right.} \right)}\\
&=:\frac{{{p_{{{\rm{Z}}_{{{\rm{A}}_{\rm{c}}}}}}}}}{{{p_{{{\rm{X}}_{{{\rm{A}}_{\rm{c}}}}}}}}}
{\mathcal{E}_{{{\rm{X}}_{{{\rm{A}}_{\rm{c}}}}} =  - }}.
\end{array}
\end{equation}
In Appendix D one can see that to upper bound $\mathcal{E}_{{{\rm{X}}_{{{\rm{A}}_{\rm{c}}}}} =  - }$ we take a worst-case scenario where we assume that all the events `${{{\rm{X}}_{{{\rm{A}}_{\rm{c}}}}} =  - }$' are detected. Note that, ${{\mathcal{E}_{{\rm{X}}{\rm{,X - error}}}}}$ denotes the expected number of click events where Alice chooses the ${\rm Z}_{\rm A_c}$ basis to measure the quantum coin and obtains the measurement outcome `1' (corresponding to the preparation of ${{{\left|{\Psi_{\rm{X}}^i}\right\rangle}_{{{\rm{A}}_{\rm{q}}},{{\rm{A}}_{\rm{p}}},{{\rm{A}}_{\rm{a}}},{{\rm{E}}_{\rm{a}}},
{\rm{E}}_{\rm{p}}}}}$) and there is an X basis error. Similarly, ${{\mathcal{E}_{{\rm{Z}}{\rm{,X - error}}}}}$ denotes the expected number of click events where Alice chooses the ${\rm Z}_{\rm A_c}$ basis to measure the quantum coin and obtains the measurement outcome `0' (corresponding to the preparation of ${{{\left|{\Psi_{\rm{Z}}^i}\right\rangle}_{{{\rm{A}}_{\rm{q}}},{{\rm{A}}_{\rm{p}}},{{\rm{A}}_{\rm{a}}},{{\rm{E}}_{\rm{a}}},
{\rm{E}}_{\rm{p}}}}}$) and there is an X basis error. The other expected numbers are defined in an analogous way. Importantly, ${{\mathcal{E}_{{\rm{Z}}{\rm{,X - error}}}}}$ is actually equal to $N_{\rm phase~error}$. The quantity ${\mathcal{E}_{{{\rm{X}}_{{{\rm{A}}_{\rm{c}}}}} =  - }}$ denotes the expected number of `$-$' among the X basis measurements on the quantum coins, and these coins are associated to Bob's click events as well as Alice's choice of the X basis for measuring the quantum coins and it can be bounded by using the overlap between the states ${{{\left|{\Psi_{\rm{Z}}^i}\right\rangle}_{{{\rm{A}}_{\rm{q}}},{{\rm{A}}_{\rm{p}}},{{\rm{A}}_{\rm{a}}},{{\rm{E}}_{\rm{a}}},
{\rm{E}}'_{\rm{p}}}}}$ and ${{{\left|{\Psi_{\rm{X}}^i}\right\rangle}_{{{\rm{A}}_{\rm{q}}},{{\rm{A}}_{\rm{p}}},{{\rm{A}}_{\rm{a}}},{{\rm{E}}_{\rm{a}}},
{\rm{E}}'_{\rm{p}}}}}$ defined above. See Appendix D for more details.
\\
\\*
\noindent{\it Finite-key Regime:}
Since Azuma's inequality can deal with any correlation among events (for instance, caused by a coherent attack by Eve), we use Azuma's inequality to relate the expected number of events to the corresponding actual number, like we did to analyze the THA against the IM. According to Azuma's inequality~\cite{azuma1967weighted}, we have that
 \begin{equation}
{\mathcal{E}_\lambda } =  {{N_\lambda}}  + {\delta _\lambda },
\end{equation}
holds except for a probability exponentially small in the number of trials, $\it i.e.$, the click events, where $\mathcal{E}_\lambda$ denotes the expected number of times that the event `$\lambda$' occurs, and ${{N_\lambda}}$ denotes the actual number of times that the event `$\lambda$' occurs. $\delta _\lambda$ represents the corresponding deviation term and it lies in an interval $[ { - \Delta _{\lambda},\widehat{\Delta} _{\lambda}} ]$ except for a small error probability $\varepsilon_{\lambda}+\widehat{\varepsilon}_{\lambda}$ where the bounds $ \Delta _{\lambda}$ and $\widehat{\Delta} _{\lambda}$ can be calculated by using Azuma's inequality. Then, if we replace all the expected numbers in Eq.~(\ref{n phase}) with the actual numbers plus the corresponding deviation terms in the actual protocol, we obtain:
\begin{equation}
\begin{array}{*{20}{l}}
&{{p_{{{\rm{Z}}_{{{\rm{A}}_{\rm{c}}}}}}}\left( {{{N_{{\rm{click}}}}}  + {\delta _{{\rm{click}}}}} \right)-2\frac{{{p_{{{\rm{Z}}_{{{\rm{A}}_{\rm{c}}}}}}}}}{{{p_{{{\rm{X}}_{{{\rm{A}}_{\rm{c}}}}}}}}} ({N_{{{\rm{X}}_{{{\rm{A}}_{\rm{c}}}}} =  - }} +\delta'_{\rm X_{{{{\rm{A}}_{\rm{c}}}}=  -}})} \\
 &\le{2\sqrt {\left( { {{N_{{\rm{X}},{\rm{X - error}}}}} + {\delta _{{\rm{X}},{\rm{X - error}}}}} \right)\left( {{{N_{{\rm{Z}},{\rm{X - error}}}}} + {\delta _{{\rm{Z}},{\rm{X - error}}}}} \right)} }\\
 &+2\sqrt {\left( {{{N_{\rm{click|X}}}} - {{N_{{\rm{X}},{\rm{X - error}}}}} + {\delta _{{\rm{X}},{\rm{No}}\;{\rm{X - error}}}}}\right)} \\&\times\sqrt{\left({{{N_{\rm{click|Z}}}}  -  {{N_{{\rm{Z}},{\rm{X - error}}}}}  + {\delta _{{\rm{Z}},{\rm{No}}\;{\rm{X - error}}}}} \right)}
\end{array}\label{Nphase}
\end{equation}
except for an exponentially small error probability $(\varepsilon'_{\rm X_{{{{\rm{A}}_{\rm{c}}}}=  -}}+\widehat{\varepsilon}'_{\rm X_{{{{\rm{A}}_{\rm{c}}}}=  -}})+\sum_\lambda  {\left( {{\varepsilon _\lambda } + {{\hat \varepsilon }_\lambda }} \right)}$, where $\lambda \in \{\rm(X,X-error),(Z,X-error),(X,No~X-error),(Z,No~X-error)\}$. Note that, although in the actual protocol there is no data corresponding to `$N_{{{\rm{X}}_{{{\rm{A}}_{\rm{c}}}}} =  - }$', this quantity can first be upper bounded by using Eq.~(\ref{NAC}) in Appendix D in the asymptotic case, and then an estimation of the actual number $N_{{{\rm{X}}_{{{\rm{A}}_{\rm{c}}}}} =  - }$ can be derived by using the Chernoff bound~\cite{chernoff1952measure} with an exponentially small error probability and $\delta'_{\rm X_{{{{\rm{A}}_{\rm{c}}}}=  -}}$ denoting the corresponding deviation term (See Appendix D for more details). In the above equation, ${{N_{\rm{click|Z(X)}}}}$ denotes the actual number of events where both Alice and Bob select the Z (X) basis for measuring system $\rm A_q$ and the incoming signal, respectively, and Bob obtains a click, $ \it i.e.$, ${{{N_{{\rm{click|Z(X)}}}}} = {{N_{{\rm{Z(X),X - error}}}}}  +  {{N_{{\rm Z(X),\rm{No}}\;{\rm{X - error}}}}} }$. ${{N_{{\rm{Z}},{\rm{X - error}}}}} $ is the quantity to be estimated.

So far all the analysis considers that Alice has a single-photon source. However, it is straightforward to adapt the analysis above to the decoy-state BB84 protocol described in Sec. II where the secret key is only distilled from the data corresponding to the signal intensity setting. For this, we only need to consider that now all the actual numbers that appear in Eq.~(\ref{Nphase}) refer to the single-photon contributions within the signal intensity setting, $\gamma^{\rm s}$.
That is, now ${ {{N_{{\rm{click}}}} ,}~ {{N_{{\rm{X}},{\rm{X - error}}}},}~ {{N_{{\rm{Z}},{\rm{X - error}}}},}~ {{N_{{\rm{click|X}}}} ,}~ {{N_{{\rm{click|Z}}}}}}$ refer to $ {{N_{{\rm{click,1,\gamma^s}}}}} ,~ {{N_{{\rm{X - error}},{\rm{1,\gamma^s|X}}}}} ,~{{N_{{\rm{X - error}},{\rm{1,\gamma^s|Z}}}}} ,~ {{N_{{\rm{click,1,\gamma^s|X}}}}}$ , $ {{N_{{\rm{click,1,\gamma^s|Z}}}}}$. One can use the same method based on linear optimization that we used in the previous sections to estimate $ {{N_{{\rm{click,1,\gamma^s}}}},}  ~{{N_{{\rm{X - error}},{\rm{1,\gamma^s|X}}}} ,}~{{N_{{\rm{X - error}},{\rm{1,\gamma^s|Z}}}} ,}~ {{N_{{\rm{click,1,\gamma^s|X}}}}}$, $ {{N_{{\rm{click,1,\gamma^s|Z}}}}}$. In so doing, one can estimate an upper bound on $ {{N_{{\rm{X - error}},{\rm{1,\gamma^s|Z}}}}}$ and thus obtain $e_{{\rm{ph}}}^{\rm{U}}$.
\section{Simulation}\label{secIV}

For illustration purposes, in the simulation we shall assume that the information leakage comes from a THA where Eve injects into Alice's transmitter high-intensity single-mode coherent states of the form $|\beta e^{i\theta}\rangle$ with $\beta$ being the limited amplitude of the input light and $\theta$ being the corresponding phase which can have an arbitrary value. Since we do not have an experimental characterization of how  Alice's modulators behave, for simplicity in this section we shall assume that the back-reflected light to Eve is still a coherent state of the form $|\beta_{k} e^{i\theta_{k}}\rangle$ with $k\in\{\rm s,v,w\}$. The subscript $k$ that appears in both the amplitude and the phase of the back-reflected state indicates that they may depend on the intensity setting selected by Alice each given time. However, note that the formalism presented in the previous section could be applied as well to any other type of back-reflected light.

In the presence of information leakage, the length $\ell'$ of the secret key is actually given by
\begin{equation} \label{key'}
\ell' \geq \mathop {\max }\limits_{{\Gamma _{\rm AB}}} \mathop {\min }\limits_{{\Gamma _{\rm E}}} \ell,
\end{equation}
where $\ell$ is given by Eq.~(\ref{key}). Here, $\Gamma_{\rm AB}$ and  $\Gamma _{\rm E}$ denote the spaces of the parameters controlled by Alice and Bob, and Eve, respectively. In the simulation, we let $\Gamma_{\rm AB}=\{\gamma^{\rm s},\gamma^{\rm v},p_{\rm Z_{A_c}},p_{\rm s},p_{\rm v},p_{\rm Z}\}$ and $\Gamma _{\rm E}=\{\theta_{{\rm v}},~\theta_{{\rm w}}\}$. In addition we assume that $\gamma^{\rm w}=5\times10^{-4}$ (which is a reasonable value for the weakest decoy state as in practice it is difficult to generate a perfect vacuum state due to the imperfect extinction ratio of the IM), and, without loss of generality, we assume that $\theta_{\rm s}=0$. In addition, we consider for simplicity a model where the amount of information revealed by Alice during the error correction step of the protocol is given by $leak_{\rm EC}=|Z^{\rm s}|f_{\rm EC}H(E_{\rm Z}^{\rm s})$. For more accurate models of the value of $leak_{\rm EC}$ as a function of $N$, we refer the reader to~\cite{tomamichel2014fundamental}. The experimental parameters considered in the simulation are listed in table~\ref{paras}.

\begin{table}[tbp]
\caption{\footnotesize {Experimental parameters used in the simulation. $e_{\rm d}$ is the intrinsic error rate due to the misalignment of the QKD system; $p_{\rm d}$ is the dark count rate of Bob's detectors, where we assume for simplicity that it is equal for all of them; $\eta_{\rm det}$ is the overall detection efficiency of Bob's receiver, i.e., this parameter already includes the detection efficiency of his detectors (which again we assume for simplicity is equal for all of them); $\alpha$ is the loss coefficient of the channel measured in dB/km; $\gamma^{\rm w}$ is the intensity of the weakest decoy state; $f_{\rm EC}$ is the efficiency of the error correction code.}}
\centering \label{paras}
\begin{tabular}{p{1.2cm}<{\centering}p{1.2cm}<{\centering}p{1.2cm}<{\centering}p{1.2cm}<{\centering}p{1.6cm}
<{\centering}p{1.2cm}c}
\hline
$e_{\rm d}$  & $p_{\rm d}$  & $\eta_{\rm det}$ & $\alpha$ & $\gamma^{\rm w}$ & $f_{\rm EC}$\\ \hline
 $0.01$& $5\times 10^{-6}$ & $0.25$ & $0.2$ & $5\times 10^{-4}$ &$1.2$\\
 \hline
\end{tabular}
\end{table}

To quantitatively show the results of our finite-key estimation method in the presence of information leakage, we simulate the secret key rates in the same three different cases considered in~\cite{tamaki2016decoy}. In each case, there is a particular model for the back-reflected light to the channel. The different values for the trace distance terms $D_{n,\{ {\gamma ^j}\} ,\{ {\gamma ^k} \},\chi}$ and $D_{n,\{ {\gamma ^j}\} ,\{ {\gamma ^k},{\gamma ^l}\} ,\chi }$ with $j,k,l\in \{\rm s,v,w\}$ for the leaked states in these three cases are given in Appendix E.

\subsection{Case 1}
Since in our simulation we assume that the back-reflected light is a coherent state of the form $|\beta_{k} e^{i\theta_{k}}\rangle$ and the phases $\theta_{k}$ can have arbitrary values controlled by Eve, the larger the value of $\beta_{k}$ is, the more information is leaked to Eve. Here, we consider a model where Alice and Bob may overestimate the intensity of the back-reflected light leaked to Eve. In particular, suppose that the intensity $ {{\beta_{k}}} ^2$ is upper bounded by a certain value $I_{\rm max}$ for all $k$. Then, we assume a conservative scenario for Alice and Bob, where
\begin{equation}
{I_{\rm max }} = { {{\beta_{{\rm s}}}} ^2} = { {{\beta_{{\rm v}}}} ^2} = { {{\beta_{{\rm w}}}}^2}. \label{Case1}
\end{equation}

\begin{figure}[!t]
 \includegraphics*[scale=0.28]{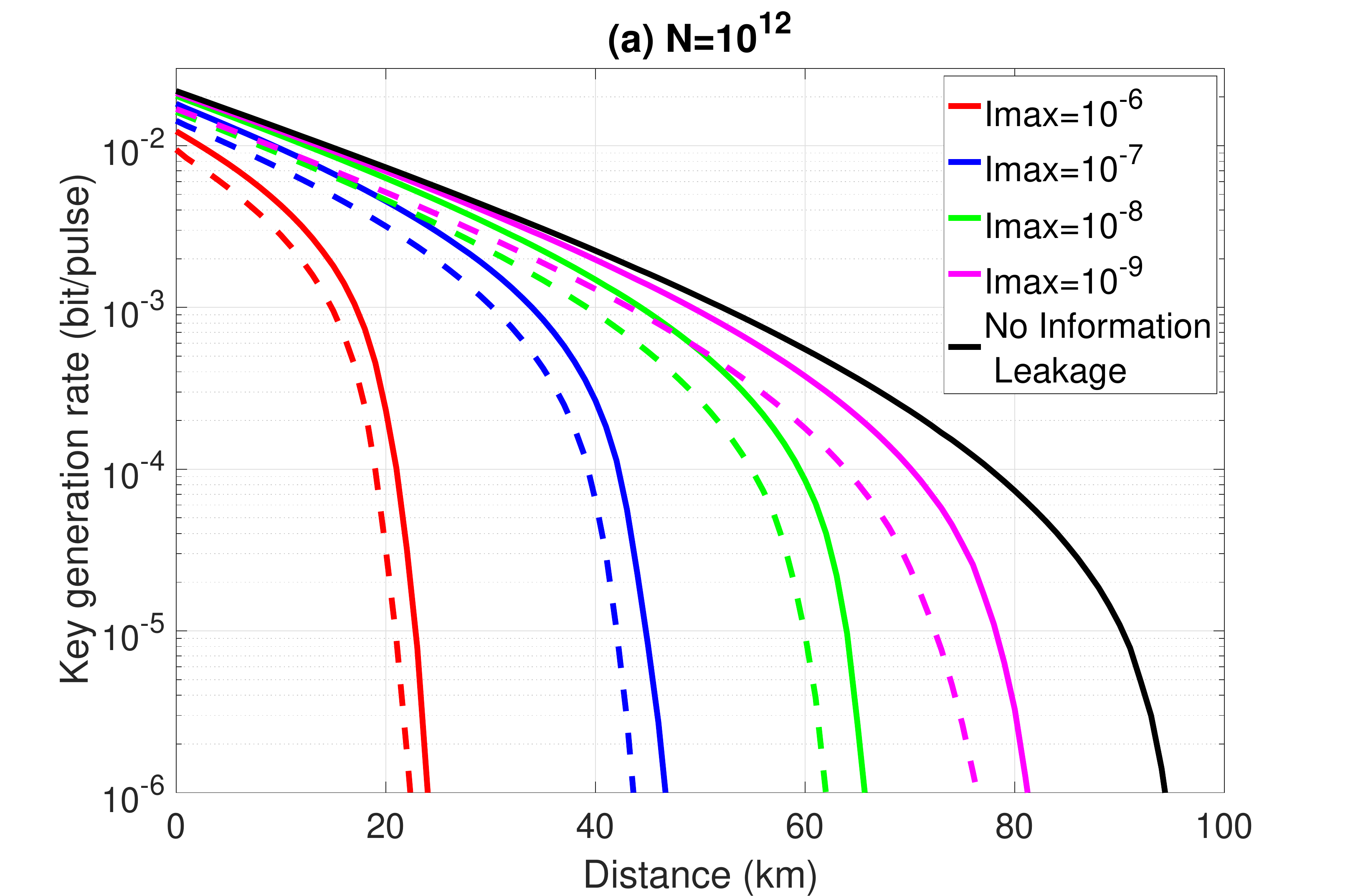}
 \includegraphics*[scale=0.28]{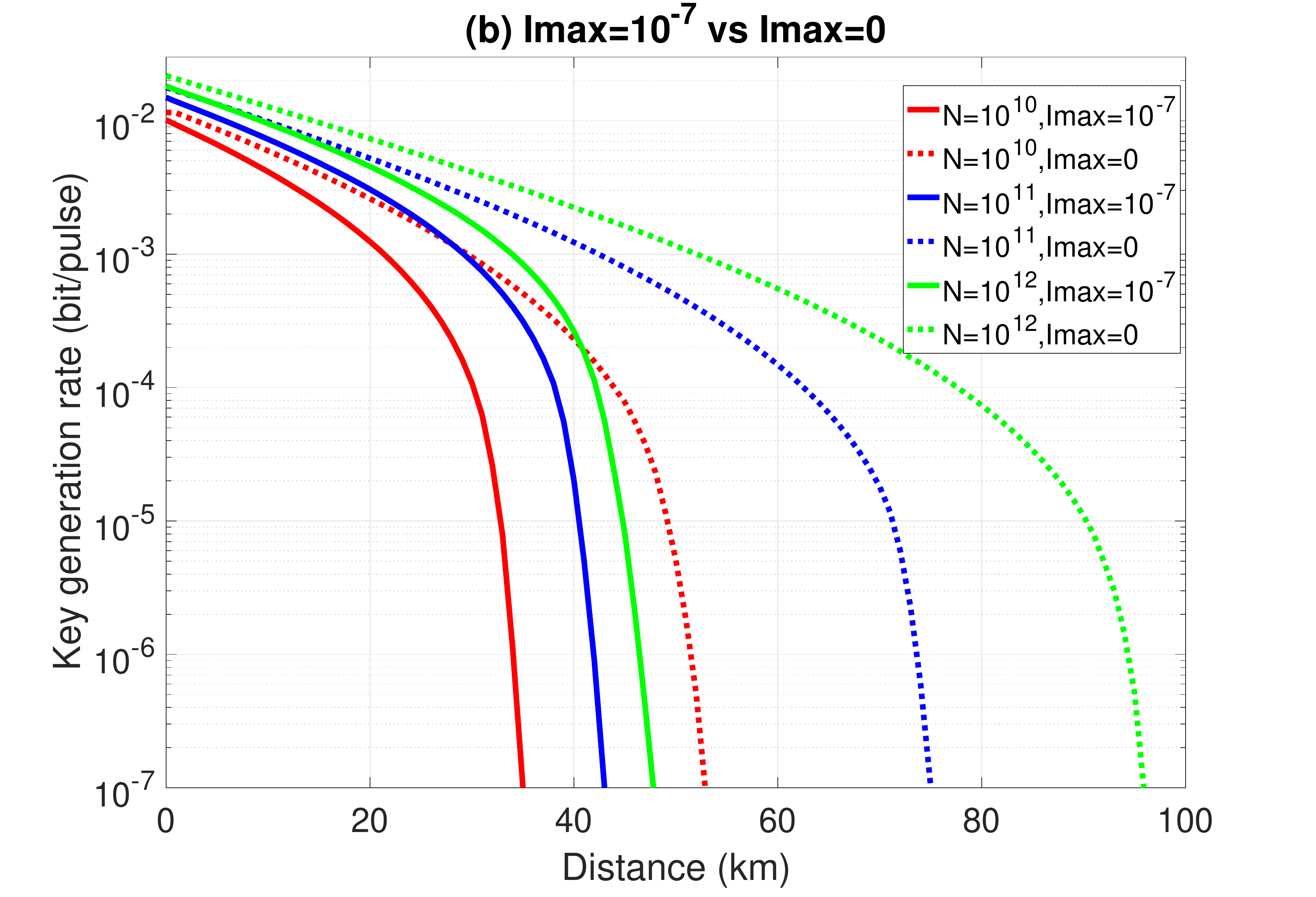}
 \caption{ \footnotesize { (a) The secret key rate in logarithmic scale as a function of the distance for a fixed value of the total number of transmitted pulses, $N=10^{12}$. The black solid line represents the perfectly isolated situation where there is no information leakage (i.e., $I_{\rm max}=0$) and the different colored lines correspond to different amounts of information leakage. More precisely, the coloured solid (dashed) lines represent the secret key rates in the presence of a THA against the IM (both the IM and PM). (b) The secret key rate in logarithmic scale as a function of the distance for two fixed values of information leakage, $I_{\rm max}=\{0,~10^{-7}\}$. Different colored lines correspond to different values of the number of transmitted pulses. Here we consider a THA against both the IM and the PM. In our simulations, for each value of the distance we maximize the secret key rate over the amplitudes $\gamma^{\rm s}$ and $\gamma^{\rm v}$, and the probabilities $p_{\rm Z_{A_c}}$, $p_{\rm s}$, $p_{\rm v}$, $p_{\rm Z}$ which are controlled by Alice and Bob, and we minimize it over the angles $\theta_k$ controlled by Eve, respectively.}} \label{figcase1}
\end{figure}

The simulation result of the secret key rate, $\ell'/N$, as a function of the transmission distance between Alice and Bob in this case is shown in Fig.~\ref{figcase1}~(a) for a fixed value of the total number of transmitted pulses, $N=10^{12}$. In this figure, the black solid line represents the perfectly isolated situation where there is no information leakage and the different colored lines correspond to different amounts of information leakage. More precisely, the coloured solid lines represent the key rates in the presence of a THA against only the IM. If we compare these results with the longest achievable distance without information leakage, which is about 96 km, we find that now the secret key rate vanishes at about 24 km when $I_{\rm max}=10^{-6}$. In addition, we also find that if $I_{\rm max}$ is as small as $I_{\rm max}=10^{-12}$, then the secret key rate is very close to the one corresponding to the ideal case where there is no information leakage. This can be seen in Fig.~\ref{ratiocase1} in Appendix F. The coloured dashed lines shown in Fig.~\ref{figcase1}~(a) represent the secret key rates in the presence of a THA against both the IM and the PM. Here we assume that the intensity of the back-reflected light from the PM is also $I_{\rm max}$. Now the secret key rates are obviously slightly lower than the ones corresponding to a THA against only the IM. This effect is more evident than in the asymptotic scenario considered in~\cite{tamaki2016decoy}, where the secret key rates considering information leakage from both the IM and the PM basically coincide with those in the presence of a THA against only the IM. This is because of the following. Although it is not explicitly written in~\cite{tamaki2016decoy}, in the asymptotic case one can consider that the probability $p_{\rm Z_{A_c}}$ is basically one, while in the finite-key case we find that the optimized value of this parameter typically lies in an interval $\left[0.7,0.95 \right]$. It is the probability with which Alice selects part of the data corresponding to the click events to be used for parameter estimation (while she discards the other part of the data corresponding to the click events with probability $1-p_{\rm Z_{A_c}}$). That is, the larger the value of $p_{\rm Z_{A_c}}$, the higher the efficiency of the protocol and thus also its secret key rate.

In Fig.~\ref{figcase1}~(b), the solid lines show the secret key rate as a function of the distance for a fixed value $I_{\rm max}=10^{-7}$ and for different values of the total number of transmitted pulses. For comparison, this figure also plots the secret key rate when $I_{\rm max}=0$, i.e., when there is no information leakage (see dotted lines). This way we can see the effect of the information leakage as a function of the number of pulses sent. For example, when $I_{\rm max}=10^{-7}$, the longest achievable distance at which the secret key rate is positive is about 48 km when the total number of transmitted pulses is $N=10^{12}$. However, when $N=10^{10}$, this distance decreases to 35 km. Additional simulation results can be found in Appendix F.

Our simulation results also suggest that when $I_{\rm max}$ increases, the resulting secret key rate drops quicker when the total number of transmitted pulses decreases. See Appendix F for more details.

\subsection{Case 2}
Here we consider a scenario where the intensity of the back-reflected light leaked to Eve depends on the intensity setting selected by Alice each given time. This could happen because Eve's light might be reflected after the IM and then on its way back to the channel it is modulated in  exactly the same way as Alice's pulses are during the state preparation stage. This is probably a more realistic scenario than that considered in the previous case, which could be a bit overpessimistic. We have therefore that now
\begin{equation}
{I_{\max }} = { {{\beta_{{\rm s}}}} ^2} = \frac{{{\gamma ^{\rm s}}}}{{{\gamma ^{\rm v}}}}{ {{\beta_{{\rm v}}}} ^2} = \frac{{{\gamma ^{\rm s}}}}{{{\gamma ^{\rm w}}}}{ {{\beta_{{\rm w}}}} ^2}. \label{Case2}
\end{equation}
That is, here we assume that ${I_{\max }} = { {{\beta_{{\rm s}}}} ^2}$, and the following conditions hold ${{\beta_{{\rm s}}}} ^2/{{\beta_{{\rm v}}}} ^2={\gamma ^{\rm s}}/{\gamma ^{\rm v}}$ and ${{\beta_{{\rm s}}}} ^2/{{\beta_{{\rm w}}}} ^2={\gamma ^{\rm s}}/{\gamma ^{\rm w}}$.

\begin{figure}[!t]
 \centering
 \includegraphics*[scale=0.28]{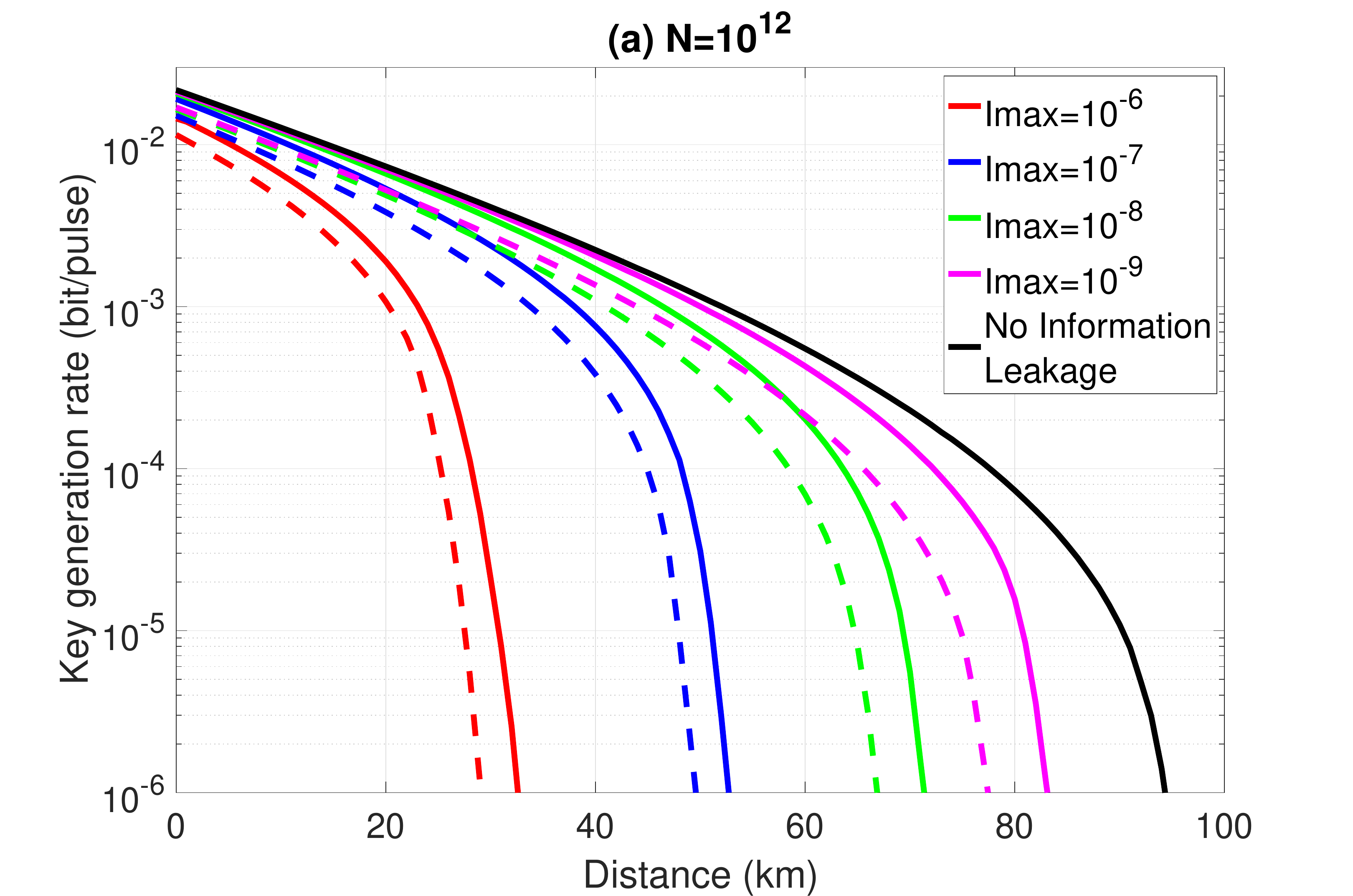}
 \includegraphics*[scale=0.28]{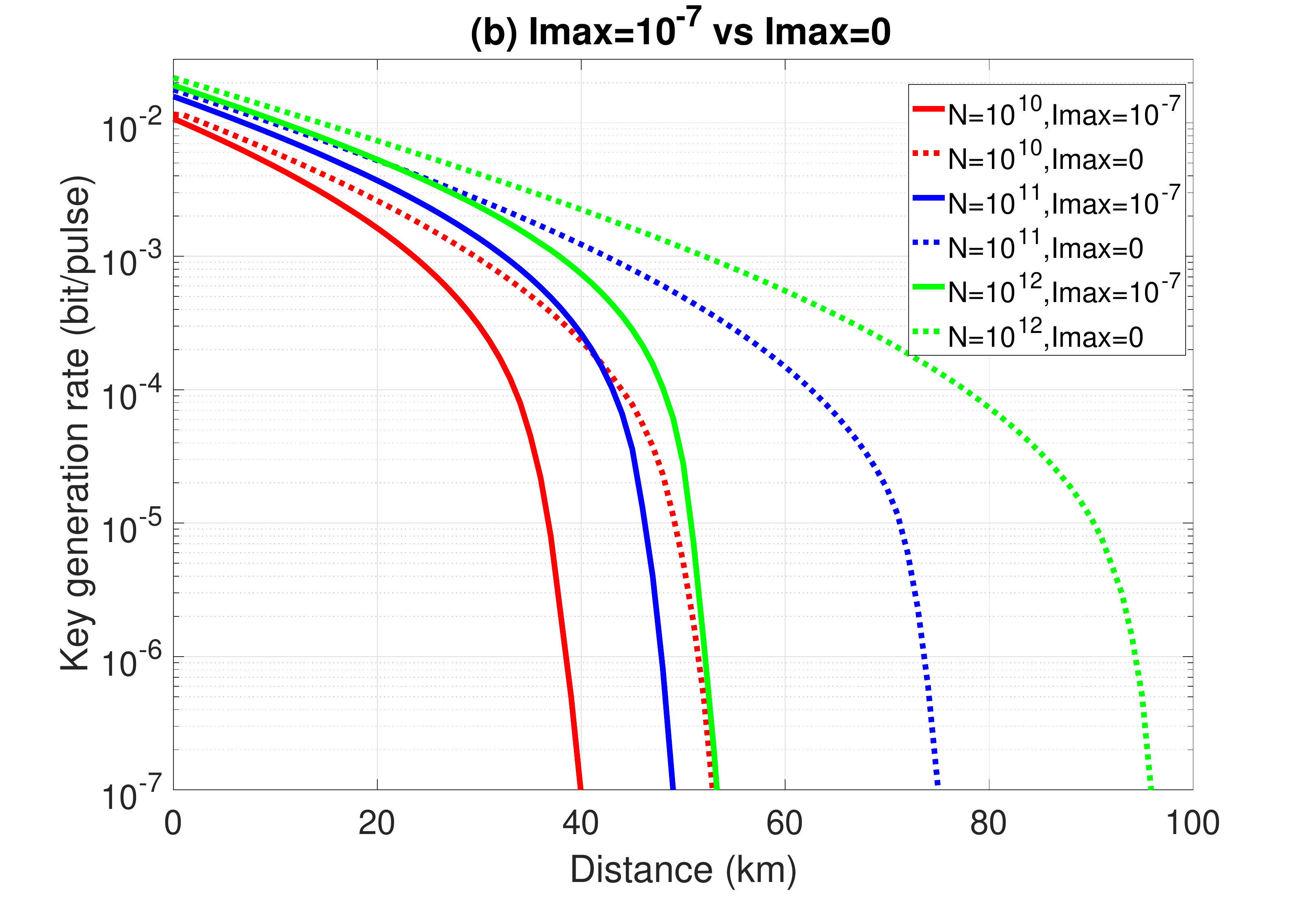}
\caption{ \footnotesize { (a) The secret key rate in logarithmic scale as a function of the distance for a fixed value of the total number of transmitted pulses, $N=10^{12}$. The black solid line represents the perfectly isolated situation where there is no information leakage (i.e., $I_{\rm max}=0$) and the different colored lines correspond to different amounts of information leakage. More precisely, the coloured solid (dashed) lines represent the secret key rates in the presence of a THA against the IM (both the IM and PM). (b) The secret key rate in logarithmic scale as a function of the distance for two fixed values of information leakage, $I_{\rm max}=\{0,~10^{-7}\}$. Different colored lines correspond to different values of the number of transmitted pulses. Here we consider a THA against both the IM and the PM. In our simulations, for each value of the distance we maximize the secret key rate over the amplitudes $\gamma^{\rm s}$ and $\gamma^{\rm v}$, and the probabilities $p_{\rm Z_{A_c}}$, $p_{\rm s}$, $p_{\rm v}$, $p_{\rm Z}$ which are controlled by Alice and Bob, and we minimize it over the angles $\theta_k$ controlled by Eve, respectively.}} \label{figcase2}
\end{figure}

The simulation result of the secret key rate as a function of the transmission distance between Alice and Bob when $N=10^{12}$ and for different values of $I_{\rm max}$ is shown in Fig.~\ref{figcase2}~(a). Fig.~\ref{figcase2}~(b), on the other hand, shows the secret key rate as a function of the distance for two fixed values of $I_{\rm max}=\{0,~10^{-7}\}$ and different values of $N$. The behaviour of different lines as well as the optimal range of $p_{{\rm Z_A}}$ is very similar to the one of case 1. The main difference is that now the cut-off points where the secret key rate is zero are, as expected, larger than the ones in the previous case. For example, now when the total number of transmitted pulses is $10^{12}$ and $I_{\rm max}=10^{-6}$, we find that the secret key is positive up to about 34 km while in case 1 this distance is 24 km when we consider information leakage only from the IM.

\subsection{Case 3}
Finally, in this case we consider that Alice applies a phase randomization step to each signal leaving her transmitter. Moreover, for simplicity, we assume that there is no information leakage about the phase selected by Alice each given time for the phase randomization step. In so doing, we have that the state of Eve's back-reflected light is of the form
\begin{equation}
{\rho _{{\gamma ^k}}} = {e^{ -(\beta_{k})^2}}\sum\limits_{n = 0}^\infty  {\frac{{(\beta_{k})^2}}{{n!}}} \left| n \right\rangle \left\langle n \right|, \label{Case3}
\end{equation}
where the amplitudes $\beta_{k}$ satisfy Eq. (\ref{Case2}). That is, in this case the information about Alice's settings can only be leaked to Eve through the amplitudes of the back-reflected light but not through the phases. This is obviously the most beneficial situation for Alice and Bob.

\begin{figure}[!t]
 \centering
 \includegraphics*[scale=0.28]{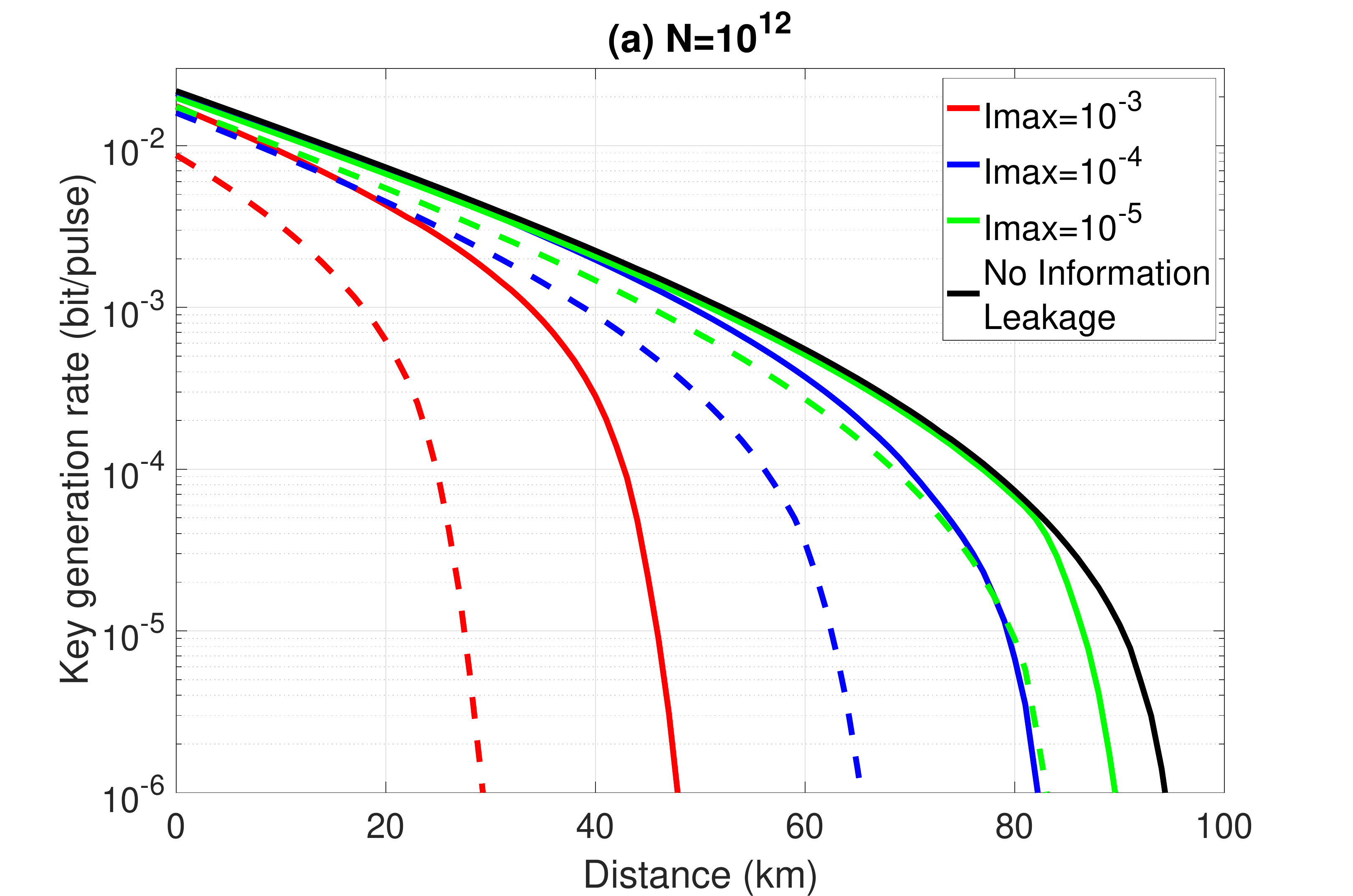}
 \includegraphics*[scale=0.28]{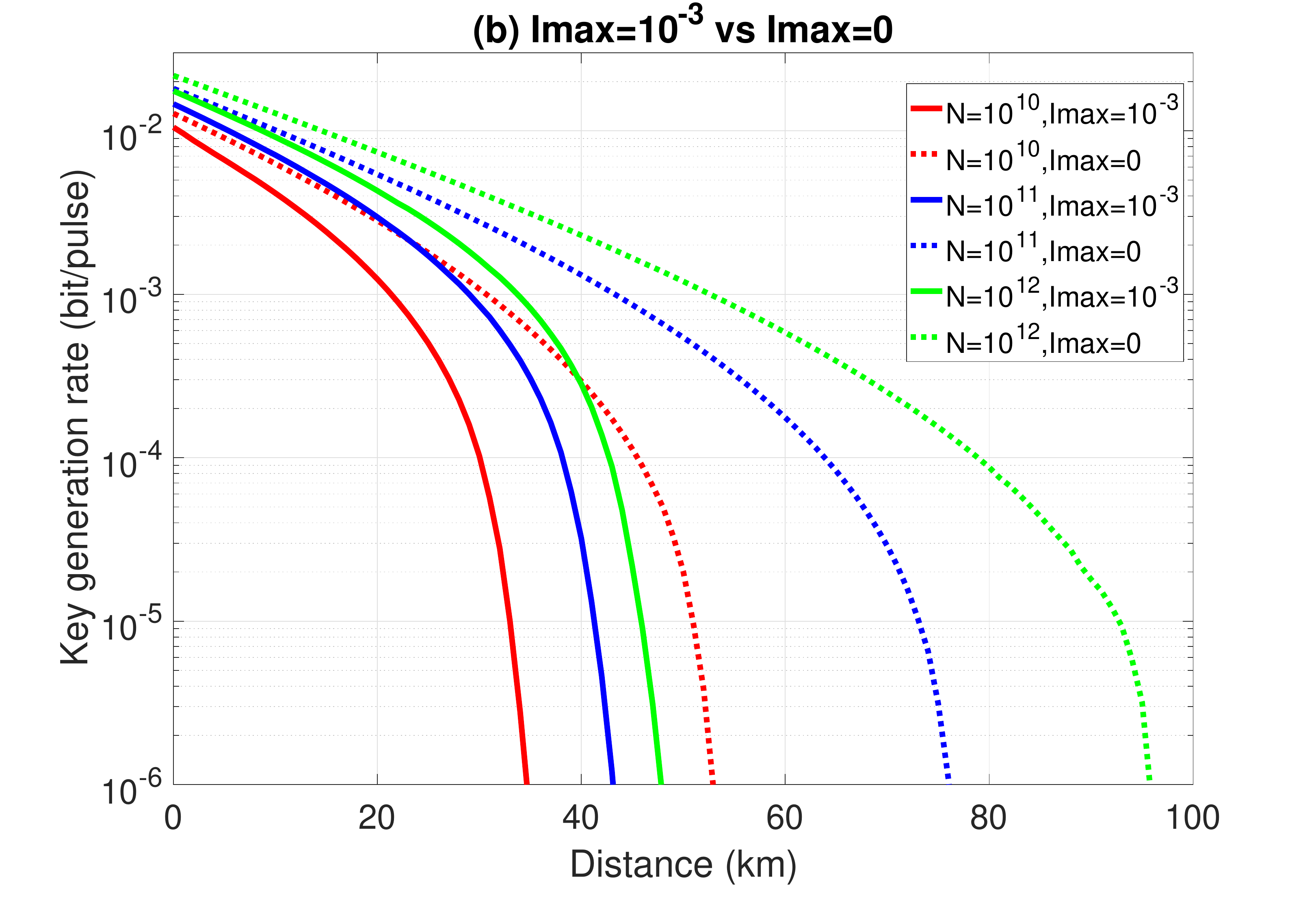}
\caption{ \footnotesize { (a) The secret key rate in logarithmic scale as a function of the distance for a fixed value of the total number of transmitted pulses, $N=10^{12}$. The black solid line represents the perfectly isolated situation where there is no information leakage (i.e., $I_{\rm max}=0$) and the different colored lines correspond to different amounts of information leakage. More precisely, the coloured solid (dashed) lines represent the secret key rates in the presence of a THA against the IM (both the IM and PM). (b) The secret key rate in logarithmic scale as a function of the distance for two fixed values of information leakage, $I_{\rm max}=\{0,~10^{-3}\}$. Different colored lines correspond to different values of the number of transmitted pulses. Here we consider a THA against both the IM and the PM. In our simulations, for each value of the distance we maximize the secret key rate over the amplitudes $\gamma^{\rm s}$ and $\gamma^{\rm v}$, and the probabilities $p_{\rm Z_{A_c}}$, $p_{\rm s}$, $p_{\rm v}$, $p_{\rm Z}$ which are controlled by Alice and Bob, and we minimize it over the angles $\theta_k$ controlled by Eve, respectively.}} \label{figcase3}
\end{figure}

The simulation result of the secret key rate as a function of the transmission distance between Alice and Bob when $N=10^{12}$ and for different values of $I_{\rm max}$ is shown in Fig.~\ref{figcase3}~(a) like in the previous two cases. Fig.~\ref{figcase3}~(b) shows the finite-key effect on the secret key rate as a function of the distance for two fixed values of $I_{\rm max}=\{0,~10^{-3}\}$ and for different values of $N$. Here, we find that the parameter $p_{\rm Z_{A_c}}$ typically lies in an interval $\left[0.75,0.95 \right]$ and the secret key rate is obviously improved compared with the ones shown in Figs~\ref{figcase1} and \ref{figcase2}. For example, when the total number of transmitted pulses is $10^{12}$ and $I_{\rm max}=10^{-4}$, the secret key rate remains positive up to about 83 km (66 km) in the presence of a THA against the IM (both the IM and the PM). Recall that in case 2 the maximum achievable distance with the same number of transmitted pulses and assuming an $I_{\rm max}$ as low as $10^{-6}$ is about 34 km (30 km), and the situation in case 1 is even worse. As already mentioned previously, this is mainly because now Eve can not get information from the phase of the output states.

In practice, however, Eve might also launch a THA to determine the random phase used by Alice each given time for phase randomization and the benefit of this step will be reduced. This last scenario could be also analyzed with the techniques presented in this paper and  we omit it here for simplicity.

\section{Conclusion}\label{secV}
In this paper, we have analyzed the finite-key security of a decoy-state QKD protocol in the presence of information leakage from the two main active devices of Alice's transmitter: the intensity modulator (IM) and the phase modulator (PM). For this, we have extended the results introduced in~\cite{tamaki2016decoy} to the finite-key regime. More precisely, we have evaluated the effect of information leakage from the IM by means of a trace distance argument which provides a relationship between the $n$-photon yields associated to different intensity settings. To take statistical fluctuations into account, we have used Azuma's inequality. This inequality considers arbitrary correlations between the different receiving pulses at Bob's side and thus is valid against general attacks. To evaluate the effect of information leakage from the PM, we have used the idea of a quantum coin. This method provides a means to study the security of a QKD system with basis dependent states. For the finite-key security proof to go through, we have included a classical post-processing step where Alice probabilistically post-selects part of the data which is associated to those detection events at Bob's side to be used for parameter estimation, while she discards the rest. This step reduces slightly the overall efficiency of the protocol; it is left as an open question whether there exist alternative procedures, probably not based on a quantum coin idea, which can handle information leakage from the PM without requiring such a random post-selection step.

For illustration purposes, we have simulated the resulting secret key rate in three practical examples of Trojan horse attacks which Eve could launch against the QKD system, as a function of the intensity of Eve's back reflected light and the total number of pulses sent by Alice. Our results demonstrate the feasibility of quantum key distribution over long distances and within a reasonable time frame given that Alice's source is sufficiently isolated. Also, for a given value of the isolation of Alice's source, we have shown that the effect that the information leakage has on the secret key rate is greater when the total number of transmitted pulses decreases.

\section*{Acknowledgements}
We thank Marco Lucamarini for valuable discussions. This work was supported by the Spanish Ministry of Economy and Competitiveness (MINECO), the Fondo Europeo de Desarrollo Regional (FEDER) through grant TEC2014-54898-R, and the European Commission (project ``QCALL"). W.W. gratefully acknowledges support from the National Natural Science Foundation of China under Grant No. 61472446. K.T. acknowledge support from JST-CREST JPMJCR 1671.

\section*{Appendix A. Azuma's inequality}
Azuma's inequality can be applied to random variables which satisfy both the martingale condition and the bounded difference condition (BDC)~\cite{azuma1967weighted}. A sequence of random variables $\{{X^{\left( 0 \right)}},{X^{\left( 1 \right)}},...,{X^{\left( k \right)}}\}$ is called a martingale if and only if the condition $E\left[ {{X^{\left( {k + 1} \right)}}\left| {{X^{\left( 0 \right)}},{X^{\left( 1 \right)}},...,{X^{\left( k \right)}}} \right.} \right] = {X^{\left( k \right)}}$ is satisfied for all $k\geq0$. Here, $E[ ~\cdot~ ]$ denotes the expectation value. Moreover, we say that the sequence $\{{X^{\left( 0 \right)}},{X^{\left( 1 \right)}},...,{X^{\left( k \right)}}\}$ fulfills the BDC if and only if there exists $c^{(k)}>0$ such that $\left| {{X^{\left( {k + 1} \right)}} - {X^{\left( k \right)}}} \right| \le {c^{\left( k \right)}}$ for all $k\geq0$. Suppose we have $N$ trials whose $k$th event is specified by a random variable ${X^{\left( k \right)}}$, then if ${X^{\left( k \right)}}$ is a martingale and satisfies the BDC with $c^{(k)} = 1$, Azuma's inequality states that~\cite{azuma1967weighted}
\begin{equation}\Pr \left[ {\left| {{X^{\left( N \right)}} - {X^{\left( 0 \right)}}} \right| > N\delta } \right] \le 2{e^{ - \frac{{N{\delta ^2}}}{2}}},
\end{equation}
holds for any $\delta \in (0,1)$.

To derive the result in the main text, let us define a random variable
\begin{equation}
{X^{\left( k \right)}}: = {\Pi ^{\left( k \right)}} - \sum\limits_{j = 1}^k {{{\rm{Pr}}^j}\left( {{\zeta _{j}}=1\left| {{\zeta _0},{\zeta _1},...,} \right.{\zeta _{j - 1}}} \right),} \label{RV}
\end{equation}
where $\Pi ^{\left( k \right)}$ denotes the actual observed number of events during the first $k$ trials, and ${{\rm Pr}^j\left( {{\zeta _{j}}=1\left| {{\zeta _1},{\zeta _2},...,} \right.{\zeta _{j - 1}}} \right)}$ is the conditional probability that in the $j$th trial the outcome is `1' given that the first $j - 1$ outcomes are ${{\zeta _0},{\zeta _1},...,{\zeta _{j - 1}}}$. It can be shown that Eq.~(\ref{RV}) is martingale and satisfies the BDC with $c^{(k)} = 1$. Therefore, according to Azuma's inequality we have that
\begin{equation}
\Pr \left[ {\left| {{\Pi ^{\left( N \right)}} - \sum\limits_{j = 1}^N {{{\rm Pr}^j}\left( {1\left| {{\zeta _0},...,} \right.{\zeta _{j - 1}}} \right)} } \right| > N\delta } \right] \le 2{e^{ - \frac{{N{\delta ^2}}}{2}}}.
\end{equation}
Equivalently, this means that
\begin{equation}
{\Pi ^{\left( N \right)}} = \sum\limits_{j = 1}^N {{{\rm Pr}^j}\left( {1\left| {{\zeta _0},...,} \right.{\zeta _{j - 1}}} \right)}  + \delta_{\rm A},
\end{equation}
except for error probability ${\varepsilon} + {\hat \varepsilon }$, where the deviation term $\delta_{\rm A}$ lies in an interval $ [ { - {\Delta},{{\hat \Delta }}} ]$ and the bounds are given by $\Delta = f( {{N },\varepsilon} )$ and $\widehat{\Delta} = f( {{N},\widehat{\varepsilon} })$, respectively, with the function $f( {x,y} ) = \sqrt {2x\ln (1/y)}$. That is, $\varepsilon $ quantifies the error probability that $\delta_{\rm A}$ is not lower bounded by $-\Delta $ and $\widehat{\varepsilon}$ quantifies the error probability that the parameter $\delta_{\rm A}$ is not upper bounded by $\widehat{\Delta}$.

\section*{Appendix B. Numerical estimation of $N^{L}_{{\rm click,0,\gamma^s}|\rm Z}$, $N^{L}_{{\rm click,1,\gamma^s}|\rm Z}$ and $e^{\rm U}_{\rm ph}$ with information leakage from the IM}

Let us start with the estimation of $N^{L}_{{\rm click,0,\gamma^s}|\rm Z}$. The first step is to reduce the infinite number of unknown variables which appear in the linear constraints derived in the main text to a finite set. Our starting point is Eq.~(\ref{Nsvw}). Due to the fact that the expected number of clicks arising from certain events is always greater than or equal to 0 but cannot be greater than the number of pulses sent associated to such events, we have that $0\leq \mathcal{E}_{{\rm click},n,\gamma^j|\chi} \leq N_{\chi}p_jp_n^j$ for all $n$ and $j \in \{\rm s,v,w\}$. This means, in particular, that
\begin{equation}
\begin{array}{*{20}{l}}\label{cut}
&\sum\limits_{n = 0}^\infty  {\left( {\frac{{{p_j}p_n^j}}{{{p_{\rm{s}}}p_n^{\rm{s}}}}{\mathcal{E}_{{\rm{click}},n,{\gamma ^{\rm{s}}}|\chi}}} \right)}\ge \sum\limits_{n = 0}^{S{\rm{cut}}} {\left( {\frac{{{p_j}p_n^j}}{{{p_{\rm{s}}}p_n^{\rm{s}}}}{\mathcal{E}_{{\rm{click}},n,{\gamma ^{\rm{s}}}|\chi}}} \right)}, \\
&\sum\limits_{n = 0}^\infty  {\left( {\frac{{{p_j}p_n^j}}{{{p_{\rm{s}}}p_n^{\rm{s}}}}{\mathcal{E}_{{\rm{click}},n,{\gamma ^{\rm{s}}}|\chi}}} \right)} \\
&\le\sum\limits_{n = 0}^{S{\rm{cut}}} {\left( {\frac{{{p_j}p_n^j}}{{{p_{\rm{s}}}p_n^{\rm{s}}}}{\mathcal{E}_{{\rm{click}},n,{\gamma ^{\rm{s}}}|\chi}}} \right)} + \sum\limits_{n = S{\rm{cut}} + 1}^\infty  {\left( {\frac{{{p_j}p_n^j}}{{{p_{\rm{s}}}p_n^{\rm{s}}}}{N_\chi }{p_{\rm{s}}}p_n^{\rm{s}}} \right)} \\
&=\sum\limits_{n = 0}^{S{\rm{cut}}} {\left( {\frac{{{p_j}p_n^j}}{{{p_{\rm{s}}}p_n^{\rm{s}}}}{\mathcal{E}_{{\rm{click}},n,{\gamma ^{\rm{s}}}|\chi}}} \right)}  + {N_\chi }{p_j}{T_{{\rm S_{cut}}}^j},
\end{array}
\end{equation}
where ${T_{{\rm S_{cut}}}^{j}} = \sum\limits_{n = S_{\rm cut} + 1}^\infty   {p_{n}^{j} = }  1 - \sum\limits_{n = 0}^{S_{\rm cut}}  {p_{n}^{j}}$ for any $S_{\rm cut} \geq0 $.
Thus, we obtain the following equations:
\begin{equation}\label{Ncut}
\begin{array}{*{20}{l}}
&{\mathcal{E}_{{\rm{click}},{\gamma ^{\rm{s}}}|\chi}} \ge \sum\limits_{n = 0}^{{S_{{\rm{cut}}}}} {{\mathcal{E}_{{\rm{click}},n,{\gamma ^{\rm s}}|\chi}}},\\
&{\mathcal{E}_{{\rm{click}},{\gamma ^{\rm{s}}}|\chi}} \le \sum\limits_{n = 0}^{{S_{{\rm{cut}}}}} {{\mathcal{E}_{{\rm{click}},n,{\gamma ^{\rm s}}|\chi}}} + N_{\chi}p_{\rm s}{T_{{\rm S_{cut}}}^{\rm s}}, \\
& {\mathcal{E}_{{\rm{click}},{\gamma ^{\rm{v}}}|\chi}} \ge \sum\limits_{n = 0}^{{S_{{\rm{cut}}}}} \frac{{{p_{\rm{v}}}p_n^{\rm{v}}}}{{{p_{\rm{s}}}p_n^{\rm{s}}}}{{\mathcal{E}_{{\rm{click}},n,{\gamma ^{\rm s}}|\chi}}} + {\Delta_{{\chi}} ^{{\rm{vs}}}} ,\\
& {{\mathcal{E}_{{\rm{click}},{\gamma ^{\rm{v}}}|\chi}} \le \sum\limits_{n = 0}^{{S_{{\rm{cut}}}}} \frac{{{p_{\rm{v}}}p_n^{\rm{v}}}}{{{p_{\rm{s}}}p_n^{\rm{s}}}}{{\mathcal{E}_{{\rm{click}},n,{\gamma ^{\rm s}}|\chi}}} + {N_{\chi}}{p_{\rm{v}}}{T_{{\rm S_{cut}}}^{\rm{v}}} + {\Delta_{\chi} ^{{\rm{vs}}}}} , \\
&{\mathcal{E}_{{\rm{click}},{\gamma ^{\rm{w}}}|\chi}} \ge \sum\limits_{n = 0}^{{S_{{\rm{cut}}}}} \frac{{{p_{\rm{w}}}p_n^{\rm{w}}}}{{{p_{\rm{s}}}p_n^{\rm{s}}}}{{\mathcal{E}_{{\rm{click}},n,{\gamma ^{\rm s}}|\chi}}} + {\Delta_{\chi} ^{{\rm{ws}}}} , \\
&{\mathcal{E}_{{\rm{click}},{\gamma ^{\rm{w}}}|\chi}}  \le \sum\limits_{n = 0}^{{S_{{\rm{cut}}}}} \frac{{{p_{\rm{w}}}p_n^{\rm{w}}}}{{{p_{\rm{s}}}p_n^{\rm{s}}}}{{\mathcal{E}_{{\rm{click}},n,{\gamma ^{\rm s}}|\chi}}} +{N_{\chi}}{p_{\rm{w}}}{T_{{\rm S_{cut}}}^{\rm{w}}} + {\Delta_{\chi} ^{{\rm{ws}}}},
\end{array}
\end{equation}
which now depend on a finite number of unknown variables.

Next, we replace the expected values $\mathcal{E}_{{\rm{click}},{\gamma ^j}|\chi}$ and $\mathcal{E}_{{\rm{click}},n,{\gamma ^j}|\chi}$ with the corresponding actual numbers plus their deviation terms by applying Eqs.~(\ref{D}) and (\ref{Nn}), and we take into account as well the constraints given by~Eq. (\ref{Dsvw}) after considering any $n\leq S_{\rm cut}$. This way, we find that to calculate $N^{L}_{{\rm click,0,\gamma^s}|\rm Z}$ one could use the following linear program:
\begin{widetext}
\begin{equation}
\begin{array}{*{20}{l}} \label{LP2}
\min {\kern 8pt}  &N_{{\rm{click}},0,{\gamma ^{\rm s}}|\rm Z} \\
s.t. {\kern 8pt} & N_{{\rm{click}},{\gamma ^{\rm s}}|\rm Z}  \ge \sum\limits_{n = 0}^{{S_{{\rm{cut}}}}} \left( { N_{{\rm{click}},n,{\gamma ^{\rm s}}|\rm Z} + \delta _{{\rm Z},n}^{\rm{s}}} \right)- \delta_{\rm Z}^{\rm s},\\
& N_{{\rm{click}},{\gamma ^{\rm s}}|\rm Z}   \le \sum\limits_{n = 0}^{{S_{{\rm{cut}}}}} \left( { N_{{\rm{click}},n,{\gamma ^{\rm s}}|\rm Z}  + \delta _{{\rm Z},n}^{\rm{s}}} \right)+ N_{\rm Z}p_{\rm s}{T_{{\rm S_{cut}}}^{\rm s}}- \delta_{\rm Z}^{\rm s}\\
&  N_{{\rm{click}},{\gamma ^{\rm v}}|\rm Z}  \ge \sum\limits_{n = 0}^{{S_{{\rm{cut}}}}} {\frac{{{p_{\rm{v}}}p_n^{\rm{v}}}}{{{p_{\rm{s}}}p_n^{\rm{s}}}}\left( { N_{{\rm{click}},n,{\gamma ^{\rm s}}|\rm Z}  + \delta _{{\rm Z},n}^{\rm{s}}} \right)}+ {\Delta_{{\rm Z}} ^{{\rm{vs}}}}- \delta_{\rm Z}^{\rm v},\\
& { N_{{\rm{click}},{\gamma ^{\rm v}}|\rm Z}  \le \sum\limits_{n = 0}^{{S_{{\rm{cut}}}}} {\frac{{{p_{\rm{v}}}p_n^{\rm{v}}}}{{{p_{\rm{s}}}p_n^{\rm{s}}}}\left( {N_{{\rm{click}},n,{\gamma ^{\rm s}}|\rm Z}  + \delta _{{\rm Z},n}^{\rm{s}}} \right)} + {N_{\rm{Z}}}{p_{\rm{v}}}{T_{{\rm S_{cut}}}^{\rm{v}}} + {\Delta_{{\rm Z}} ^{{\rm{vs}}}}} - \delta_{\rm Z}^{\rm v}, \\
& N_{{\rm{click}},{\gamma ^{\rm w}}|\rm Z}  \ge \sum\limits_{n = 0}^{{S_{{\rm{cut}}}}} {\frac{{{p_{\rm{w}}}p_n^{\rm{w}}}}{{{p_{\rm{s}}}p_n^{\rm{s}}}}\left( { N_{{\rm{click}},n,{\gamma ^{\rm s}}|\rm Z} + \delta _{{\rm Z},n}^{\rm{s}}} \right)} + {\Delta_{{\rm Z}} ^{{\rm{ws}}}}- \delta_{\rm Z}^{\rm w}, \\
&N_{{\rm{click}},{\gamma ^{\rm w}}|\rm Z} \le \sum\limits_{n = 0}^{{S_{{\rm{cut}}}}} {\frac{{{p_{\rm{w}}}p_n^{\rm{w}}}}{{{p_{\rm{s}}}p_n^{\rm{s}}}}\left( {N_{{\rm{click}},n,{\gamma ^{\rm s}}|\rm Z}  + \delta _{{\rm Z},n}^{\rm{s}}} \right)} +{N_{\rm{Z}}}{p_{\rm{w}}}{T_{{\rm S_{cut}}}^{\rm{w}}}+ {\Delta_{{\rm Z}} ^{{\rm{ws}}}} - \delta_{\rm Z}^{\rm w}, \\
& 0\leq N_{{\rm{click}},n,{\gamma ^{\rm s}}|\rm Z}+ \delta _{{\rm Z},n}^{\rm{s}}\leq N_{\rm Z}p_{\rm s}p_n^{\rm s},~~\forall n \leq S_{\rm cut},\\
&-\Delta _{{\rm Z}}^j\leq \delta_{{\rm Z}}^j \leq \widehat{\Delta}^j _{{\rm Z}},~-\Delta _{{\rm Z},n}^{\rm s}\leq \delta_{{\rm Z},n}^{\rm s} \leq \widehat{\Delta}^{\rm s} _{{\rm Z},n},~~j\in \{\rm s,v,w\}\\
& -\sum\limits_{n = 0}^{\infty}{p_{k}p_n^{k}N_{\rm Z}D_{n,\gamma^k,\gamma^{\rm {s}}}}\leq \Delta_{{\rm Z}}^{k\rm s}\leq \sum\limits_{n = 0}^{\infty}{p_{k}p_n^{k}N_{\rm Z}D_{n,\gamma^k,\gamma^{\rm{s}}}},~~k\in \{\rm v,w\} \\
& -D_{n,\{\gamma^{\rm{s}}\},\{\gamma^{\rm{v}},\gamma^{\rm{w}}\}}\leq \frac{q_{n\rm {vw}}}{p_{\rm s}p_n^{\rm s}N_{\rm Z}}\frac{p_{\rm s}p_n^{\rm s}}{p_{\rm v}p_n^{\rm v}}\Delta_{\rm Z}^{\rm vs}+\frac{\left( {1 - {q_{n\rm vw}}} \right)}{p_{\rm s}p_n^{\rm s}N_{\rm Z}}\frac{p_{\rm s}p_n^{\rm s}}{p_{\rm w}p_n^{\rm w}}\Delta_{\rm Z}^{\rm ws} \leq D_{n,\{\gamma^{\rm{s}}\},\{\gamma^{\rm{v}},\gamma^{\rm{w}}\}},~~\forall n\leq S_{\rm cut}\\
& -D_{n,\{\gamma^{\rm{v}}\},\{\gamma^{\rm{s}},\gamma^{\rm{w}}\}}\leq \frac{1}{p_{\rm v}p_n^{\rm v}N_{\rm Z}}\Delta_{\rm Z}^{\rm vs}-\frac{\left( {1 - {q_{n\rm sw}}} \right)}{p_{\rm v}p_n^{\rm v}N_{\rm Z}}\frac{p_{\rm v}p_n^{\rm v}}{p_{\rm w}p_n^{\rm w}}\Delta_{\rm Z}^{\rm ws} \leq D_{n,\{\gamma^{\rm{v}}\},\{\gamma^{\rm{s}},\gamma^{\rm{w}}\}},~~\forall n\leq S_{\rm cut}\\
& -D_{n,\{\gamma^{\rm{w}}\},\{\gamma^{\rm{s}},\gamma^{\rm{v}}\}}\leq \frac{1}{p_{\rm w}p_n^{\rm w}N_{\rm Z}}\Delta_{\rm Z}^{\rm ws}-\frac{\left( {1 - {q_{n\rm sv}}} \right)}{p_{\rm w}p_n^{\rm w}N_{\rm Z}}\frac{p_{\rm w}p_n^{\rm w}}{p_{\rm v}p_n^{\rm v}}\Delta_{\rm Z}^{\rm vs} \leq D_{n,\{\gamma^{\rm{w}}\},\{\gamma^{\rm{s}},\gamma^{\rm{v}}\}},~~\forall n\leq S_{\rm cut}.
\end{array}
\end{equation}
\end{widetext}
The linear program contains $2\times(S_{\rm cut} + 1) + 5$ unknown variables: $ N_{{\rm{click}},n,{\gamma ^{\rm s}}|\rm Z} $, $\delta _{{\rm Z} ,n}^{\rm{s}}$, $\Delta_{\rm Z}^{\rm vs}$, $\Delta_{\rm Z}^{\rm ws}$, $\delta _{\rm Z} ^{\rm s}$, $\delta _{\rm Z} ^{\rm v}$, and $\delta _{\rm Z} ^{\rm w}$. The calculation of the parameters $D_{n,\{\gamma^{j}\},\{\gamma^{k},\gamma^{l}\}}$ with $j,k,l\in \{\rm s,v,w\}$ is provided in Appendix E and is taken from Ref.~\cite{tamaki2016decoy}. The solution to the linear program above is directly $N^{L}_{{\rm click,0,\gamma^s}|\rm Z}$, being the total error probability equal to
\begin{equation}
{\varepsilon _{{\rm{Z}},{\rm{0}}}} = \sum\limits_{j = {\rm{s}},{\rm{v}},{\rm{w}}} {( {\varepsilon _{\rm{Z}}^j + \hat \varepsilon _{\rm{Z}}^j} )}  + \sum\limits_{n = 0}^{{S_{{\rm{cut}}}}} {\left( {\varepsilon _{{\rm{Z}},n}^{\rm{s}} + \hat \varepsilon _{{\rm{Z}},n}^{\rm{s}}} \right)} ,
\end{equation}
where $\varepsilon _{\rm Z}^{j}$, $\hat \varepsilon _{\rm Z}^{j}$ are the error probabilities associated to the estimation of the bounds on ${\delta_{\rm Z}^{j}} $ and $\varepsilon _{{\rm{Z}},n}^{\rm{s}}$, $\hat \varepsilon _{{\rm{Z}},n}^{\rm{s}}$ are the error probabilities associated to the estimation of the bounds on ${\delta_{{\rm Z},n}^{\rm s}}$, respectively.

The procedure to obtain $N^{L}_{{\rm click,1,\gamma^s}|\rm Z}$ is analogous. In particular, one can basically use the same linear program given by Eq.~ (\ref{LP2}) after replacing ``min $ N_{{\rm{click}},0,{\gamma ^{\rm s}}|\rm Z} $'' with ``min $ N_{{\rm{click}},1,{\gamma ^{\rm s}}|\rm Z} $''.

To calculate $e^{\rm U}_{\rm ph}$, first we can obtain expressions for error events which are similar to the ones given by  Eq.~(\ref{Ncut}) by utilizing the fact that $0\leq \mathcal{E}_{{\rm error},n,\gamma^j|\chi} \leq N_{\chi}p_jp_n^j$ for all $n$ and $j \in \{\rm s,v,w\}$, where $\mathcal{E}_{{\rm error},n,\gamma^j|\chi}$ denotes the expected number of events where Alice selects the intensity $\gamma^j$ and sends Bob an $n$-photon pulse, and Bob obtains an error given that both Alice and Bob select the $\chi$ basis.
Then we estimate a lower bound on the number of single-photon click events in the X basis with intensity $\gamma^{\rm s}$, which we denote by $N^{L}_{{\rm click,1,\gamma^s}|\rm X}$, as well as an upper bound on the corresponding number of errors, which we denote by $N^{U}_{{\rm error,1,\gamma^s}|\rm X}$. To obtain $N^{L}_{{\rm click,1,\gamma^s}|\rm X}$, we can use the linear program which has been  used to estimate $N^{L}_{{\rm click,1,\gamma^s}|\rm Z}$ after replacing all the parameters and variables in the Z basis with those in the X basis. Similarly, we can also modify the program for estimating $N^{L}_{{\rm click,1,\gamma^s}|\rm X}$ to calculate $N^{U}_{{\rm error,1,\gamma^s}|\rm X}$. For this, we simply replace all the numbers that refer to click events with those referring to error events. In addition, we replace ``min $N_{{\rm 1,\gamma^s}|\rm X}$'' with ``min $-N_{{\rm error,1,\gamma^s}|\rm X}$'', which is equivalent to maximizing $N_{{\rm error,1,\gamma^s}|\rm X}$. Finally, given the values of $N^{L}_{{\rm click,1,\gamma^s}|\rm Z}$, $N^{L}_{{\rm click,1,\gamma^s}|\rm X}$ and $N^{U}_{{\rm error,1,\gamma^s}|\rm X}$, we use random sampling without replacement to estimate $e^{\rm U}_{\rm ph}$. For this, we use the method employed in~\cite{tomamichel2012tight,curty2014finite}, which relies on Serfling's inequality~\cite{serfling1974probability}. We obtain that
\begin{widetext}
\begin{equation}
\begin{array}{*{20}{l}}
e^{\rm U}_{\rm ph} =\frac{1}{N^{L}_{{\rm click,1,\gamma^s}|\rm Z}}\min \Bigg\{ \Bigg[ &{{N^{L}_{{\rm click,1,\gamma^s}|\rm Z}}} \frac{{{N^{U}_{{\rm error,1,\gamma^s}|\rm X}}}}{{{N^{L}_{{\rm click,1,\gamma^s}|\rm X}}}} + \left( {{N^{L}_{{\rm click,1,\gamma^s}|\rm Z}}} + {N^{L}_{{\rm click,1,\gamma^s}|\rm X}} \right)\\
& \times \Upsilon \left( {N^{L}_{{\rm click,1,\gamma^s}|\rm Z}},{N^{L}_{{\rm click,1,\gamma^s}|\rm X}},{\varepsilon' } \right) \Bigg] ,{N^{L}_{{\rm click,1,\gamma^s}|\rm Z}} \Bigg\},
\end{array}
\end{equation}
\end{widetext}
except for error probability
\begin{equation}
{\varepsilon _{{\rm{ph}},{\rm{1}}}} \le {\varepsilon'} + {\varepsilon _{{\rm{X}}{\rm{,1}}}} + {\varepsilon _{{{\rm{E}}_{\rm{X}}}{\rm{,1}}}},
\end{equation}
where the function $\Upsilon \left( x,y,z \right)$ is defined as $\Upsilon \left( {x,y,z} \right) = \sqrt {\left( {x + 1} \right)\ln \left( {{z^{ - 1}}} \right)/\left[ {2y\left( {x + y} \right)} \right]}$, and ${\varepsilon _{{\rm{X}}{\rm{,1}}}}$ and ${\varepsilon _{{{\rm{E}}_{\rm{X}}}{\rm{,1}}}}$ are the error probabilities associated to the estimation of $N^{L}_{{\rm click,1,\gamma^s}|\rm X}$ and $N^{U}_{{\rm error,1,\gamma^s}|\rm X}$, respectively.
\section*{Appendix C. Quantum systems,~symbols, random variables and sets}
In this section, we summarize all the quantum systems, symbols, random variables and sets that appear in the main text in tables~\ref{Qsys},~\ref{Rvs}.

\begin{table*}[tbp]
\centering\label{Qsys}
\caption{Quantum systems and symbols}
\begin{tabular}{|m{2.1cm}<{\centering}|m{13.9cm}<{\centering}|}
\hline
   $E_{\rm p}$ & Eve's probe system\\
   \hline
   $E_{\rm a}$ & Eve's ancillary system\\
   \hline
   $E'_{\rm p}$ & Back-reflected light from $E_{\rm p}$\\
   \hline
   $\rho _{n,\chi}^{{S_{\alpha},i} }$ & Normalized joint state of Alice's $n$-photon pulse in the $i$th trial when she selects an intensity in the subset $S_{\alpha}$ and the basis $\chi$ and Eve's systems $E_{\rm a}$, $E'_{\rm p}$\\
\hline
   ${\rm A}_{\rm q}$ & Virtual qubit, which contains Alice's bit value choice\\
   \hline
   ${\rm A}_{\rm p}$ & Alice's photonic system that she sends to Bob via a quantum channel\\
   \hline
   ${\rm A}_{\rm a}$ & Ancillary system in Alice's hands to account for the loss in her transmitter\\
   \hline
   ${\rm A}_{\rm c}$ & Alice's quantum coin\\
  \hline
   $p_{\rm Z}$ ($p_{\rm X}$) & Probability for Alice/Bob to choose the Z (X) basis to prepare her state\\
   \hline
   $p_{\rm Z_{A_c}}$ ($p_{\rm X_{A_c}}$) & Probability for Alice to choose the fictitious basis ${\rm Z}_{\rm c}$ (${\rm X}_{\rm c})$ \\
   \hline
   $N^j_{\rm Z}$ ($N^j_{\rm X}$) & Predetermined threshold value for the set $\widehat{Z}^j$ ($\widehat{X}^j$)\\
   \hline
   $N^j_{\rm Z}$ ($N^j_{\rm X}$) & Predetermined threshold value for the set $\widehat{Z}^j$ ($\widehat{X}^j$)\\
   \hline
   $E_{\rm Z}^{\rm s}$ & Predetermined quantum bit error rate\\
   \hline
   $\varepsilon_{\rm sec}$ & Secrecy parameter of the protocol\\
   \hline
   $\varepsilon_{\rm cor}$ & Correctness parameter of the protocol\\
   \hline
   $\ell'$ & Length of the final key\\
   \hline
   $N$ & Total number of pulses sent by Alice\\
   \hline
   $p_l$ & Probability for Alice to choose the intensity setting $l$\\
   \hline
   $p_n^l$ & Conditional probability that Alice emits a pulse with $n$ photons given that she chooses the intensity setting $l$\\
   \hline
   $\beta$ & Limited amplitude of Eve's input light\\
   \hline
   $\theta$ & Phase of Eve's input light\\
   \hline
   $\beta_k$ & Amplitude of Eve's back-reflected light\\
   \hline
   $\theta_k$ & Phase of Eve's back-reflected light\\
   \hline
   $I_{\rm max}$ & Upper bound on the intensity of Eve's back-reflected light\\
\hline
\end{tabular}
\end{table*}

\begin{table*}[tbp]
\caption{Random variables and sets}
\centering \label{Rvs}
\begin{tabular}{|m{2cm}<{\centering}|m{14cm}<{\centering}|}
\hline
   $\chi_{\rm A}$ & Alice's basis choice with $\chi_{\rm A}\in\{Z,X\}$\\\hline
   $\chi_{\rm B}$ & Bob's basis choice with $\chi_{\rm B}\in\{Z,X\}$\\\hline
   $\chi$ & The same basis choice by Alice and Bob\\\hline
   ${\rm Z}_{\rm c}$ (${\rm X}_{\rm c}$) & Alice's fictitious basis choice\\\hline
   $\gamma^j$ & Alice's intensity setting with $j \in\{\rm s,v,w\}$\\\hline
   $\widehat{Z}^j$ ($\widehat{X}^j$) &  Set of indexes that identifies the click events where Alice chose the ${\rm Z}_{\rm c}$ basis and the intensity $\gamma^j$ and both Alice and Bob chose the basis Z (X)\\\hline
   ${Z}^j$ (${X}^j$) & Post-selected set with $|Z^j|=N^j_{\rm Z}$ ($|X^j|=N^j_{\rm X}$)\\\hline
   $N^{L}_{{\rm click,0,\gamma^s}|\rm Z}$ ($N^{L}_{{\rm click,1,\gamma^s}|\rm Z}$) & Lower bound on the number of vacuum (single-photon) events in the sifted key data identified by the set ${Z}^{\rm s}$. \\\hline
   $e^{\rm U}_{\rm ph}$ &  Upper bound on the single-photon phase error rate of the sifted key data indexed by ${Z}^{\rm s}$\\\hline
   $leak_{\rm EC}$ & Amount of syndrome information revealed by Alice in error correction step\\\hline
   $S$ & Alice's intensity set $\{\gamma^1,\gamma^2,...\gamma^m\}$\\\hline
   $S_{\alpha}$ & Subset of $S$\\\hline
   $D_{n,{S_{\rm{1}}},{S_{\rm{2}},\chi}}^i$ & Trace distance between the states $\rho _{n,\chi}^{{S_{\rm{1}}},i}$ and $\rho _{n,\chi}^{{S_{\rm{2}}},i}$\\\hline
      $\mathcal{E}_{\lambda}$ & Expected number of times that event `$\lambda$' happens\\\hline
   $N_{\lambda}$ & Actual number of times that event `$\lambda$' happens\\\hline
   ${\Delta_{\chi,n} ^{jk}}$ & Deviation term between $\mathcal{E}_{{\rm click},n,\gamma^j|\chi}$ and $\mathcal{E}_{{\rm click},n,\gamma^k|\chi}$ due to information leakage from the IM\\\hline
  $\delta_{\lambda}$ & Deviation term between the expected number and actual number of times that event
   `$\lambda$' happens when using Azuma's inequality\\\hline
   $\varepsilon_{\lambda}$ ($\widehat \varepsilon_{\lambda}$)& Error probability when using Azuma's inequality to estimate a lower (upper) bound on $\delta_{\lambda}$\\\hline
   $N^{L}_{{\rm click,1,\gamma^s}|\rm X}$ & Lower bound on the number of single-photon click events in the X basis with intensity $\gamma^{\rm s}$\\\hline
   $N^{U}_{{\rm error,1,\gamma^s}|\rm X}$ & Upper bound on the of errors of single-photon click events in the X basis with intensity $\gamma^{\rm s}$\\\hline
   $N_{{{\rm error},\gamma^j}|\rm X}$ & Actual number of errors given that Alice sends Bob a pulse with intensity $\gamma^j$ and both Alice and Bob select the X basis\\\hline
   $N_{{{\rm error},n,\gamma^j}|\rm X}$ & Actual number of errors given that Alice selects the intensity $\gamma^j$ and sends Bob an $n$-photon pulse and both Alice and Bob select the X basis\\\hline
   $\delta'_{\rm X_{{{{\rm{A}}_{\rm{c}}}}=  -}}$ & Deviation term when using Chernoff bound to estimate $N_{\rm X_{{\rm{A}}_{\rm{c}}}=-}$\\\hline
   $\varepsilon'_{\rm X_{{{{\rm{A}}_{\rm{c}}}}=  -}}$ ($\widehat \varepsilon '_{{{\rm{X}}_{{{\rm{A}}_{\rm{c}}}{\rm{ =  - }}}}}$) & Error probability when using Chernoff bound to estimate a lower (upper) bound on $\delta'_{\rm X_{{{{\rm{A}}_{\rm{c}}}}=  -}}$\\\hline
\end{tabular}
\end{table*}

\section*{Appendix D. Estimation of $e^{\rm U}_{\rm ph}$ in the presence of a THA against the IM and the PM}
First, as what we did in Sec. III 2 in the main text, for simplicity we focus on the single-photon scenario. For this, let us consider a single-photon BB84 QKD protocol. The steps of this protocol can be directly obtained from those of the decoy-state BB84 protocol defined in Sec. II in the main text. In particular, this protocol is defined as follows (where, for simplicity, here we have not included the parameter estimation, information reconciliation and privacy amplification steps):
\begin{enumerate}
  \item The first two steps of the protocol are repeated $N$ times. In each round, Alice probabilistically selects a basis $\chi_{\rm A} \in\{\rm Z,~\rm X\}$ with probabilities $p_{\rm Z}$ and $p_{\rm X}=1-p_{\rm Z}$, respectively. Then she prepares a single-photon state in the selected basis and sends it to Bob via a quantum channel.
  \item For each incoming signal, Bob selects a measurement basis $\chi_{\rm B} \in\{\rm Z,~\rm X\}$ with probabilities $p_{\rm Z}$ and $p_{\rm X}=1-p_{\rm Z}$, respectively. He employs the selected basis to measure it and takes note if he obtains a click event.
  \item Then, Bob announces which events result in a click. Alice and Bob discard all the data associated to the non-click events.
  \item  For each click event, Alice selects the ${\rm Z}_{\rm A_c}$ or ${\rm X}_{\rm A_c}$ basis with probabilities $p_{\rm Z_{A_c}}$ and $p_{\rm X_{A_c}}$, respectively (we shall call this Alice's coin basis choice), and she announces her choice.
  \item If Alice's coin basis choice is ${\rm X}_{\rm A_c}$, Bob declares his measurement basis choice but Alice does not declare her state preparation basis choice. Afterwards, they discard the corresponding system and data. If Alice's choice is ${\rm Z}_{\rm A_c}$, Alice and Bob declare their state preparation and measurement basis choices. If Alice's and Bob's basis choices disagree, they discard the corresponding systems and data. If they match, then Alice and Bob keep the corresponding data.
  \item Finally, Alice and Bob announce all the data corresponding to the case where Alice selected the coin basis ${\rm Z}_{\rm A_c}$ and Alice selected the ${\rm X}$ basis to prepare her state and Bob selects the ${\rm X}$ basis to measurement it.
\end{enumerate}

As already pointed out in the main text, in the presence of a THA against the PM, the joint state of the single-photon states emitted by Alice together with Eve's back-reflected light could be basis dependent. To estimate the phase error rate in this scenario, we use the following fictitious protocol:
\begin{enumerate}
  \item The first two steps of the protocol are repeated $N$ times. In each round, Alice prepares a state of the form:
  \begin{equation} \label{state}
  \begin{array}{*{20}{l}}
&{\left| {{\Psi ^i}} \right\rangle _{{{\rm{A}}_c},{{\rm{A}}_{\rm{q}}},{{\rm{A}}_{\rm{p}}},{{\rm{A}}_{\rm{a}}},{{\rm{E}}^\prime }_{\rm{p}},{{\rm{E}}_{\rm{a}}}}}\\ &=\sqrt {{p_{{{\rm{Z}}}}}} {\left| 0 \right\rangle _{{{\rm{A}}_{\rm{c}}}}}{\left| {\Psi _{\rm{Z}}^i} \right\rangle _{{{\rm{A}}_{\rm{q}}},{{\rm{A}}_{\rm{p}}},{{\rm{A}}_{\rm{a}}},{{\rm{E}}^\prime }_{\rm{p}},{{\rm{E}}_{\rm{a}}}}}\\& + \sqrt {{p_{{{\rm{X}}}}}} {\left| 1 \right\rangle _{{{\rm{A}}_{\rm{c}}}}}{\left| {\Psi _{\rm{X}}^i} \right\rangle _{{{\rm{A}}_{\rm{q}}},{{\rm{A}}_{\rm{p}}},{{\rm{A}}_{\rm{a}}},{{\rm{E}}^\prime }_{\rm{p}},{\rm{E}}_{\rm{a}}}},
\end{array}
\end{equation}
where ${\rm A}_{\rm c}$ is the so-called quantum coin. She keeps the systems ${\rm A}_{\rm c}$, $ {\rm A}_{\rm q}$ and $ {\rm A}_{\rm a}$ in a quantum memory, and sends system $ {\rm A}_{\rm p}$ to Bob.
  \item Bob performs a quantum nondemolition (QND) measurement on each received signal and takes note if he obtains a click event. For each click event, Bob selects the measurement basis between the ${\rm Z}$ and ${\rm X}$ basis with probabilities $p_{\rm Z}$ and $p_{\rm X}$, respectively.
  \item Then Bob announces which events result in a click. Alice and Bob discard all the systems and data associated to non-click events.
    \item For each click event, Alice selects the ${\rm Z}_{\rm A_c}$ or ${\rm X}_{\rm A_c}$ basis with probabilities $p_{\rm Z_{A_c}}$ and $p_{\rm X_{A_c}}$, respectively, and measures the quantum coin in the selected basis. Note that when Alice selects the ${\rm Z}_{\rm A_c}$ basis, the result of her measurement on the quantum coin directly determines her state preparation basis choice (i.e., ${\rm Z}_{\rm A_c}=0$ implies the $\rm Z$ basis and ${\rm Z}_{\rm A_c}=1$ implies the $\rm X$ basis). If her choice is the ${\rm X}_{\rm A_c}$ basis, then Alice selects the ${\rm Z}$ or ${\rm X}$ basis with probabilities $p_{\rm Z}$ and $p_{\rm X}$, respectively. Then she announces her coin basis choice.
  \item If Alice's coin basis choice is ${\rm X}_{\rm A_c}$, Bob declares his measurement basis choice but Alice does not declare her state preparation basis choice. If Alice's coin basis choice is ${\rm Z}_{\rm A_c}$, Alice and Bob declare their state preparation and measurement basis choices. Moreover, if Alice's and Bob's basis choices disagree, they discard the corresponding systems and data. If they match, both Alice and Bob measure their systems in the ${\rm X}$ basis, rather than in the original selected basis.
  \item Finally, Alice and Bob announce all the data corresponding to the case where Alice's ${\rm Z}_{\rm A_c}$ basis measurement outcome on the quantum coin was `1' and Bob selected the ${\rm X}$ basis in step 2. Also, Alice counts the number of events ``$\rm X_{\rm{A_{c}}}=-$'' in $N$ trials.
\end{enumerate}

Importantly, note that from Eve's point of view the fictitious protocol defined above is equivalent to the actual protocol in the sense that the quantum states and the classical information interchanged between Alice and Bob are the same in both protocols.

Then, by applying the Bloch sphere bound in terms of probabilities~\cite{tamaki2003unconditionally} to this virtual scenario, one obtains
\begin{equation}
\begin{array}{*{20}{l}}
&{1 - 2{{\Pr }^i}\left( {{{\rm{X}}_{{{\rm{A}}_{\rm{c}}}}} =  - \left| {{\rm{click}}{\rm{,sb}}{\rm{,X - error}},{{\rm{X}}_{{{\rm{A}}_{\rm{c}}}}}} \right.} \right)}\\
 &\le2\sqrt {{{\Pr }^i}\left( {{{\rm{Z}}_{{{\rm{A}}_{\rm{c}}}}} = 1\left| {{\rm{click}}{\rm{,sb}}{\rm{,X - error}},{{\rm{Z}}_{{{\rm{A}}_{\rm{c}}}}}} \right.} \right)}\\
 &\times\sqrt{ {1 - {{\Pr }^i}\left( {{{\rm{Z}}_{{{\rm{A}}_{\rm{c}}}}} = 1\left| {{\rm{click}}{\rm{,sb}}{\rm{,X - error}},{{\rm{Z}}_{{{\rm{A}}_{\rm{c}}}}}} \right.} \right)} } ,
\end{array}\label{Xerror}
\end{equation}
\begin{equation}
 \begin{array}{*{20}{l}}
&{1 - 2{{\Pr }^i}\left( {{{\rm{X}}_{{{\rm{A}}_{\rm{c}}}}} =  - \left| {{\rm{click}}{\rm{,sb}}{\rm{,No}}\;{\rm{X - error}},{{\rm{X}}_{{{\rm{A}}_{\rm{c}}}}}} \right.} \right)}\\
&\le2\sqrt {{{\Pr }^i}\left( {{{\rm{Z}}_{{{\rm{A}}_{\rm{c}}}}} = 1\left| {{\rm{click}}{\rm{,sb}}{\rm{,No}}\;{\rm{X - error}},{{\rm{Z}}_{{{\rm{A}}_{\rm{c}}}}}} \right.} \right)}\\
 &\times\sqrt{ {1 - {{\Pr }^i}\left( {{{\rm{Z}}_{{{\rm{A}}_{\rm{c}}}}} = 1\left| {{\rm{click}}{\rm{,sb}}{\rm{,No}}\;{\rm{X - error}},{{\rm{Z}}_{{{\rm{A}}_{\rm{c}}}}}} \right.} \right)} }.
\end{array} \label{NoXerror}
\end{equation}

In order to upper bound the phase error rate, our strategy is first to convert each probability appearing in Eqs.~(\ref{Xerror}) and (\ref{NoXerror}) into a conditional probability conditioned only on a click event. Next, we obtain a relationship among all the conditional probabilities by adding the converted inequalities. Note that these probabilities are related to the events we obtain in the trials after step 4 in the fictitious protocol, and most importantly the number of the trials is fixed to the number of the click events. Thanks to this, we can apply Azuma's inequality and Chernoff bound (note that these inequalities assume a fixed probability space) to those conditional probabilities to convert the relationship in terms of the number.

Let us first convert each probability appearing in Eqs. (43) and (44) into conditional probabilities conditioned only on the click event. Due to the fact that the events `${{{\rm{X}}_{{{\rm{A}}_{\rm{c}}}}}}$' and `${{{\rm{Z}}_{{{\rm{A}}_{\rm{c}}}}}}$' are independent of the events `click' and `sb' and, also, we have that the event `click' is independent of the event `sb', we find that
\begin{equation}
\begin{array}{*{20}{l}}
&{\Pr ^i}\left( {{\rm{sb}},{\rm{X - error}}\left| {{\rm{click}},{{\rm{X}}_{{{\rm{A}}_{\rm{c}}}}}} \right.} \right)\\
 &= {\Pr ^i}\left( {{\rm{sb}},{\rm{X - error}}\left| {{\rm{click}},{{\rm{Z}}_{{{\rm{A}}_{\rm{c}}}}}} \right.} \right)\\
&={\Pr ^i}\left( {{\rm{sb}},{\rm{X - error}}}\left|\rm{click}\right.\right),\\
&{\Pr ^i}\left( {{\rm{sb}},{\rm{No\;X - error}}\left| {{\rm{click}},{{\rm{X}}_{{{\rm{A}}_{\rm{c}}}}}} \right.} \right)\\
& ={\Pr ^i}\left( {{\rm{sb}},{\rm{No\;X - error}}\left| {{\rm{click}},{{\rm{Z}}_{{{\rm{A}}_{\rm{c}}}}}} \right.} \right)\\
&={\Pr ^i}\left( {{\rm{sb}},{\rm{No\;X - error}}} \left|\rm{click}\right.\right).
\end{array}
\end{equation}
By multiplying Eq.~(\ref{Xerror}) by ${{\Pr }^i}\left( { {\rm{sb}},{\rm{X - error}}\left| {{\rm{click}},{{\rm{X}}_{{{\rm{A}}_{\rm{c}}}}}} \right.} \right)$ and multiplying Eq.~(\ref{NoXerror}) by ${{\Pr }^i}\left( { {\rm{sb}}{\rm{,No}}\;{\rm{X - error}}\left| {{\rm{click}},{{\rm{X}}_{{{\rm{A}}_{\rm{c}}}}}} \right.} \right)$ and adding them together, we obtain
\begin{equation}
\begin{array}{*{20}{l}} \label{pclick}
&{\Pr ^i}\left( {{\rm{sb}}} \right) - {\Pr ^i}\left( {{\rm{sb}}} \right){\Pr ^i}\left( {{{\rm{X}}_{{{\rm{A}}_{\rm{c}}}}} =  - \left| {{\rm{click}}{\rm{,}}{{\rm{X}}_{{{\rm{A}}_{\rm{c}}}}}} \right.} \right)\\
& \le2\sqrt {{{\Pr }^i}\left( {{{\rm{Z}}_{{{\rm{A}}_{\rm{c}}}}} = 1,{\rm{sb}}{\rm{,X - error}}\left| {{\rm{click}},{{\rm{Z}}_{{{\rm{A}}_{\rm{c}}}}}} \right.} \right)}\\
  &\times\sqrt{{{{\Pr }^i}\left( {{{\rm{Z}}_{{{\rm{A}}_{\rm{c}}}}} = 0,{\rm{sb}}{\rm{,X - error}}\left| {{\rm{click}},{{\rm{Z}}_{{{\rm{A}}_{\rm{c}}}}}} \right.} \right)}}  \\
 &+2\sqrt {{{\Pr }^i}\left( {{{\rm{Z}}_{{{\rm{A}}_{\rm{c}}}}} = 1,{\rm{sb}}{\rm{,No}}\;{\rm{X - error}}\left| {{\rm{click}},{{\rm{Z}}_{{{\rm{A}}_{\rm{c}}}}}} \right.} \right)}\\
 &\times\sqrt{ {{{\Pr }^i}\left( {{{\rm{Z}}_{{{\rm{A}}_{\rm{c}}}}} = 0,{\rm{sb}}{\rm{,No}}\;{\rm{X - error}}\left| {{\rm{click}},{{\rm{Z}}_{{{\rm{A}}_{\rm{c}}}}}} \right.} \right)}} ,
\end{array}
\end{equation}
where we have taken into account that
\begin{equation}
\begin{array}{*{20}{l}}
&{\Pr ^i}\left( {{\rm{sb}},{\rm{X - error}}\left| {{\rm{click}},{{\rm{X}}_{{{\rm{A}}_{\rm{c}}}}}} \right.} \right)\\& + {\Pr ^i}\left( {{\rm{sb}},{\rm{No}}\;{\rm{X - error}}\left| {{\rm{click}},{{\rm{X}}_{{{\rm{A}}_{\rm{c}}}}}} \right.} \right)\\
 &={\Pr ^i}\left( {{\rm{sb}},{\rm{X - error}}\left| {{\rm{click}}} \right.} \right)\\& + {\Pr ^i}\left( {{\rm{sb}},{\rm{No}}\;{\rm{X - error}}\left| {{\rm{click}}} \right.} \right)\\
 &={\Pr ^i}\left( {{\rm{sb}}\left| {{\rm{click}}} \right.} \right) = {\Pr ^i}\left( {{\rm{sb}}} \right),\\
&{\Pr ^i}\left( {{{\rm{X}}_{{{\rm{A}}_{\rm{c}}}}} =  - ,{\rm{sb}},{\rm{X - error}}\left| {{\rm{click}},{{\rm{X}}_{{{\rm{A}}_{\rm{c}}}}}} \right.} \right)\\& + {\Pr ^i}\left( {{{\rm{X}}_{{{\rm{A}}_{\rm{c}}}}} =  - ,{\rm{sb}},{\rm{No}}\;{\rm{X - error}}\left| {{\rm{click}},{{\rm{X}}_{{{\rm{A}}_{\rm{c}}}}}} \right.} \right)\\
 &={\Pr ^i}\left( {{{\rm{X}}_{{{\rm{A}}_{\rm{c}}}}} =  - ,{\rm{sb}},\left| {{\rm{click}},{{\rm{X}}_{{{\rm{A}}_{\rm{c}}}}}} \right.} \right)\\
&= {\Pr ^i}\left( {{\rm{sb}},\left| {{\rm{click}},{{\rm{X}}_{{{\rm{A}}_{\rm{c}}}}}} \right.} \right){\Pr ^i}\left( {{{\rm{X}}_{{{\rm{A}}_{\rm{c}}}}} =  - \left| {{\rm{sb}}{\rm{,click}},{{\rm{X}}_{{{\rm{A}}_{\rm{c}}}}}} \right.} \right)\\
 &={\Pr ^i}\left( {{\rm{sb}},\left| {{\rm{click}}} \right.} \right){\Pr ^i}\left( {{{\rm{X}}_{{{\rm{A}}_{\rm{c}}}}} =  - \left| {{\rm{click}},{{\rm{X}}_{{{\rm{A}}_{\rm{c}}}}}} \right.} \right) \\
&={\Pr ^i}\left( {{\rm{sb}}} \right){\Pr ^i}\left( {{{\rm{X}}_{{{\rm{A}}_{\rm{c}}}}} =  - \left| {{\rm{click}},{{\rm{X}}_{{{\rm{A}}_{\rm{c}}}}}} \right.} \right).
\end{array}
\end{equation}

Here, ${\Pr ^i}\left( {{{\rm{X}}_{{{\rm{A}}_{\rm{c}}}}} =  - \left| {{\rm{click}},{{\rm{X}}_{{{\rm{A}}_{\rm{c}}}}}} \right.} \right)$ is the conditional probability that the measurement outcome on the quantum coin is `$-$' given that Bob obtains a click and Alice performs the $\rm X_{A_c}$ basis measurement on the quantum coin, and ${{\Pr }^i}\left( {{{\rm{Z}}_{{{\rm{A}}_{\rm{c}}}}} = 1,{\rm{sb}}{\rm{,X - error}}\left| {{\rm{click}},{{\rm{Z}}_{{{\rm{A}}_{\rm{c}}}}}} \right.} \right)$ is the joint conditional probability that the measurement outcome on the quantum coin is `1', Alice and Bob select the same basis and there is an X basis error given that Bob obtains a click and Alice performs the $\rm Z_{A_c}$ basis measurement on the quantum coin. The other joint conditional probabilities are defined similarly.

Note that
\begin{equation}
\begin{array}{*{20}{l}}
&{\Pr ^i}\left( {{{\rm{Z}}_{{{\rm{A}}_{\rm{c}}}}} = 1,{\rm{sb}}{\rm{,X - error}}\left| {{\rm{click}},{{\rm{Z}}_{{{\rm{A}}_{\rm{c}}}}}} \right.} \right)\\
&={\Pr ^i}\left( {{\rm{sb}}{\rm{,}}\left| {{\rm{click}},{{\rm{Z}}_{{{\rm{A}}_{\rm{c}}}}}} \right.} \right)\\&\times{\Pr ^i}\left( {{{\rm{Z}}_{{{\rm{A}}_{\rm{c}}}}} = 1,{\rm{X - error}}\left| {{\rm{click}},{{\rm{Z}}_{{{\rm{A}}_{\rm{c}}}}}{\rm{,sb}}} \right.} \right)\\
&={\Pr ^i}\left( {{\rm{sb}}} \right){\Pr ^i}\left( {{{\rm{Z}}_{{{\rm{A}}_{\rm{c}}}}} = 1,{\rm{X - error}}\left| {{\rm{click}},{{\rm{Z}}_{{{\rm{A}}_{\rm{c}}}}}} \right.} \right),
\end{array}
\end{equation}
and a similar relation also holds for the other terms in Eq.~(\ref{pclick}). This means, in particular, that we can rewrite Eq.~(\ref{pclick}) as

\begin{equation}
\begin{array}{*{20}{l}}
&{1 - 2{{\Pr }^i}\left( {{{\rm{X}}_{{{\rm{A}}_{\rm{c}}}}} =  - \left| {{\rm{click}},{{\rm{X}}_{{{\rm{A}}_{\rm{c}}}}}} \right.} \right)}\\
 &\le2\sqrt {{{\Pr }^i}\left( {{{\rm{Z}}_{{{\rm{A}}_{\rm{c}}}}} = 1,{\rm{X - error}}\left| {{\rm{click}},{{\rm{Z}}_{{{\rm{A}}_{\rm{c}}}}}} \right.} \right)}\\
 &\times \sqrt {{{\Pr }^i}\left( {{{\rm{Z}}_{{{\rm{A}}_{\rm{c}}}}} = 0,{\rm{X - error}}\left| {{\rm{click}}} \right.,{{\rm{Z}}_{{{\rm{A}}_{\rm{c}}}}}} \right)} \\
 &+2\sqrt {{{\Pr }^i}\left( {{{\rm{Z}}_{{{\rm{A}}_{\rm{c}}}}} = 1,{\rm{No}}\;{\rm{X - error}}\left| {{\rm{click}}} \right.,{{\rm{Z}}_{{{\rm{A}}_{\rm{c}}}}}} \right)}\\
 &\times \sqrt {{{\Pr }^i}\left( {{{\rm{Z}}_{{{\rm{A}}_{\rm{c}}}}} = 0,{\rm{No}}\;{\rm{X - error}}\left| {{\rm{click}}} \right.,{{\rm{Z}}_{{{\rm{A}}_{\rm{c}}}}}} \right)} ,
\end{array}\label{Pclick}
\end{equation}
where we have already canceled the common factor `${\Pr ^i}\left( {{\rm{sb}}} \right)$' that appears on both sides of Eq.~(\ref{Pclick}).

Then by multiplying both sides of Eq.~(\ref{Pclick}) by ${{{\Pr }^i}\left( {{{\rm{Z}}_{{{\rm{A}}_{\rm{c}}}}}\left| {{\rm{click}}} \right.} \right)}$, we obtain

\begin{equation}
\begin{array}{*{20}{l}}
&{{{\Pr }^i}\left( {{{\rm{Z}}_{{{\rm{A}}_{\rm{c}}}}}\left| {{\rm{click}}} \right.} \right)} \\&- {{{\Pr }^i}\left( {{{\rm{Z}}_{{{\rm{A}}_{\rm{c}}}}}\left| {{\rm{click}}} \right.} \right){{\Pr }^i}\left( {{{\rm{X}}_{{{\rm{A}}_{\rm{c}}}}} =  - \left| {{\rm{click}},{{\rm{X}}_{{{\rm{A}}_{\rm{c}}}}}} \right.} \right)}\\
 &\le2\sqrt {{{\Pr }^i}\left( {{{\rm{Z}}_{{{\rm{A}}_{\rm{c}}}}} = 1,{\rm{X - error}},{{\rm{Z}}_{{{\rm{A}}_{\rm{c}}}}}\left| {{\rm{click}}} \right.} \right)}\\
 &\times \sqrt {{{{\Pr }^i}\left( {{{\rm{Z}}_{{{\rm{A}}_{\rm{c}}}}} = 0,{\rm{X - error}},{{\rm{Z}}_{{{\rm{A}}_{\rm{c}}}}}\left| {{\rm{click}}} \right.} \right)} } \\
 &+2\sqrt {{{\Pr }^i}\left( {{{\rm{Z}}_{{{\rm{A}}_{\rm{c}}}}} = 1,{\rm{No}}\;{\rm{X - error}},{{\rm{Z}}_{{{\rm{A}}_{\rm{c}}}}}\left| {{\rm{click}}} \right.} \right) }\\
 &\times \sqrt {{{{\Pr }^i}\left( {{{\rm{Z}}_{{{\rm{A}}_{\rm{c}}}}} = 0,{\rm{No}}\;{\rm{X - error}},{{\rm{Z}}_{{{\rm{A}}_{\rm{c}}}}}\left| {{\rm{click}}} \right.} \right)} } .
\end{array}
\end{equation}

Due to the concavity of the square root function, we take the sum over $i \in \{1,2,...,N_{\rm click}\}$ and obtain

\begin{equation}
\begin{array}{*{20}{l}}
{}&{\sum\limits_{i = 1}^{{N_{{\rm{click}}}}} {{{\Pr }^i}\left( {{{\rm{Z}}_{{{\rm{A}}_{\rm{c}}}}}\left| {{\rm{click}}} \right.} \right)}}\\&  -{ \sum\limits_{i = 1}^{{N_{{\rm{click}}}}} {{{\Pr }^i}\left( {{{\rm{Z}}_{{{\rm{A}}_{\rm{c}}}}}\left| {{\rm{click}}} \right.} \right){{\Pr }^i}\left( {{{\rm{X}}_{{{\rm{A}}_{\rm{c}}}}} =  - \left| {{\rm{click}},{{\rm{X}}_{{{\rm{A}}_{\rm{c}}}}}} \right.} \right)} }\\
 &\le2\sqrt {\sum\limits_{i = 1}^{{N_{{\rm{click}}}}} {{{\Pr }^i}\left( {{{\rm{Z}}_{{{\rm{A}}_{\rm{c}}}}} = 1,{\rm{X - error}},{{\rm{Z}}_{{{\rm{A}}_{\rm{c}}}}}\left| {{\rm{click}}} \right.} \right)} }\\
 &\times \sqrt {\sum\limits_{i = 1}^{{N_{{\rm{click}}}}}{{{\Pr }^i}\left( {{{\rm{Z}}_{{{\rm{A}}_{\rm{c}}}}} = 0,{\rm{X - error}},{{\rm{Z}}_{{{\rm{A}}_{\rm{c}}}}}\left| {{\rm{click}}} \right.} \right)} }\\
 &+2\sqrt {\sum\limits_{i = 1}^{{N_{{\rm{click}}}}} {{{\Pr }^i}\left( {{{\rm{Z}}_{{{\rm{A}}_{\rm{c}}}}} = 1,{\rm{No}}\;{\rm{X - error}},{{\rm{Z}}_{{{\rm{A}}_{\rm{c}}}}}\left| {{\rm{click}}} \right.} \right)} }\\
 &\times \sqrt {\sum\limits_{i = 1}^{{N_{{\rm{click}}}}}{{{\Pr }^i}\left( {{{\rm{Z}}_{{{\rm{A}}_{\rm{c}}}}} = 0,{\rm{No}}\;{\rm{X - error}},{{\rm{Z}}_{{{\rm{A}}_{\rm{c}}}}}\left| {{\rm{click}}} \right.} \right)} } ,
\end{array}\label{Nph}
\end{equation}

Let us denote the expected values after $N_{\rm click}$ trials by
\begin{equation}
\begin{array}{*{20}{l}}
&\sum\limits_{i = 1}^{{N_{{\rm{click}}}}} {{{\Pr }^i}\left( {{{\rm{Z}}_{{{\rm{A}}_{\rm{c}}}}} = 1,{\rm{X - error}},{{\rm{Z}}_{{{\rm{A}}_{\rm{c}}}}}\left| {{\rm{click}}} \right.} \right)} \\& = {\mathcal{E}_{{\rm{X}},{\rm{X - error}}}},\\
&\sum\limits_{i = 1}^{{N_{{\rm{click}}}}} {{{\Pr }^i}\left( {{{\rm{Z}}_{{{\rm{A}}_{\rm{c}}}}} = 0,{\rm{X - error}},{{\rm{Z}}_{{{\rm{A}}_{\rm{c}}}}}\left| {{\rm{click}}} \right.} \right)} \\&  = {\mathcal{E}_{{\rm{Z}},{\rm{X - error}}}},\\
&\sum\limits_{i = 1}^{{N_{{\rm{click}}}}} {{{\Pr }^i}\left( {{{\rm{Z}}_{{{\rm{A}}_{\rm{c}}}}} = 1,{\rm{No}}\;{\rm{X - error}},{{\rm{Z}}_{{{\rm{A}}_{\rm{c}}}}}\left| {{\rm{click}}} \right.} \right)} \\&  = {\mathcal{E}_{{\rm{X}},{\rm{No}}\;{\rm{X - error}}}},\\
&\sum\limits_{i = 1}^{{N_{{\rm{click}}}}} {{{\Pr }^i}\left( {{{\rm{Z}}_{{{\rm{A}}_{\rm{c}}}}} = 0,{\rm{No}}\;{\rm{X - error}},{{\rm{Z}}_{{{\rm{A}}_{\rm{c}}}}}\left| {{\rm{click}}} \right.} \right)} \\&  = {\mathcal{E}_{{\rm{Z}},{\rm{No}}\;{\rm{X - error}}}}.
\end{array}
\end{equation}
Then Eq.~ (\ref{Nph}) can be rewritten as:
\begin{equation}
\begin{array}{*{20}{l}}
&{p_{{{\rm{Z}}_{{{\rm{A}}_{\rm{c}}}}}}}{N_{{\rm{click}}}} - 2\frac{{{p_{{{\rm{Z}}_{{{\rm{A}}_{\rm{c}}}}}}}}}{{{p_{{{\rm{X}}_{{{\rm{A}}_{\rm{c}}}}}}}}}{\mathcal{E}
_{{{\rm{X}}_{{{\rm{A}}_{\rm{c}}}}} =  - }}\\
 &\le 2\sqrt {{\mathcal{E}_{{\rm{X}},{\rm{X - error}}}}{\mathcal{E}_{{\rm{Z}},{\rm{X - error}}}}}\\
 &  + 2\sqrt {{\mathcal{E}_{{\rm{X}},{\rm{No}}\;{\rm{X - error}}}}{\mathcal{E}_{{\rm{Z}},{\rm{No}}\;{\rm{X - error}}}}}\label{nphase},
\end{array}
\end{equation}
where we have also used the fact that
\begin{equation}
\sum\limits_{i = 1}^{{N_{{\rm{click}}}}} {{{\Pr }^i}\left( {{{\rm{Z}}_{{{\rm{A}}_{\rm{c}}}}}\left| {{\rm{click}}} \right.} \right)=p_{\rm Z_{A_c}}\sum\limits_{i = 1}^{{N_{{\rm{click}}}}}1=p_{{{\rm{Z}}_{{{\rm{A}}_{\rm{c}}}}}}}{N_{{\rm{click}}}},
\end{equation}
and
\begin{equation}
\begin{array}{*{20}{l}}
&\sum\limits_{i = 1}^{{N_{{\rm{click}}}}} {{{\Pr }^i}\left( {{{\rm{Z}}_{{{\rm{A}}_{\rm{c}}}}}\left| {{\rm{click}}} \right.} \right){{\Pr }^i}\left( {{{\rm{X}}_{{{\rm{A}}_{\rm{c}}}}} =  - \left| {{\rm{click}},{{\rm{X}}_{{{\rm{A}}_{\rm{c}}}}}} \right.} \right)}\\
&={\rm{ }}\sum\limits_{i = 1}^{{N_{{\rm{click}}}}}{{p_{{{\rm{Z}}_{{{\rm{A}}_{\rm{c}}}}}}}}\frac{{{{\Pr }^i}\left( {{{\rm{X}}_{{{\rm{A}}_{\rm{c}}}}} =  - ,{{\rm{X}}_{{{\rm{A}}_{\rm{c}}}}}\left| {{\rm{click}}} \right.} \right)}}{{{{\Pr }^i}\left( {{{\rm{X}}_{{{\rm{A}}_{\rm{c}}}}}\left| {{\rm{click}}} \right.} \right)}}  \\
&=\frac{{{p_{{{\rm{Z}}_{{{\rm{A}}_{\rm{c}}}}}}}}}{{{p_{{{\rm{X}}_{{{\rm{A}}_{\rm{c}}}}}}}}}\sum\limits_{i = 1}^{{N_{{\rm{click}}}}} {{{\Pr }^i}\left( {{{\rm{X}}_{{{\rm{A}}_{\rm{c}}}}} =  - ,{{\rm{X}}_{{{\rm{A}}_{\rm{c}}}}}\left| {{\rm{click}}} \right.} \right)} \\
&=\frac{{{p_{{{\rm{Z}}_{{{\rm{A}}_{\rm{c}}}}}}}}}{{{p_{{{\rm{X}}_{{{\rm{A}}_{\rm{c}}}}}}}}}{\mathcal{E}_{{{\rm{X}}_
{{{\rm{A}}_{\rm{c}}}}} =  - }}.
\end{array}
\end{equation}
Importantly, note that ${{\mathcal{E}_{{\rm{Z}}{\rm{,X - error}}}}}$ is actually equal to $\mathcal{E}_{\rm phase~error}$. The quantity ${\mathcal{E}_{{{\rm{X}}_{{{\rm{A}}_{\rm{c}}}}} =  - }}$ denotes the expected number of click events where Alice chooses the X basis to measure the quantum coin and obtains the outcome `$-$'. Since the number of events that we obtain from the $\rm{ X_{A_c}}$ basis measurement only on the click events can never be larger than the total number of events that we could have obtained if we had measured all the $N$ quantum coins along the $\rm{ X_{A_c}}$ basis, in the asymptotic limit we have that
\begin{equation}
\begin{array}{*{20}{l}}
{\mathcal{E}_{{{\rm{X}}_{{{\rm{A}}_{\rm{c}}}}} =  - }}&\le Np_{{{\rm{X}}_{{{\rm{A}}_{\rm{c}}}}}}{{\rm{P}}{{\rm{r}}}\left( {{X_{{{\rm{A}}_{\rm{c}}}}} =  - |{{\rm{X}}_{{{\rm{A}}_{\rm{c}}}}}} \right)}
\\
&\le{\frac{1}{2}Np_{{{\rm{X}}_{{{\rm{A}}_{\rm{c}}}}}}\left[ {1 - \mathop {\min }\limits_i \sqrt {{p_{{{\rm{Z}}}}}{p_{{{\rm{X}}}}}} \cos {\phi ^i}\left| {\left\langle {\Psi _{\rm{Z}}^i} \right.\left| {\Psi _{\rm{X}}^i} \right\rangle } \right|} \right],}
\end{array} \label{NAC}
\end{equation}
where for simplicity we denote ${_{{{\rm{A}}_{\rm{q}}},{{\rm{A}}_{\rm{p}}},{{\rm{A}}_{\rm{a}}},{{\rm{E}}^\prime }_{\rm{p}},{{\rm{E}}_{\rm{a}}}}\left\langle {\Psi _{\rm{Z}}^i} \right.{{\left| {\Psi _{\rm{X}}^i} \right\rangle }_{{{\rm{A}}_{\rm{q}}},{{\rm{A}}_{\rm{p}}},{{\rm{A}}_{\rm{a}}},{{\rm{E}}^\prime }_{\rm{p}},{{\rm{E}}_{\rm{a}}}}}}$ by ${\left\langle {\Psi _{\rm{Z}}^i} \right.\left| {\Psi _{\rm{X}}^i} \right\rangle }$, and $\phi^i$ denotes the angle between the states ${\left| {\Psi _{\rm{Z}}^i} \right\rangle }$ and ${\left| {\Psi _{\rm{X}}^i} \right\rangle }$.

Next, we apply Azuma's inequality~\cite{azuma1967weighted} to all the terms of Eq.~(\ref{nphase}) except for ${\mathcal{E}_{{{\rm{X}}_{{{\rm{A}}_{\rm{c}}}}} =  - }}$, and we employ Chernoff bound to ${\mathcal{E}_{{{\rm{X}}_{{{\rm{A}}_{\rm{c}}}}} =  - }}$. As a result, we obtain the following inequality:
\begin{widetext}
\begin{equation}
\begin{array}{*{20}{l}}
&{{p_{{{\rm{Z}}_{{{\rm{A}}_{\rm{c}}}}}}}\left( { {{N_{{\rm{click}}}}} + {\delta _{{\rm{click}}}}} \right)-2\frac{{{p_{{{\rm{Z}}_{{{\rm{A}}_{\rm{c}}}}}}}}}{{{p_{{{\rm{X}}_{{{\rm{A}}_{\rm{c}}}}}}}}} ({N_{{{\rm{X}}_{{{\rm{A}}_{\rm{c}}}}} =  - }} +\delta'_{\rm X_{{{{\rm{A}}_{\rm{c}}}}=  -}})} \\
& \le{2\sqrt {\left( { {{N_{{\rm{X}},{\rm{X - error}}}}}  + {\delta _{{\rm{X}},{\rm{X - error}}}}} \right)\left( {  {{N_{{\rm{Z}},{\rm{X - error}}}}}  + {\delta _{{\rm{Z}},{\rm{X - error}}}}} \right)} }\\
 &+{2\sqrt {\left( {{{N_{\rm{click|X}}}} -  {{N_{{\rm{X}},{\rm{X - error}}}}}  + {\delta _{{\rm{X}},{\rm{No}}\;{\rm{X - error}}}}} \right)}{\left( { {{N_{\rm{click|Z}}}}  -  {{N_{{\rm{Z}},{\rm{X - error}}}}}  + {\delta _{{\rm{Z}},{\rm{No}}\;{\rm{X - error}}}}} \right)} }
\end{array},\label{N phase}
\end{equation}
\end{widetext}
except for an exponentially small error probability $(\varepsilon'_{\rm X_{{{{\rm{A}}_{\rm{c}}}}=  -}}+\widehat{\varepsilon}'_{\rm X_{{{{\rm{A}}_{\rm{c}}}}=  -}})+\sum_\lambda  {\left( {{\varepsilon _\lambda } + {{\hat \varepsilon }_\lambda }} \right)}$, where $\lambda \in \{\rm (click), (X,X-error),(Z,X-error),(X,No~X-error),(Z,No~X-error)\}$.

In the above equation, $ {{N_\lambda}} $ denotes the actual number of times that the event `$\lambda$' occurs. The quantities $\delta_{\lambda}$ denote the deviation terms due to the use of Azuma's inequality to estimate ${{N_\lambda}} $ and they are bounded by $\delta_{\lambda}\in [-\Delta_{\lambda},\hat \Delta_{\lambda}]$ with the bounds being given by $\Delta_{\lambda}=f\left(N_{\rm click},\varepsilon_{\lambda}\right)$ and $\hat \Delta_{\lambda}=f\left(N_{\rm click},\hat \varepsilon_{\lambda}\right)$. The expected number $\mathcal{E}_{{{\rm{X}}_{{{\rm{A}}_{\rm{c}}}}} =  - }$ can first be upper bounded by using Eq.~(\ref{NAC}) in the asymptotic case. Then we estimate the actual number $N_{{{\rm{X}}_{{{\rm{A}}_{\rm{c}}}}} =  - }$ by using the Chernoff bound~\cite{chernoff1952measure}, which guarantees that $N_{{{\rm{X}}_{{{\rm{A}}_{\rm{c}}}}} =  - }=\mathcal{E}_{{{\rm{X}}_{{{\rm{A}}_{\rm{c}}}}} =  - }+\delta'_{\rm X_{{{{\rm{A}}_{\rm{c}}}}=  -}}$ except for an exponentially small error probability $\varepsilon'_{\rm X_{{{{\rm{A}}_{\rm{c}}}}=  -}}+\widehat{\varepsilon}'_{\rm X_{{{{\rm{A}}_{\rm{c}}}}=  -}}$. The corresponding fluctuation deviation term $\delta'_{\rm X_{{{{\rm{A}}_{\rm{c}}}}=  -}}$ lies in an interval $ [ { - {\Delta'}_{\rm X_{{{{\rm{A}}_{\rm{c}}}}=  -}},{{\hat \Delta '}}}_{\rm X_{{{{\rm{A}}_{\rm{c}}}}=  -}} ]$ and the bounds are given by $\Delta'_{{{\rm{X}}_{{{\rm{A}}_{\rm{c}}}}} =  - } = g( {{_{{{\rm{X}}_{{{\rm{A}}_{\rm{c}}}}} =  - }},\varepsilon'_{\rm X_{{{{\rm{A}}_{\rm{c}}}}=  -}}} )$ and $\widehat{\Delta'}_{{{\rm{X}}_{{{\rm{A}}_{\rm{c}}}}} =  - } = \widehat{g}( {{N_{{{\rm{X}}_{{{\rm{A}}_{\rm{c}}}}} =  - }},\widehat{\varepsilon}'_{\rm X_{{{{\rm{A}}_{\rm{c}}}}=  -}} })$, respectively, with the functions $g( {x,y} ) = \sqrt {2x\ln (1/y)}$ and $\widehat{g}( {x,y} ) = \sqrt {3x\ln (1/y)}$~\cite{mizutani2015finite}. $ {{N_{\rm{click|Z(X)}}}} $ denotes the actual number of events where both Alice and Bob select the Z (X) basis and Bob obtains a click, i.e, ${{{N_{{\rm{click|Z(X)}}}}}  =  {{N_{{\rm{Z(X),X - error}}}}} +  {{N_{{\rm Z(X),\rm{No}}\;{\rm{X - error}}}}} }$. Finally, $ {{N_{{\rm{Z}},{\rm{X - error}}}}} $ is the quantity that we wish to estimate.

So far, the analysis above considers the case where Alice has a single-photon source at her disposal. To adapt it to the decoy-state BB84 protocol described in Sec. II, which is based on the use of WCPs, is however straightforward. In particular, since the secret key is only distilled from data associated to the signal intensity setting, now all the actual numbers that appear in Eq.~(\ref{N phase}) refer to the single-photon contributions within the signal intensity setting. $\gamma^{\rm s}$.
That is, now ${ {{N_{{\rm{click}}}}} ,~ {{N_{{\rm{X}},{\rm{X - error}}}}} ,~ {{N_{{\rm{Z}},{\rm{X - error}}}}},~ {{N_{{\rm{click|X}}}}} ,~ {{N_{{\rm{click|Z}}}}}}$ refer to $ {{N_{{\rm{click,1,\gamma^s}}}}} ,~ {{N_{{\rm{X - error}},{\rm{1,\gamma^s|X}}}}} ,~{{N_{{\rm{X - error}},{\rm{1,\gamma^s|Z}}}}} ,~ {{N_{{\rm{click,1,\gamma^s|X}}}}}$ ,$ {{N_{{\rm{click,1,\gamma^s|Z}}}}}$. One can use the same method based on linear optimization that we used in the previous sections to estimate $ {{N_{{\rm{click,1,\gamma^s}}}}} ,~ {{N_{{\rm{X - error}},{\rm{1,\gamma^s|X}}}}} ,~{{N_{{\rm{X - error}},{\rm{1,\gamma^s|Z}}}}} ,~ {{N_{{\rm{click,1,\gamma^s|X}}}}}$ ,$ {{N_{{\rm{click,1,\gamma^s|Z}}}}}$. In so doing, one can estimate an upper bound on $ {{N_{{\rm{X - error}},{\rm{1,\gamma^s|Z}}}}}$ and thus obtain $e_{{\rm{ph}}}^{\rm{U}}$.

\section*{Appendix E. Parameters $D_{n,\{ {\gamma ^j}\} ,\{ {\gamma ^k} \},\chi}$ and $D_{n,\{ {\gamma ^j}\} ,\{ {\gamma ^k},{\gamma ^l}\} ,\chi }$}
The parameters $D_{n,\{ {\gamma ^j}\} ,\{ {\gamma ^k} \},\chi}$ and $D_{n,\{ {\gamma ^j}\} ,\{ {\gamma ^k},{\gamma ^l}\} ,\chi }$ with $j,k,l\in \{\rm s,v,w\}$ which are needed to estimate $N^{L}_{{\rm click,0,\gamma^s}|\rm Z}$, $N^{L}_{{\rm click,1,\gamma^s}|\rm Z}$ and $e^{\rm U}_{\rm ph}$ have been calculated in Ref.~\cite{tamaki2016decoy}. For the special case of an individual THA where there is no quantum correlation between Alice's and Eve's systems, i.e., $\rho _{n,\chi }^{{\gamma ^k},i} = \left| n \right\rangle {\left\langle n \right|_\chi } \otimes \rho _\chi ^{{\gamma ^k},i}$ for all $i$ and $n$, where $\left| n \right\rangle _\chi $ is a Fock state with $n$ photons prepared in the basis $\chi$, and $\rho _\chi ^{{\gamma ^k}} = \left| {{\beta _k}{e^{i{\theta _k}}}} \right\rangle \left\langle {{\beta _k}{e^{i{\theta _k}}}} \right|$ is the state of Eve's back-reflected light. In this case, $D_{n,\{ {\gamma ^j}\} ,\{ {\gamma ^k} \},\chi}$ and $D_{n,\{ {\gamma ^j}\} ,\{ {\gamma ^k},{\gamma ^l}\} ,\chi }$ do not depend on $\chi$ and $D_{n,\{ {\gamma ^j}\} ,\{ {\gamma ^k} \},\chi}$ do not depend on $n$. So we shall denote them by $D_{\{ {\gamma ^j}\} ,\{ {\gamma ^k} \}}$ and $D_{n,\{ {\gamma ^j}\} ,\{ {\gamma ^k},{\gamma ^l}\} }$, respectively. For completeness, below we include the values of these parameters in different cases~\cite{tamaki2016decoy}.

\subsection{Case 1}
We have that
\begin{equation}
\begin{array}{l}
{D_{\{\gamma^{\rm{v}}\},\{\gamma^{\rm{s}}\}}} = \sqrt {1 - {e^{2{I_{\max }}\left[ {\cos \left( {{\theta _{\rm{v}}}} \right) - 1} \right]}}} ,\\
{D_{\{\gamma^{\rm{w}}\},\{\gamma^{\rm{s}}\}}} = \sqrt {1 - {e^{2{I_{\max }}\left[ {\cos \left( {{\theta _{\rm{w}}}} \right) - 1} \right]}}} .
\end{array}
\end{equation}

To calculate the parameters $D_{n,\{ {\gamma ^j}\} ,\{ {\gamma ^k},{\gamma ^l}\}}$, we have that
\begin{equation}
{D_{n,\{ {\gamma ^j}\} ,\{ {\gamma ^k},{\gamma ^l}\}  }} =\frac{1}{2}\sum\limits_i {\left| {{\lambda _i}} \right|},
\end{equation}
where $\lambda_i$ are the eigenvalues of a $\rm {3\times3}$ matrix $\Lambda$ whose elements are given by
\begin{equation}
\begin{array}{*{20}{l}}
{\Lambda _{a,b}} = &{\delta _{a,1}}\left\langle {{\beta {e^{{\rm{i}}{\theta _j}}}}}
 \mathrel{\left | {\vphantom {{\beta {e^{{\rm{i}}{\theta _1}}}} {\beta {e^{{\rm{i}}{\theta _b}}}}}}
 \right. \kern-\nulldelimiterspace}
 {{\beta {e^{{\rm{i}}{\theta _b}}}}} \right\rangle  - {q_{nkl}}{\delta _{a,2}}\left\langle {{\beta {e^{{\rm{i}}{\theta _k}}}}}
 \mathrel{\left | {\vphantom {{\beta {e^{{\rm{i}}{\theta _k}}}} {\beta {e^{{\rm{i}}{\theta _b}}}}}}
 \right. \kern-\nulldelimiterspace}
 {{\beta {e^{{\rm{i}}{\theta _b}}}}} \right\rangle\\
  & -\left( {1 - {q_{nkl}}} \right){\delta _{a,3}}\left\langle {{\beta {e^{{\rm{i}}{\theta _l}}}}}
 \mathrel{\left | {\vphantom {{\beta {e^{{\rm{i}}{\theta _l}}}} {\beta {e^{{\rm{i}}{\theta _b}}}}}}
 \right. \kern-\nulldelimiterspace}
 {{\beta {e^{{\rm{i}}{\theta _b}}}}} \right\rangle,
\end{array}
\end{equation}
where $\beta=\sqrt{I_{\rm max}}$, $\theta_{\rm 1}=\theta_j$, $\theta_{\rm 2}=\theta_k$, $\theta_{\rm 3}=\theta_l$, respectively, $a,b\in \{1,2,3\}$, and $\delta_{a,b}$ is the Kronecker delta.

\subsection{Case 2}
Here, we have that
\begin{equation}
\begin{array}{*{20}{l}}
{D_{\{\gamma^{\rm{v}}\},\{\gamma^{\rm{s}}\}}} = \sqrt {1 - {e^{ - \frac{{{I_{\max }}}}{{{\gamma ^{\rm{s}}}}}\left[ {{\gamma ^{\rm{s}}} + {\gamma ^{\rm{v}}} - 2\sqrt {{\gamma ^{\rm{s}}}{\gamma ^{\rm{v}}}} \cos \left( {{\theta _{\rm{v}}}} \right)} \right]}}} ,\\
{D_{\{\gamma^{\rm{w}}\},\{\gamma^{\rm{s}}\}}}= \sqrt {1 - {e^{ - \frac{{{I_{\max }}}}{{{\gamma ^{\rm{s}}}}}\left[ {{\gamma ^{\rm{s}}} + {\gamma ^{\rm{w}}} - 2\sqrt {{\gamma ^{\rm{s}}}{\gamma ^{\rm{w}}}} \cos \left( {{\theta _{\rm{w}}}} \right)} \right]}}} .
\end{array}
\end{equation}

And the parameters $D_{n,\{ {\gamma ^j}\} ,\{ {\gamma ^k},{\gamma ^l}\}}$ are given by
\begin{equation}
{D_{n,\{ {\gamma ^j}\} ,\{ {\gamma ^k},{\gamma ^l}\}  }} =\frac{1}{2}\sum\limits_i {\left| {{\lambda _i}} \right|},
\end{equation}
where $\lambda_i$ are the eigenvalues of a $\rm {3\times3}$ matrix $\Lambda$ whose elements are given by
\begin{equation}
\begin{array}{*{20}{l}}
{\Lambda _{a,b}} = &{\delta _{a,1}}\left\langle {{\beta_j {e^{{\rm{i}}{\theta _j}}}}}
 \mathrel{\left | {\vphantom {{\beta {e^{{\rm{i}}{\theta _1}}}} {\beta {e^{{\rm{i}}{\theta _b}}}}}}
 \right. \kern-\nulldelimiterspace}
 {{\beta_b {e^{{\rm{i}}{\theta _b}}}}} \right\rangle  - {q_{nkl}}{\delta _{a,2}}\left\langle {{\beta_k {e^{{\rm{i}}{\theta _k}}}}}
 \mathrel{\left | {\vphantom {{\beta {e^{{\rm{i}}{\theta _2}}}} {\beta {e^{{\rm{i}}{\theta _b}}}}}}
 \right. \kern-\nulldelimiterspace}
 {{\beta_b {e^{{\rm{i}}{\theta _b}}}}} \right\rangle \\
  &- \left( {1 - {q_{nkl}}} \right){\delta _{a,3}}\left\langle {{\beta_l {e^{{\rm{i}}{\theta _l}}}}}
 \mathrel{\left | {\vphantom {{\beta {e^{{\rm{i}}{\theta _3}}}} {\beta {e^{{\rm{i}}{\theta _b}}}}}}
 \right. \kern-\nulldelimiterspace}
 {{\beta_b {e^{{\rm{i}}{\theta _b}}}}} \right\rangle,
\end{array}
\end{equation}
where $\beta_{\alpha}=\sqrt{I_{\rm max}\gamma^{\alpha}/\gamma^{\rm s}}$ with $\alpha \in \{j,k,l\}$, $\theta_{\rm 1}=\theta_j$, $\theta_{\rm 2}=\theta_k$, $\theta_{\rm 3}=\theta_l$, respectively, $a,b\in \{1,2,3\}$, and $\delta_{a,b}$ is the Kronecker delta.

\subsection{Case 3}
In this case we have that when $I_{\rm max}\leq \rm{log}2$ and $\gamma^{\rm w}\leq \gamma^{\rm v}\leq \gamma^{\rm s}$, for any $P_{\rm{cut}}\ge 1$, the parameters $D_{n,\{ {\gamma ^j}\} ,\{ {\gamma ^k} \},\chi}$ and $D_{n,\{ {\gamma ^j}\} ,\{ {\gamma ^k},{\gamma ^l}\} ,\chi }$ have the form:
\begin{widetext}
\begin{equation}
\begin{array}{*{20}{l}}
{D_{\{\gamma^{\rm{v}}\},\{\gamma^{\rm{s}}\}}} &\le \frac{1}{2} - \frac{{{e^{ - {I_{\max }}}}}}{2}\sum\limits_{n = 0}^{{P_{{\rm{cut}}}}} {\frac{{I_{\max }^n}}{{n!}}\left[ {1 - \left| {1 - {e^{{I_{\max }}\left( {1 - {\gamma ^{\rm{v}}}/{\gamma ^{\rm{s}}}} \right)}}{{\left( {\frac{{{\gamma ^{\rm{v}}}}}{{{\gamma ^{\rm{s}}}}}} \right)}^n}} \right|} \right]} ,\\
{D_{\{\gamma^{\rm{w}}\},\{\gamma^{\rm{s}}\}}} &\le \frac{1}{2} - \frac{{{e^{ - {I_{\max }}}}}}{2}\sum\limits_{n = 0}^{{P_{{\rm{cut}}}}} {\frac{{I_{\max }^n}}{{n!}}\left[ {1 - \left| {1 - {e^{{I_{\max }}\left( {1 - {\gamma ^{\rm{w}}}/{\gamma ^{\rm{s}}}} \right)}}{{\left( {\frac{{{\gamma ^{\rm{w}}}}}{{{\gamma ^{\rm{s}}}}}} \right)}^n}} \right|} \right]} ,\\
{D_{n,\{ {\gamma ^{\rm{s}}}\} ,\{ {\gamma ^{\rm{v}}},{\gamma ^{\rm{w}}}\} }} &\le \frac{1}{2}\left\{ {1 - \sum\limits_{n = 0}^{{P_{{\rm{cut}}}}} {{e^{ - {I_{\max }}}}\frac{{{I_{{{\max }^n}}}}}{{n!}}\left[ {1 - \left| {1 - {q_{n{\rm{vw}}}}{e^{{I_{\max }}\left( {1 - {\gamma ^{\rm{v}}}/{\gamma ^{\rm{s}}}} \right)}}} \right.} \right.} } \right.\\
&\left. {\left. { \times {{\left( {\frac{{{\gamma ^{\rm{v}}}}}{{{\gamma ^{\rm{s}}}}}} \right)}^n} - \left( {1 - {q_{n{\rm{vw}}}}} \right){e^{{I_{\max }}\left( {1 - {\gamma ^{\rm{w}}}/{\gamma ^{\rm{s}}}} \right)}}\left. {{{\left( {\frac{{{\gamma ^{\rm{w}}}}}{{{\gamma ^{\rm{s}}}}}} \right)}^n}} \right|} \right]} \right\},\\
{D_{n,\{ {\gamma ^{\rm{v}}}\} ,\{ {\gamma ^{\rm{s}}},{\gamma ^{\rm{w}}}\} }} &\le \frac{1}{2}\left\{ {{q_{n{\rm{sw}}}} + \left( {1 - {q_{n{\rm{sw}}}}} \right){e^{{I_{\max }}}} - \sum\limits_{n = 0}^{{P_{{\rm{cut}}}}} {{e^{ - {I_{\max }}}}\frac{{{I_{{{\max }^n}}}}}{{n!}}\left[ {{q_{n{\rm{sw}}}} + \left( {1 - {q_{n{\rm{sw}}}}} \right){e^{{I_{\max }}}}} \right.} } \right.\\
&\left. {\left. { - \left| {{e^{{I_{\max }}\left( {1 - {\gamma ^{\rm{v}}}/{\gamma ^{\rm{s}}}} \right)}}{{\left( {\frac{{{\gamma ^{\rm{v}}}}}{{{\gamma ^{\rm{s}}}}}} \right)}^n} - {q_{n{\rm{sw}}}} - \left( {1 - {q_{n{\rm{sw}}}}} \right){e^{{I_{\max }}\left( {1 - {\gamma ^{\rm{w}}}/{\gamma ^{\rm{s}}}} \right)}}{{\left( {\frac{{{\gamma ^{\rm{w}}}}}{{{\gamma ^{\rm{s}}}}}} \right)}^n}} \right|} \right]} \right\},\\
{D_{n,\{ {\gamma ^{\rm{w}}}\} ,\{ {\gamma ^{\rm{s}}},{\gamma ^{\rm{v}}}\} }} &\le \frac{1}{2}\left\{ {{q_{n{\rm{sv}}}} + \left( {1 - {q_{n{\rm{sv}}}}} \right){e^{{I_{\max }}}} - \sum\limits_{n = 0}^{{P_{{\rm{cut}}}}} {{e^{ - {I_{\max }}}}\frac{{{I_{{{\max }^n}}}}}{{n!}}\left[ {{q_{n{\rm{sv}}}} + \left( {1 - {q_{n{\rm{sv}}}}} \right){e^{{I_{\max }}}}} \right.} } \right.\\
&\left. {\left. { - \left| {{e^{{I_{\max }}\left( {1 - {\gamma ^{\rm{w}}}/{\gamma ^{\rm{s}}}} \right)}}{{\left( {\frac{{{\gamma ^{\rm{w}}}}}{{{\gamma ^{\rm{s}}}}}} \right)}^n} - {q_{n{\rm{sv}}}} - \left( {1 - {q_{n{\rm{sv}}}}} \right){e^{{I_{\max }}\left( {1 - {\gamma ^{\rm{v}}}/{\gamma ^{\rm{s}}}} \right)}}{{\left( {\frac{{{\gamma ^{\rm{v}}}}}{{{\gamma ^{\rm{s}}}}}} \right)}^n}} \right|} \right]} \right\}.
\end{array}
\end{equation}
\end{widetext}

\section*{Appendix F. Simulation results for the secret key rate ratio}
\begin{figure}[!t]
 \includegraphics*[scale=0.28]{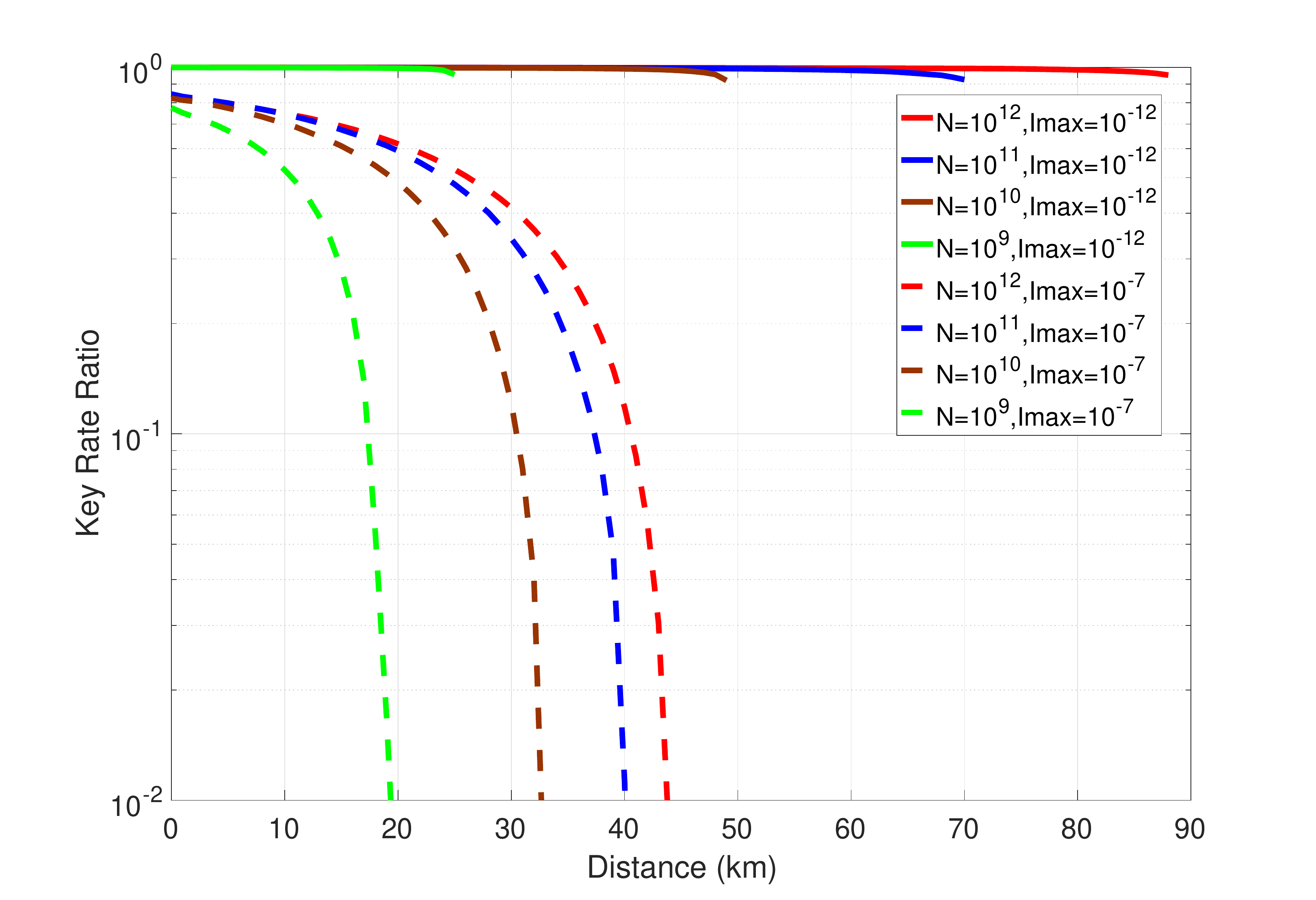}
\caption{ \footnotesize {The ratio ($\ell'_{I_{\rm max}>0}/\ell'_{I_{\rm max}=0}$) between the secret key rates in logarithmic scale with and without information leakage as a function of the distance for two fixed positive values of $I_{\rm max}=\{10^{-7},~10^{-12}\}$. The solid (dashed) lines represent the case $I_{\rm max}=10^{-12}$ ($I_{\rm max}=10^{-7}$). Different coloured lines correspond to different values of $N$. }} \label{ratiocase1}
\end{figure}
To further illustrate the effect that the information leakage has on the secret key rate as a function of the finite number of transmitted pulses, in Fig.~\ref{ratiocase1} we plot the ratio between the secret key rates for two fixed values of information leakage, $I_{\rm max}=\{10^{-7},~10^{-12}\}$ and when $I_{\rm max}=0$ (i.e., when there is no information leakage). Here we consider the scenario analyzed in Case 1 in the main text and, for simplicity, we disregard the information leakage from the PM. The simulation results for the other two cases are analogous. In particular, we see that when $I_{\rm max}$ is as small as $10^{-12}$, this ratio is always very close to one at all achievable distances independently of the value of $N$. This means that the effect of information leakage can be basically neglected. However, when the amount of information leakage increases to $I_{\rm max}=10^{-7}$, then the ratio is considerably less than one and drops quicker as the distance increases. In addition, our simulation results seem to indicate that for a fixed value of $I_{\rm max}$ (say $I_{\rm max}=10^{-7}$) and for a fixed distance, the smaller the value of $N$ is, the lower the key rate ratio is. That is, the effect of information leakage increases when $N$ decreases, as expected.

\bibliographystyle{apsrev4-1}
\bibliography{Bib}

%merlin.mbs apsrev4-1.bst 2010-07-25 4.21a (PWD, AO, DPC) hacked
%Control: key (0)
%Control: author (72) initials jnrlst
%Control: editor formatted (1) identically to author
%Control: production of article title (-1) disabled
%Control: page (0) single
%Control: year (1) truncated
%Control: production of eprint (0) enabled
\begin{thebibliography}{41}%
\makeatletter
\providecommand \@ifxundefined [1]{%
 \@ifx{#1\undefined}
}%
\providecommand \@ifnum [1]{%
 \ifnum #1\expandafter \@firstoftwo
 \else \expandafter \@secondoftwo
 \fi
}%
\providecommand \@ifx [1]{%
 \ifx #1\expandafter \@firstoftwo
 \else \expandafter \@secondoftwo
 \fi
}%
\providecommand \natexlab [1]{#1}%
\providecommand \enquote  [1]{``#1''}%
\providecommand \bibnamefont  [1]{#1}%
\providecommand \bibfnamefont [1]{#1}%
\providecommand \citenamefont [1]{#1}%
\providecommand \href@noop [0]{\@secondoftwo}%
\providecommand \href [0]{\begingroup \@sanitize@url \@href}%
\providecommand \@href[1]{\@@startlink{#1}\@@href}%
\providecommand \@@href[1]{\endgroup#1\@@endlink}%
\providecommand \@sanitize@url [0]{\catcode `\\12\catcode `\$12\catcode
  `\&12\catcode `\#12\catcode `\^12\catcode `\_12\catcode `\%12\relax}%
\providecommand \@@startlink[1]{}%
\providecommand \@@endlink[0]{}%
\providecommand \url  [0]{\begingroup\@sanitize@url \@url }%
\providecommand \@url [1]{\endgroup\@href {#1}{\urlprefix }}%
\providecommand \urlprefix  [0]{URL }%
\providecommand \Eprint [0]{\href }%
\providecommand \doibase [0]{http://dx.doi.org/}%
\providecommand \selectlanguage [0]{\@gobble}%
\providecommand \bibinfo  [0]{\@secondoftwo}%
\providecommand \bibfield  [0]{\@secondoftwo}%
\providecommand \translation [1]{[#1]}%
\providecommand \BibitemOpen [0]{}%
\providecommand \bibitemStop [0]{}%
\providecommand \bibitemNoStop [0]{.\EOS\space}%
\providecommand \EOS [0]{\spacefactor3000\relax}%
\providecommand \BibitemShut  [1]{\csname bibitem#1\endcsname}%
\let\auto@bib@innerbib\@empty
%</preamble>
\bibitem [{\citenamefont {Tamaki}\ \emph {et~al.}(2016)\citenamefont {Tamaki},
  \citenamefont {Curty},\ and\ \citenamefont {Lucamarini}}]{tamaki2016decoy}%
  \BibitemOpen
  \bibfield  {author} {\bibinfo {author} {\bibfnamefont {K.}~\bibnamefont
  {Tamaki}}, \bibinfo {author} {\bibfnamefont {M.}~\bibnamefont {Curty}}, \
  and\ \bibinfo {author} {\bibfnamefont {M.}~\bibnamefont {Lucamarini}},\
  }\href@noop {} {\bibfield  {journal} {\bibinfo  {journal} {New Journal of
  Physics}\ }\textbf {\bibinfo {volume} {18}},\ \bibinfo {pages} {065008}
  (\bibinfo {year} {2016})}\BibitemShut {NoStop}%
\bibitem [{\citenamefont {Bennett}\ and\ \citenamefont
  {Brassard}(1984)}]{bennett1984quantum}%
  \BibitemOpen
  \bibfield  {author} {\bibinfo {author} {\bibfnamefont {C.~H.}\ \bibnamefont
  {Bennett}}\ and\ \bibinfo {author} {\bibfnamefont {G.}~\bibnamefont
  {Brassard}},\ }in\ \href@noop {} {\emph {\bibinfo {booktitle} {International
  Conference on Computer System and Signal Processing, IEEE}}}\ (\bibinfo
  {year} {1984})\ pp.\ \bibinfo {pages} {175--179}\BibitemShut {NoStop}%
\bibitem [{\citenamefont {Scarani}\ \emph {et~al.}(2009)\citenamefont
  {Scarani}, \citenamefont {Bechmann-Pasquinucci}, \citenamefont {Cerf},
  \citenamefont {Du{\v{s}}ek}, \citenamefont {L{\"u}tkenhaus},\ and\
  \citenamefont {Peev}}]{scarani2009security}%
  \BibitemOpen
  \bibfield  {author} {\bibinfo {author} {\bibfnamefont {V.}~\bibnamefont
  {Scarani}}, \bibinfo {author} {\bibfnamefont {H.}~\bibnamefont
  {Bechmann-Pasquinucci}}, \bibinfo {author} {\bibfnamefont {N.~J.}\
  \bibnamefont {Cerf}}, \bibinfo {author} {\bibfnamefont {M.}~\bibnamefont
  {Du{\v{s}}ek}}, \bibinfo {author} {\bibfnamefont {N.}~\bibnamefont
  {L{\"u}tkenhaus}}, \ and\ \bibinfo {author} {\bibfnamefont {M.}~\bibnamefont
  {Peev}},\ }\href@noop {} {\bibfield  {journal} {\bibinfo  {journal} {Reviews
  of Modern Physics}\ }\textbf {\bibinfo {volume} {81}},\ \bibinfo {pages}
  {1301} (\bibinfo {year} {2009})}\BibitemShut {NoStop}%
\bibitem [{\citenamefont {Lo}\ \emph {et~al.}(2014)\citenamefont {Lo},
  \citenamefont {Curty},\ and\ \citenamefont {Tamaki}}]{lo2014secure}%
  \BibitemOpen
  \bibfield  {author} {\bibinfo {author} {\bibfnamefont {H.-K.}\ \bibnamefont
  {Lo}}, \bibinfo {author} {\bibfnamefont {M.}~\bibnamefont {Curty}}, \ and\
  \bibinfo {author} {\bibfnamefont {K.}~\bibnamefont {Tamaki}},\ }\href@noop {}
  {\bibfield  {journal} {\bibinfo  {journal} {Nature Photonics}\ }\textbf
  {\bibinfo {volume} {8}},\ \bibinfo {pages} {595} (\bibinfo {year}
  {2014})}\BibitemShut {NoStop}%
\bibitem [{\citenamefont {Gisin}\ \emph {et~al.}(2006)\citenamefont {Gisin},
  \citenamefont {Fasel}, \citenamefont {Kraus}, \citenamefont {Zbinden},\ and\
  \citenamefont {Ribordy}}]{gisin2006trojan}%
  \BibitemOpen
  \bibfield  {author} {\bibinfo {author} {\bibfnamefont {N.}~\bibnamefont
  {Gisin}}, \bibinfo {author} {\bibfnamefont {S.}~\bibnamefont {Fasel}},
  \bibinfo {author} {\bibfnamefont {B.}~\bibnamefont {Kraus}}, \bibinfo
  {author} {\bibfnamefont {H.}~\bibnamefont {Zbinden}}, \ and\ \bibinfo
  {author} {\bibfnamefont {G.}~\bibnamefont {Ribordy}},\ }\href@noop {}
  {\bibfield  {journal} {\bibinfo  {journal} {Physical Review A}\ }\textbf
  {\bibinfo {volume} {73}},\ \bibinfo {pages} {022320} (\bibinfo {year}
  {2006})}\BibitemShut {NoStop}%
\bibitem [{\citenamefont {Jain}\ \emph {et~al.}(2015)\citenamefont {Jain},
  \citenamefont {Stiller}, \citenamefont {Khan}, \citenamefont {Makarov},
  \citenamefont {Marquardt},\ and\ \citenamefont {Leuchs}}]{jain2015risk}%
  \BibitemOpen
  \bibfield  {author} {\bibinfo {author} {\bibfnamefont {N.}~\bibnamefont
  {Jain}}, \bibinfo {author} {\bibfnamefont {B.}~\bibnamefont {Stiller}},
  \bibinfo {author} {\bibfnamefont {I.}~\bibnamefont {Khan}}, \bibinfo {author}
  {\bibfnamefont {V.}~\bibnamefont {Makarov}}, \bibinfo {author} {\bibfnamefont
  {C.}~\bibnamefont {Marquardt}}, \ and\ \bibinfo {author} {\bibfnamefont
  {G.}~\bibnamefont {Leuchs}},\ }\href@noop {} {\bibfield  {journal} {\bibinfo
  {journal} {IEEE Journal of Selected Topics in Quantum Electronics}\ }\textbf
  {\bibinfo {volume} {21}},\ \bibinfo {pages} {168} (\bibinfo {year}
  {2015})}\BibitemShut {NoStop}%
\bibitem [{\citenamefont {Lucamarini}\ \emph {et~al.}(2015)\citenamefont
  {Lucamarini}, \citenamefont {Choi}, \citenamefont {Ward}, \citenamefont
  {Dynes}, \citenamefont {Yuan},\ and\ \citenamefont
  {Shields}}]{lucamarini2015practical}%
  \BibitemOpen
  \bibfield  {author} {\bibinfo {author} {\bibfnamefont {M.}~\bibnamefont
  {Lucamarini}}, \bibinfo {author} {\bibfnamefont {I.}~\bibnamefont {Choi}},
  \bibinfo {author} {\bibfnamefont {M.~B.}\ \bibnamefont {Ward}}, \bibinfo
  {author} {\bibfnamefont {J.~F.}\ \bibnamefont {Dynes}}, \bibinfo {author}
  {\bibfnamefont {Z.}~\bibnamefont {Yuan}}, \ and\ \bibinfo {author}
  {\bibfnamefont {A.~J.}\ \bibnamefont {Shields}},\ }\href@noop {} {\bibfield
  {journal} {\bibinfo  {journal} {Physical Review X}\ }\textbf {\bibinfo
  {volume} {5}},\ \bibinfo {pages} {031030} (\bibinfo {year}
  {2015})}\BibitemShut {NoStop}%
\bibitem [{\citenamefont {Lo}\ and\ \citenamefont
  {Preskill}(2007)}]{lo2007security}%
  \BibitemOpen
  \bibfield  {author} {\bibinfo {author} {\bibfnamefont {H.-K.}\ \bibnamefont
  {Lo}}\ and\ \bibinfo {author} {\bibfnamefont {J.}~\bibnamefont {Preskill}},\
  }\href@noop {} {\bibfield  {journal} {\bibinfo  {journal} {Quantum
  Information \& Computation}\ }\textbf {\bibinfo {volume} {7}},\ \bibinfo
  {pages} {431} (\bibinfo {year} {2007})}\BibitemShut {NoStop}%
\bibitem [{\citenamefont {Hwang}(2003)}]{hwang2003quantum}%
  \BibitemOpen
  \bibfield  {author} {\bibinfo {author} {\bibfnamefont {W.-Y.}\ \bibnamefont
  {Hwang}},\ }\href@noop {} {\bibfield  {journal} {\bibinfo  {journal}
  {Physical Review Letters}\ }\textbf {\bibinfo {volume} {91}},\ \bibinfo
  {pages} {057901} (\bibinfo {year} {2003})}\BibitemShut {NoStop}%
\bibitem [{\citenamefont {Lo}\ \emph {et~al.}(2005{\natexlab{a}})\citenamefont
  {Lo}, \citenamefont {Ma},\ and\ \citenamefont {Chen}}]{lo2005decoy}%
  \BibitemOpen
  \bibfield  {author} {\bibinfo {author} {\bibfnamefont {H.-K.}\ \bibnamefont
  {Lo}}, \bibinfo {author} {\bibfnamefont {X.}~\bibnamefont {Ma}}, \ and\
  \bibinfo {author} {\bibfnamefont {K.}~\bibnamefont {Chen}},\ }\href@noop {}
  {\bibfield  {journal} {\bibinfo  {journal} {Physical Review Letters}\
  }\textbf {\bibinfo {volume} {94}},\ \bibinfo {pages} {230504} (\bibinfo
  {year} {2005}{\natexlab{a}})}\BibitemShut {NoStop}%
\bibitem [{\citenamefont {Wang}(2005)}]{wang2005beating}%
  \BibitemOpen
  \bibfield  {author} {\bibinfo {author} {\bibfnamefont {X.-B.}\ \bibnamefont
  {Wang}},\ }\href@noop {} {\bibfield  {journal} {\bibinfo  {journal} {Physical
  Review Letters}\ }\textbf {\bibinfo {volume} {94}},\ \bibinfo {pages}
  {230503} (\bibinfo {year} {2005})}\BibitemShut {NoStop}%
\bibitem [{\citenamefont {Tomamichel}\ \emph {et~al.}(2012)\citenamefont
  {Tomamichel}, \citenamefont {Lim}, \citenamefont {Gisin},\ and\ \citenamefont
  {Renner}}]{tomamichel2012tight}%
  \BibitemOpen
  \bibfield  {author} {\bibinfo {author} {\bibfnamefont {M.}~\bibnamefont
  {Tomamichel}}, \bibinfo {author} {\bibfnamefont {C.~C.~W.}\ \bibnamefont
  {Lim}}, \bibinfo {author} {\bibfnamefont {N.}~\bibnamefont {Gisin}}, \ and\
  \bibinfo {author} {\bibfnamefont {R.}~\bibnamefont {Renner}},\ }\href@noop {}
  {\bibfield  {journal} {\bibinfo  {journal} {Nature Communications}\ }\textbf
  {\bibinfo {volume} {3}},\ \bibinfo {pages} {634} (\bibinfo {year}
  {2012})}\BibitemShut {NoStop}%
\bibitem [{\citenamefont {Hayashi}\ and\ \citenamefont
  {Tsurumaru}(2012)}]{hayashi2012concise}%
  \BibitemOpen
  \bibfield  {author} {\bibinfo {author} {\bibfnamefont {M.}~\bibnamefont
  {Hayashi}}\ and\ \bibinfo {author} {\bibfnamefont {T.}~\bibnamefont
  {Tsurumaru}},\ }\href@noop {} {\bibfield  {journal} {\bibinfo  {journal} {New
  Journal of Physics}\ }\textbf {\bibinfo {volume} {14}},\ \bibinfo {pages}
  {093014} (\bibinfo {year} {2012})}\BibitemShut {NoStop}%
\bibitem [{\citenamefont {Curty}\ \emph {et~al.}(2014)\citenamefont {Curty},
  \citenamefont {Xu}, \citenamefont {Cui}, \citenamefont {Lim}, \citenamefont
  {Tamaki},\ and\ \citenamefont {Lo}}]{curty2014finite}%
  \BibitemOpen
  \bibfield  {author} {\bibinfo {author} {\bibfnamefont {M.}~\bibnamefont
  {Curty}}, \bibinfo {author} {\bibfnamefont {F.}~\bibnamefont {Xu}}, \bibinfo
  {author} {\bibfnamefont {W.}~\bibnamefont {Cui}}, \bibinfo {author}
  {\bibfnamefont {C.~C.~W.}\ \bibnamefont {Lim}}, \bibinfo {author}
  {\bibfnamefont {K.}~\bibnamefont {Tamaki}}, \ and\ \bibinfo {author}
  {\bibfnamefont {H.-K.}\ \bibnamefont {Lo}},\ }\href@noop {} {\bibfield
  {journal} {\bibinfo  {journal} {Nature communications}\ }\textbf {\bibinfo
  {volume} {5}},\ \bibinfo {pages} {3732} (\bibinfo {year} {2014})}\BibitemShut
  {NoStop}%
\bibitem [{\citenamefont {Lim}\ \emph {et~al.}(2014)\citenamefont {Lim},
  \citenamefont {Curty}, \citenamefont {Walenta}, \citenamefont {Xu},\ and\
  \citenamefont {Zbinden}}]{lim2014concise}%
  \BibitemOpen
  \bibfield  {author} {\bibinfo {author} {\bibfnamefont {C.~C.~W.}\
  \bibnamefont {Lim}}, \bibinfo {author} {\bibfnamefont {M.}~\bibnamefont
  {Curty}}, \bibinfo {author} {\bibfnamefont {N.}~\bibnamefont {Walenta}},
  \bibinfo {author} {\bibfnamefont {F.}~\bibnamefont {Xu}}, \ and\ \bibinfo
  {author} {\bibfnamefont {H.}~\bibnamefont {Zbinden}},\ }\href@noop {}
  {\bibfield  {journal} {\bibinfo  {journal} {Physical Review A}\ }\textbf
  {\bibinfo {volume} {89}},\ \bibinfo {pages} {022307} (\bibinfo {year}
  {2014})}\BibitemShut {NoStop}%
\bibitem [{\citenamefont {Mizutani}\ \emph {et~al.}(2015)\citenamefont
  {Mizutani}, \citenamefont {Curty}, \citenamefont {Lim}, \citenamefont
  {Imoto},\ and\ \citenamefont {Tamaki}}]{mizutani2015finite}%
  \BibitemOpen
  \bibfield  {author} {\bibinfo {author} {\bibfnamefont {A.}~\bibnamefont
  {Mizutani}}, \bibinfo {author} {\bibfnamefont {M.}~\bibnamefont {Curty}},
  \bibinfo {author} {\bibfnamefont {C.~C.~W.}\ \bibnamefont {Lim}}, \bibinfo
  {author} {\bibfnamefont {N.}~\bibnamefont {Imoto}}, \ and\ \bibinfo {author}
  {\bibfnamefont {K.}~\bibnamefont {Tamaki}},\ }\href@noop {} {\bibfield
  {journal} {\bibinfo  {journal} {New Journal of Physics}\ }\textbf {\bibinfo
  {volume} {17}},\ \bibinfo {pages} {093011} (\bibinfo {year}
  {2015})}\BibitemShut {NoStop}%
\bibitem [{\citenamefont {Lo}\ \emph {et~al.}(2005{\natexlab{b}})\citenamefont
  {Lo}, \citenamefont {Chau},\ and\ \citenamefont
  {Ardehali}}]{lo2005efficient}%
  \BibitemOpen
  \bibfield  {author} {\bibinfo {author} {\bibfnamefont {H.-K.}\ \bibnamefont
  {Lo}}, \bibinfo {author} {\bibfnamefont {H.~F.}\ \bibnamefont {Chau}}, \ and\
  \bibinfo {author} {\bibfnamefont {M.}~\bibnamefont {Ardehali}},\ }\href@noop
  {} {\bibfield  {journal} {\bibinfo  {journal} {Journal of Cryptology}\
  }\textbf {\bibinfo {volume} {18}},\ \bibinfo {pages} {133} (\bibinfo {year}
  {2005}{\natexlab{b}})}\BibitemShut {NoStop}%
\bibitem [{\citenamefont {Wei}\ \emph {et~al.}(2013)\citenamefont {Wei},
  \citenamefont {Wang}, \citenamefont {Zhang}, \citenamefont {Gao},
  \citenamefont {Ma},\ and\ \citenamefont {Ma}}]{wei2013decoy}%
  \BibitemOpen
  \bibfield  {author} {\bibinfo {author} {\bibfnamefont {Z.}~\bibnamefont
  {Wei}}, \bibinfo {author} {\bibfnamefont {W.}~\bibnamefont {Wang}}, \bibinfo
  {author} {\bibfnamefont {Z.}~\bibnamefont {Zhang}}, \bibinfo {author}
  {\bibfnamefont {M.}~\bibnamefont {Gao}}, \bibinfo {author} {\bibfnamefont
  {Z.}~\bibnamefont {Ma}}, \ and\ \bibinfo {author} {\bibfnamefont
  {X.}~\bibnamefont {Ma}},\ }\href@noop {} {\bibfield  {journal} {\bibinfo
  {journal} {Scientific Reports}\ }\textbf {\bibinfo {volume} {3}},\ \bibinfo
  {pages} {2453} (\bibinfo {year} {2013})}\BibitemShut {NoStop}%
\bibitem [{\citenamefont {Zhao}\ \emph {et~al.}(2006)\citenamefont {Zhao},
  \citenamefont {Qi}, \citenamefont {Ma}, \citenamefont {Lo},\ and\
  \citenamefont {Qian}}]{zhao2006experimental}%
  \BibitemOpen
  \bibfield  {author} {\bibinfo {author} {\bibfnamefont {Y.}~\bibnamefont
  {Zhao}}, \bibinfo {author} {\bibfnamefont {B.}~\bibnamefont {Qi}}, \bibinfo
  {author} {\bibfnamefont {X.}~\bibnamefont {Ma}}, \bibinfo {author}
  {\bibfnamefont {H.-K.}\ \bibnamefont {Lo}}, \ and\ \bibinfo {author}
  {\bibfnamefont {L.}~\bibnamefont {Qian}},\ }\href@noop {} {\bibfield
  {journal} {\bibinfo  {journal} {Physical Review Letters}\ }\textbf {\bibinfo
  {volume} {96}},\ \bibinfo {pages} {070502} (\bibinfo {year}
  {2006})}\BibitemShut {NoStop}%
\bibitem [{\citenamefont {Peng}\ \emph {et~al.}(2007)\citenamefont {Peng},
  \citenamefont {Zhang}, \citenamefont {Yang}, \citenamefont {Gao},
  \citenamefont {Ma}, \citenamefont {Yin}, \citenamefont {Zeng}, \citenamefont
  {Yang}, \citenamefont {Wang},\ and\ \citenamefont
  {Pan}}]{peng2007experimental}%
  \BibitemOpen
  \bibfield  {author} {\bibinfo {author} {\bibfnamefont {C.-Z.}\ \bibnamefont
  {Peng}}, \bibinfo {author} {\bibfnamefont {J.}~\bibnamefont {Zhang}},
  \bibinfo {author} {\bibfnamefont {D.}~\bibnamefont {Yang}}, \bibinfo {author}
  {\bibfnamefont {W.-B.}\ \bibnamefont {Gao}}, \bibinfo {author} {\bibfnamefont
  {H.-X.}\ \bibnamefont {Ma}}, \bibinfo {author} {\bibfnamefont
  {H.}~\bibnamefont {Yin}}, \bibinfo {author} {\bibfnamefont {H.-P.}\
  \bibnamefont {Zeng}}, \bibinfo {author} {\bibfnamefont {T.}~\bibnamefont
  {Yang}}, \bibinfo {author} {\bibfnamefont {X.-B.}\ \bibnamefont {Wang}}, \
  and\ \bibinfo {author} {\bibfnamefont {J.-W.}\ \bibnamefont {Pan}},\
  }\href@noop {} {\bibfield  {journal} {\bibinfo  {journal} {Physical Review
  Letters}\ }\textbf {\bibinfo {volume} {98}},\ \bibinfo {pages} {010505}
  (\bibinfo {year} {2007})}\BibitemShut {NoStop}%
\bibitem [{\citenamefont {Schmitt-Manderbach}\ \emph
  {et~al.}(2007)\citenamefont {Schmitt-Manderbach} \emph
  {et~al.}}]{schmitt2007experimental}%
  \BibitemOpen
  \bibfield  {author} {\bibinfo {author} {\bibnamefont {Schmitt-Manderbach}}
  \emph {et~al.},\ }\href@noop {} {\bibfield  {journal} {\bibinfo  {journal}
  {Physical Review Letters}\ }\textbf {\bibinfo {volume} {98}},\ \bibinfo
  {pages} {010504} (\bibinfo {year} {2007})}\BibitemShut {NoStop}%
\bibitem [{\citenamefont {Yuan}\ \emph {et~al.}(2007)\citenamefont {Yuan},
  \citenamefont {Sharpe},\ and\ \citenamefont
  {Shields}}]{yuan2007unconditionally}%
  \BibitemOpen
  \bibfield  {author} {\bibinfo {author} {\bibfnamefont {Z.}~\bibnamefont
  {Yuan}}, \bibinfo {author} {\bibfnamefont {A.}~\bibnamefont {Sharpe}}, \ and\
  \bibinfo {author} {\bibfnamefont {A.}~\bibnamefont {Shields}},\ }\href@noop
  {} {\bibfield  {journal} {\bibinfo  {journal} {Applied Physics Letters}\
  }\textbf {\bibinfo {volume} {90}},\ \bibinfo {pages} {011118} (\bibinfo
  {year} {2007})}\BibitemShut {NoStop}%
\bibitem [{\citenamefont {Rosenberg}\ \emph {et~al.}(2007)\citenamefont
  {Rosenberg}, \citenamefont {Harrington}, \citenamefont {Rice}, \citenamefont
  {Hiskett}, \citenamefont {Peterson}, \citenamefont {Hughes}, \citenamefont
  {Lita}, \citenamefont {Nam},\ and\ \citenamefont
  {Nordholt}}]{rosenberg2007long}%
  \BibitemOpen
  \bibfield  {author} {\bibinfo {author} {\bibfnamefont {D.}~\bibnamefont
  {Rosenberg}}, \bibinfo {author} {\bibfnamefont {J.~W.}\ \bibnamefont
  {Harrington}}, \bibinfo {author} {\bibfnamefont {P.~R.}\ \bibnamefont
  {Rice}}, \bibinfo {author} {\bibfnamefont {P.~A.}\ \bibnamefont {Hiskett}},
  \bibinfo {author} {\bibfnamefont {C.~G.}\ \bibnamefont {Peterson}}, \bibinfo
  {author} {\bibfnamefont {R.~J.}\ \bibnamefont {Hughes}}, \bibinfo {author}
  {\bibfnamefont {A.~E.}\ \bibnamefont {Lita}}, \bibinfo {author}
  {\bibfnamefont {S.~W.}\ \bibnamefont {Nam}}, \ and\ \bibinfo {author}
  {\bibfnamefont {J.~E.}\ \bibnamefont {Nordholt}},\ }\href@noop {} {\bibfield
  {journal} {\bibinfo  {journal} {Physical Review Letters}\ }\textbf {\bibinfo
  {volume} {98}},\ \bibinfo {pages} {010503} (\bibinfo {year}
  {2007})}\BibitemShut {NoStop}%
\bibitem [{\citenamefont {Liu}\ \emph {et~al.}(2010)\citenamefont {Liu} \emph
  {et~al.}}]{liu2010decoy}%
  \BibitemOpen
  \bibfield  {author} {\bibinfo {author} {\bibfnamefont {Y.}~\bibnamefont
  {Liu}} \emph {et~al.},\ }\href@noop {} {\bibfield  {journal} {\bibinfo
  {journal} {Optics Express}\ }\textbf {\bibinfo {volume} {18}},\ \bibinfo
  {pages} {8587} (\bibinfo {year} {2010})}\BibitemShut {NoStop}%
\bibitem [{\citenamefont {Fr{\"o}hlich}\ \emph {et~al.}(2017)\citenamefont
  {Fr{\"o}hlich}, \citenamefont {Lucamarini}, \citenamefont {Dynes},
  \citenamefont {Comandar}, \citenamefont {Tam}, \citenamefont {Plews},
  \citenamefont {Sharpe}, \citenamefont {Yuan},\ and\ \citenamefont
  {Shields}}]{frohlich2017long}%
  \BibitemOpen
  \bibfield  {author} {\bibinfo {author} {\bibfnamefont {B.}~\bibnamefont
  {Fr{\"o}hlich}}, \bibinfo {author} {\bibfnamefont {M.}~\bibnamefont
  {Lucamarini}}, \bibinfo {author} {\bibfnamefont {J.~F.}\ \bibnamefont
  {Dynes}}, \bibinfo {author} {\bibfnamefont {L.~C.}\ \bibnamefont {Comandar}},
  \bibinfo {author} {\bibfnamefont {W.~W.-S.}\ \bibnamefont {Tam}}, \bibinfo
  {author} {\bibfnamefont {A.}~\bibnamefont {Plews}}, \bibinfo {author}
  {\bibfnamefont {A.~W.}\ \bibnamefont {Sharpe}}, \bibinfo {author}
  {\bibfnamefont {Z.}~\bibnamefont {Yuan}}, \ and\ \bibinfo {author}
  {\bibfnamefont {A.~J.}\ \bibnamefont {Shields}},\ }\href@noop {} {\bibfield
  {journal} {\bibinfo  {journal} {Optica}\ }\textbf {\bibinfo {volume} {4}},\
  \bibinfo {pages} {163} (\bibinfo {year} {2017})}\BibitemShut {NoStop}%
\bibitem [{\citenamefont {Azuma}(1967)}]{azuma1967weighted}%
  \BibitemOpen
  \bibfield  {author} {\bibinfo {author} {\bibfnamefont {K.}~\bibnamefont
  {Azuma}},\ }\href@noop {} {\bibfield  {journal} {\bibinfo  {journal} {Tohoku
  Mathematical Journal, Second Series}\ }\textbf {\bibinfo {volume} {19}},\
  \bibinfo {pages} {357} (\bibinfo {year} {1967})}\BibitemShut {NoStop}%
\bibitem [{\citenamefont {Gottesman}\ \emph {et~al.}(2004)\citenamefont
  {Gottesman}, \citenamefont {Lo}, \citenamefont {L{\"u}tkenhaus},\ and\
  \citenamefont {Preskill}}]{gottesman2004security}%
  \BibitemOpen
  \bibfield  {author} {\bibinfo {author} {\bibfnamefont {D.}~\bibnamefont
  {Gottesman}}, \bibinfo {author} {\bibfnamefont {H.-K.}\ \bibnamefont {Lo}},
  \bibinfo {author} {\bibfnamefont {N.}~\bibnamefont {L{\"u}tkenhaus}}, \ and\
  \bibinfo {author} {\bibfnamefont {J.}~\bibnamefont {Preskill}},\ }\href@noop
  {} {\bibfield  {journal} {\bibinfo  {journal} {Quantum Information \&
  Computation}\ }\textbf {\bibinfo {volume} {5}} (\bibinfo {year}
  {2004})}\BibitemShut {NoStop}%
\bibitem [{\citenamefont {Koashi}(2005)}]{koashi2005simple}%
  \BibitemOpen
  \bibfield  {author} {\bibinfo {author} {\bibfnamefont {M.}~\bibnamefont
  {Koashi}},\ }\href@noop {} {\bibfield  {journal} {\bibinfo  {journal} {arXiv
  preprint quant-ph/0505108}\ } (\bibinfo {year} {2005})}\BibitemShut {NoStop}%
\bibitem [{\citenamefont {Huttner}\ \emph {et~al.}(1995)\citenamefont
  {Huttner}, \citenamefont {Imoto}, \citenamefont {Gisin},\ and\ \citenamefont
  {Mor}}]{huttner1995quantum}%
  \BibitemOpen
  \bibfield  {author} {\bibinfo {author} {\bibfnamefont {B.}~\bibnamefont
  {Huttner}}, \bibinfo {author} {\bibfnamefont {N.}~\bibnamefont {Imoto}},
  \bibinfo {author} {\bibfnamefont {N.}~\bibnamefont {Gisin}}, \ and\ \bibinfo
  {author} {\bibfnamefont {T.}~\bibnamefont {Mor}},\ }\href@noop {} {\bibfield
  {journal} {\bibinfo  {journal} {Physical Review A}\ }\textbf {\bibinfo
  {volume} {51}},\ \bibinfo {pages} {1863} (\bibinfo {year}
  {1995})}\BibitemShut {NoStop}%
\bibitem [{\citenamefont {Brassard}\ \emph {et~al.}(2000)\citenamefont
  {Brassard}, \citenamefont {L{\"u}tkenhaus}, \citenamefont {Mor},\ and\
  \citenamefont {Sanders}}]{brassard2000limitations}%
  \BibitemOpen
  \bibfield  {author} {\bibinfo {author} {\bibfnamefont {G.}~\bibnamefont
  {Brassard}}, \bibinfo {author} {\bibfnamefont {N.}~\bibnamefont
  {L{\"u}tkenhaus}}, \bibinfo {author} {\bibfnamefont {T.}~\bibnamefont {Mor}},
  \ and\ \bibinfo {author} {\bibfnamefont {B.~C.}\ \bibnamefont {Sanders}},\
  }\href@noop {} {\bibfield  {journal} {\bibinfo  {journal} {Physical Review
  Letters}\ }\textbf {\bibinfo {volume} {85}},\ \bibinfo {pages} {1330}
  (\bibinfo {year} {2000})}\BibitemShut {NoStop}%
\bibitem [{\citenamefont {Pfister}\ \emph {et~al.}(2016)\citenamefont
  {Pfister}, \citenamefont {L{\"u}tkenhaus}, \citenamefont {Wehner},\ and\
  \citenamefont {Coles}}]{pfister2016sifting}%
  \BibitemOpen
  \bibfield  {author} {\bibinfo {author} {\bibfnamefont {C.}~\bibnamefont
  {Pfister}}, \bibinfo {author} {\bibfnamefont {N.}~\bibnamefont
  {L{\"u}tkenhaus}}, \bibinfo {author} {\bibfnamefont {S.}~\bibnamefont
  {Wehner}}, \ and\ \bibinfo {author} {\bibfnamefont {P.~J.}\ \bibnamefont
  {Coles}},\ }\href@noop {} {\bibfield  {journal} {\bibinfo  {journal} {New
  Journal of Physics}\ }\textbf {\bibinfo {volume} {18}},\ \bibinfo {pages}
  {053001} (\bibinfo {year} {2016})}\BibitemShut {NoStop}%
\bibitem [{\citenamefont {Tamaki}\ \emph {et~al.}(2018)\citenamefont {Tamaki},
  \citenamefont {Lo}, \citenamefont {Mizutani}, \citenamefont {Kato},
  \citenamefont {Lim}, \citenamefont {Azuma},\ and\ \citenamefont
  {Curty}}]{tamaki2018security}%
  \BibitemOpen
  \bibfield  {author} {\bibinfo {author} {\bibfnamefont {K.}~\bibnamefont
  {Tamaki}}, \bibinfo {author} {\bibfnamefont {H.-K.}\ \bibnamefont {Lo}},
  \bibinfo {author} {\bibfnamefont {A.}~\bibnamefont {Mizutani}}, \bibinfo
  {author} {\bibfnamefont {G.}~\bibnamefont {Kato}}, \bibinfo {author}
  {\bibfnamefont {C.~C.~W.}\ \bibnamefont {Lim}}, \bibinfo {author}
  {\bibfnamefont {K.}~\bibnamefont {Azuma}}, \ and\ \bibinfo {author}
  {\bibfnamefont {M.}~\bibnamefont {Curty}},\ }\href@noop {} {\bibfield
  {journal} {\bibinfo  {journal} {Quantum Science and Technology}\ }\textbf
  {\bibinfo {volume} {3}},\ \bibinfo {pages} {014002} (\bibinfo {year}
  {2018})}\BibitemShut {NoStop}%
\bibitem [{Note1()}]{Note1}%
  \BibitemOpen
  \bibinfo {note} {Note that Alice and Bob could also probabilistically
  post-select $N^j_{\protect \rm Z}$ ($N^j_{\protect \rm X}$) events from each
  set indexed by $Z^j$ ($X^j$) to fix the sizes of the sifted data sets. This
  might be convenient for some post-processing steps of the protocol, like
  information reconciliation and privacy amplification. Here, for simplicity,
  we do not consider such a post-selection. However, note that our analysis
  could be adapted as well to the situation where the the sizes of the sifted
  data sets are fixed $\protect \it a~priori$.}\BibitemShut {Stop}%
\bibitem [{\citenamefont {Ben-Or}\ \emph {et~al.}(2005)\citenamefont {Ben-Or},
  \citenamefont {Horodecki}, \citenamefont {Leung}, \citenamefont {Mayers},\
  and\ \citenamefont {Oppenheim}}]{ben2005universal}%
  \BibitemOpen
  \bibfield  {author} {\bibinfo {author} {\bibfnamefont {M.}~\bibnamefont
  {Ben-Or}}, \bibinfo {author} {\bibfnamefont {M.}~\bibnamefont {Horodecki}},
  \bibinfo {author} {\bibfnamefont {D.~W.}\ \bibnamefont {Leung}}, \bibinfo
  {author} {\bibfnamefont {D.}~\bibnamefont {Mayers}}, \ and\ \bibinfo {author}
  {\bibfnamefont {J.}~\bibnamefont {Oppenheim}},\ }in\ \href@noop {} {\emph
  {\bibinfo {booktitle} {TCC}}},\ Vol.~\bibinfo {volume} {5}\ (\bibinfo
  {organization} {Springer},\ \bibinfo {year} {2005})\ pp.\ \bibinfo {pages}
  {386--406}\BibitemShut {NoStop}%
\bibitem [{\citenamefont {Renner}\ and\ \citenamefont
  {K{\"o}nig}(2005)}]{renner2005universally}%
  \BibitemOpen
  \bibfield  {author} {\bibinfo {author} {\bibfnamefont {R.}~\bibnamefont
  {Renner}}\ and\ \bibinfo {author} {\bibfnamefont {R.}~\bibnamefont
  {K{\"o}nig}},\ }in\ \href@noop {} {\emph {\bibinfo {booktitle} {Theory of
  Cryptography Conference}}}\ (\bibinfo {organization} {Springer},\ \bibinfo
  {year} {2005})\ pp.\ \bibinfo {pages} {407--425}\BibitemShut {NoStop}%
\bibitem [{\citenamefont {Ma}\ \emph {et~al.}(2005)\citenamefont {Ma},
  \citenamefont {Qi}, \citenamefont {Zhao},\ and\ \citenamefont
  {Lo}}]{ma2005practical}%
  \BibitemOpen
  \bibfield  {author} {\bibinfo {author} {\bibfnamefont {X.}~\bibnamefont
  {Ma}}, \bibinfo {author} {\bibfnamefont {B.}~\bibnamefont {Qi}}, \bibinfo
  {author} {\bibfnamefont {Y.}~\bibnamefont {Zhao}}, \ and\ \bibinfo {author}
  {\bibfnamefont {H.-K.}\ \bibnamefont {Lo}},\ }\href@noop {} {\bibfield
  {journal} {\bibinfo  {journal} {Physical Review A}\ }\textbf {\bibinfo
  {volume} {72}},\ \bibinfo {pages} {012326} (\bibinfo {year}
  {2005})}\BibitemShut {NoStop}%
\bibitem [{\citenamefont {Nielsen}\ and\ \citenamefont
  {Chuang}(2000)}]{nielsen2000quantum}%
  \BibitemOpen
  \bibfield  {author} {\bibinfo {author} {\bibfnamefont {M.~A.}\ \bibnamefont
  {Nielsen}}\ and\ \bibinfo {author} {\bibfnamefont {I.~L.}\ \bibnamefont
  {Chuang}},\ }\href@noop {} {\bibfield  {journal} {\bibinfo  {journal}
  {Cambridge: Cambridge University Press}\ } (\bibinfo {year}
  {2000})}\BibitemShut {NoStop}%
\bibitem [{\citenamefont {Tamaki}\ \emph {et~al.}(2003)\citenamefont {Tamaki},
  \citenamefont {Koashi},\ and\ \citenamefont
  {Imoto}}]{tamaki2003unconditionally}%
  \BibitemOpen
  \bibfield  {author} {\bibinfo {author} {\bibfnamefont {K.}~\bibnamefont
  {Tamaki}}, \bibinfo {author} {\bibfnamefont {M.}~\bibnamefont {Koashi}}, \
  and\ \bibinfo {author} {\bibfnamefont {N.}~\bibnamefont {Imoto}},\
  }\href@noop {} {\bibfield  {journal} {\bibinfo  {journal} {Physical Review
  Letters}\ }\textbf {\bibinfo {volume} {90}},\ \bibinfo {pages} {167904}
  (\bibinfo {year} {2003})}\BibitemShut {NoStop}%
\bibitem [{\citenamefont {Chernoff}(1952)}]{chernoff1952measure}%
  \BibitemOpen
  \bibfield  {author} {\bibinfo {author} {\bibfnamefont {H.}~\bibnamefont
  {Chernoff}},\ }\href@noop {} {\bibfield  {journal} {\bibinfo  {journal} {The
  Annals of Mathematical Statistics}\ }\textbf {\bibinfo {volume} {23}},\
  \bibinfo {pages} {493} (\bibinfo {year} {1952})}\BibitemShut {NoStop}%
\bibitem [{\citenamefont {Tomamichel}\ \emph {et~al.}(2014)\citenamefont
  {Tomamichel}, \citenamefont {Martinez-Mateo}, \citenamefont {Pacher},\ and\
  \citenamefont {Elkouss}}]{tomamichel2014fundamental}%
  \BibitemOpen
  \bibfield  {author} {\bibinfo {author} {\bibfnamefont {M.}~\bibnamefont
  {Tomamichel}}, \bibinfo {author} {\bibfnamefont {J.}~\bibnamefont
  {Martinez-Mateo}}, \bibinfo {author} {\bibfnamefont {C.}~\bibnamefont
  {Pacher}}, \ and\ \bibinfo {author} {\bibfnamefont {D.}~\bibnamefont
  {Elkouss}},\ }in\ \href@noop {} {\emph {\bibinfo {booktitle} {IEEE
  International Symposium on Information Theory (ISIT),}}}\ (\bibinfo
  {organization} {IEEE},\ \bibinfo {year} {2014})\ pp.\ \bibinfo {pages}
  {1469--1473}\BibitemShut {NoStop}%
\bibitem [{\citenamefont {Serfling}(1974)}]{serfling1974probability}%
  \BibitemOpen
  \bibfield  {author} {\bibinfo {author} {\bibfnamefont {R.~J.}\ \bibnamefont
  {Serfling}},\ }\href@noop {} {\bibfield  {journal} {\bibinfo  {journal} {The
  Annals of Statistics}\ }\textbf {\bibinfo {volume} {2}},\ \bibinfo {pages}
  {39} (\bibinfo {year} {1974})}\BibitemShut {NoStop}%
\end{thebibliography}%


%merlin.mbs apsrev4-1.bst 2010-07-25 4.21a (PWD, AO, DPC) hacked
%Control: key (0)
%Control: author (72) initials jnrlst
%Control: editor formatted (1) identically to author
%Control: production of article title (-1) disabled
%Control: page (0) single
%Control: year (1) truncated
%Control: production of eprint (0) enabled
%

\end{document}